\documentclass{emulateapj}
\usepackage{apjfonts}

\newcommand\kms{km~s$^{-1}$}

\newcommand\etal{{et~al.}} 
\newcommand\mM{\ifmmode(m{-}M)\else$(m{-}M)$\fi}
\newcommand\msun{\ifmmode{\hbox{M$_\odot$}}\else{M$_\odot$}\fi}

\newcommand\hst{{\it HST}}

\newcommand\zacs{\ifmmode z_{850}\else$z_{850}$\fi}
\newcommand\iacs{\ifmmode i_{775}\else$i_{775}$\fi}
\newcommand\racs{\ifmmode r_{625}\else$r_{625}$\fi}
\newcommand\vacs{\ifmmode V_{606}\else$V_{606}$\fi}
\newcommand\ricolor{{\ifmmode{(r_{625}{-}i_{775})}\else$(r_{625}{-}i_{775})$\fi}}
\newcommand\rzcolor{{\ifmmode(r_{625}{-}z_{850})\else$(r_{625}{-}z_{850})$\fi}}
\newcommand\izcolor{{\ifmmode (i_{775}{-}z_{850}) \else $(i_{775}{-}z_{850})$ \fi}}
\newcommand\vzcolor{{\ifmmode(V_{606}{-}z_{850})\else$(V_{606}{-}z_{850})$\fi}}
\newcommand\vicolor{{\ifmmode(V_{606}{-}i_{775})\else$(V_{606}{-}i_{775})$\fi}}
\newcommand\izacs{\ifmmode(i_{775}{-}z_{850})\else$(i_{775}{-}z_{850})$\fi}
\newcommand\ubrest{\ifmmode(U{-}B)_z\else$(U{-}B)_z$\fi}
\newcommand\uvrest{\ifmmode(U{-}V)_z\else$(U{-}V)_z$\fi}
\newcommand\bn{$B$--$n$}
\newcommand\riclr{{\ifmmode{r_{625}{-}i_{775}}\else$r_{625}{-}i_{775}$\fi}}
\newcommand\rzclr{{\ifmmode{r_{625}{-}z_{850}}\else$r_{625}{-}z_{850}$\fi}}
\newcommand\izclr{{\ifmmode{i_{775}{-}z_{850}}\else$i_{775}{-}z_{850}$\fi}}
\newcommand\vzclr{{\ifmmode{V_{606}{-}z_{850}}\else$V_{606}{-}z_{850}$\fi}}
\newcommand\viclr{{\ifmmode{V_{606}{-}i_{775}}\else$V_{606}{-}i_{775}$\fi}}
\newcommand\msname{MS\,1054--03}
\newcommand\rxname{RX\,J0152.7--1357}
\newcommand\clname{\rxname}
\newcommand\zf{\ifmmode z_{\rm f}\else$z_{\rm f}$\fi}
\newcommand\lta{\lesssim}
\newcommand\gta{\gtrsim}
\newcommand\taul{\ifmmode\tau_{L}\else$\tau_{L}$\fi}
\def\txitxo{Ben\'{\i}tez}
\def\magauto{{\sc mag\_auto}}
\shortauthors{Blakeslee et al.}
\shorttitle{Clusters at Half Hubble Time}

\begin{document}
\slugcomment{ApJ, in press. Scheduled for 10 June 2006 issue.}

\title{Clusters at Half Hubble Time: Galaxy Structure and Colors in \\
RX\,J0152.7$-$1357 and MS\,1054$-$03\altaffilmark{1}}

\author{John~P.~Blakeslee\altaffilmark{2,3},
B.~P.~Holden\altaffilmark{4},
M.~Franx\altaffilmark{5},
P.~Rosati\altaffilmark{6},
R.~J.~Bouwens\altaffilmark{4},
R.~Demarco\altaffilmark{3},
H.~C.~Ford\altaffilmark{3},
N.~L.~Homeier\altaffilmark{3},
G.~D.~Illingworth\altaffilmark{4},
M.~J.~Jee\altaffilmark{3},
S.~Mei\altaffilmark{3},
F.~Menanteau\altaffilmark{3},
G.~R.~Meurer\altaffilmark{3},
M.~Postman\altaffilmark{7},
and
Kim-Vy H. Tran\altaffilmark{8,9}
}

\altaffiltext{1}{Based on observations made with the NASA/ESA Hubble
Space Telescope, obtained from the Space Telescope Science Institute,
which is operated by the Association of Universities for Research in
Astronomy, Inc., under NASA contract NAS\,5-26555.
These observations are associated with programs \#9290 and \#9919.}

\altaffiltext{2}{Department of Physics \& Astronomy, Washington State University, Pullman, WA 99164-2814 \,({\it Go Cougs!});~ jblakes@wsu.edu}
\altaffiltext{3}{Department of Physics \& Astronomy, Johns Hopkins University, Baltimore, MD 21218}
\altaffiltext{4}{Lick Observatory, University of California, Santa Cruz, CA 95064}
\altaffiltext{5}{Leiden Observatory, P.O. Box 9513, 2300 Leiden, The Netherlands}
\altaffiltext{6}{European Southern Observatory, Karl-Schwarzschild-Str. 2, D-85748 Garching, Germany}
\altaffiltext{7}{Space Telescope Science Institute, 3700 San Martin Drive, Baltimore, MD 21218}
\altaffiltext{8}{Harvard-Smithsonian Center for Astrophysics, 60 Garden Street,	Cambridge, MA 02138}
\altaffiltext{9}{NSF Astronomy \& Astrophysics Fellow}

\begin{abstract}
We study the photometric and structural properties of spectroscopically
confirmed members in the two massive X-ray--selected $z{\,\approx\,}0.83$
galaxy clusters \msname\ and \clname\ using three-band mosaic imaging
with the {\it Hubble Space Telescope} Advanced Camera for Surveys.
The samples include 105 and 140 members of \rxname\ and \msname,
respectively, with ACS F775W magnitude $\iacs < 24.0$.
We fit the 2-D galaxy light profiles to determine effective radii and
S\'ersic indices; deviations from the smooth profiles are quantified
by the ratio of the rms residuals to the mean of the galaxy model. 
Galaxies are then classified according to a combination of this
rms/mean ratio and the S\'ersic index; the resulting classes
correlate well with visually classified morphological types,
but are less affected by orientation.
We find the size--surface brightness relations in the two clusters to
be very similar, supporting recent results on the evolution of this
relationship with redshift.
We examine in detail the color--magnitude relations in these clusters 
and systematic effects on the residuals with respect to these relations.
The color-magnitude residuals correlate with the local
density, as measured from both galaxy numbers and weak lensing.
These correlations are
strongest for the full galaxy samples (commensurate with
the morphology--density relation), but are also present at lower 
significance levels for the early- and late-type samples individually.
Weaker correlations are found
with cluster radius, resulting from the more fundamental dependence
on local density. We identify a threshold surface
mass density of $\Sigma \approx 0.1$, in units of the critical density,
above which there are relatively few blue (star-forming) galaxies.
In \rxname, there is an offset of $0.006\pm0.002$ in the mean redshifts
of the early- and late-type galaxies, which produces a trend in the
color residuals with velocity and may result from an infalling foreground
association of late-type galaxies.
Comparison of the color--color diagrams for these clusters to stellar population
models implies that a range of star formation time-scales are needed to
reproduce the locus of galaxy colors.
We also identify two galaxies, one in each cluster,
whose colors can only be explained by large amounts, $A_V{\,\sim\,}$1 mag,
of internal dust extinction.  Converting to rest-frame bandpasses,
we find elliptical galaxy color scatters of $0.03\pm0.01$ mag in 
$(U{-}B)$ and  $0.07\pm0.01$ mag in $(U{-}V)$, indicating mean ages
of $\sim\,$3.5 Gyr, similar to the estimates from the mean
colors and  corresponding to formation at $z\approx2.2$.
Thus, when the universe was half its present age, 
cluster ellipticals were half the age of the universe at that epoch;
the same is coincidentally true of the median ages of ellipticals today. 
However, the most massive local cluster ellipticals have ages $\gta\,$10~Gyr,
consistent with our results for their likely progenitors at $z\gta0.8$.
\end{abstract}

\section{Introduction}


Galaxy clusters have played a central role in the development of
our understanding of the hierarchical growth of structure and the
interplay of stars, hot gas, cold dark matter, and active galactic
nuclei (AGN).  They are also valuable laboratories for studying
the evolution of galaxies, including galaxy mass assembly,
morphological transformation, gas depletion, and luminosity fading.
With recent advances in both ground- and space-based observatories,
we are now able to find and study galaxy clusters to redshifts well beyond
unity, and to lookback times approaching 70\% the age of the universe
(e.g., Rosati, Borgani, \& Norman 2002; Mullis \etal\ 2005).
The emerging picture is one in which the dominant early-type galaxy
population and the enriched intracluster medium formed early, 
at redshifts $z>2$, but the clusters themselves were still actively
assembling from smaller components at $z\sim1$, and lower mass 
late-type galaxies have continued to infall from the field and undergo 
cessation of star formation and morphological transformation down 
to the present day (e.g., Dressler \etal\ 1997; Balogh \etal\ 1999; 
Kodama \& Bower 2001; van Dokkum \& Franx 2001; Ettori \etal\ 2004;
Kodama \etal\ 2004; Poggianti \etal\ 2004; Tran \etal\ 2005b).
Poggianti (2004) gives a recent review of observations relevant to cluster
galaxy evolution and implications for the various proposed evolutionary mechanisms.

The ACS Intermediate Redshift Cluster Survey (Blakeslee \etal\ 2003a;
Ford \etal\ 2005)
is a study of galaxy clusters in the redshift range $0.8<z<1.3$ 
using the Wide Field Channel (WFC) of the Advanced Camera for Surveys
(ACS; Ford \etal\ 1998) on the {\it Hubble Space Telescope} (\hst),
as part of the ACS Guaranteed Time Observation (GTO) program.
The 3\farcm3 field-of-view of the ACS/WFC is well-suited to 
the size of galaxy clusters at these redshifts.
We are studying the galaxy colors, morphologies, mass spectra, and
star formation properties from ACS photometry and ground-based
spectroscopy, as well as the cluster structure and assembly from
weak lensing, at $z{\,\sim\,}1$. 
The overall goal of the survey is to understand the evolution of cluster 
galaxies and their relationship to the evolving large-scale 
cluster environment over most of the history of the universe.

Previous papers from our survey include analyses of the
color-magnitude relations in the $z{\,>\,}1$ X-ray selected clusters
RDCS J1252.9$-$2927 and RDCS J0910+5422 
(Blakeslee \etal\ 2003a; Mei \etal\ 2006), the morphology--density
relations in all the program clusters (Postman \etal\ 2005); 
the size-surface brightness relations in \rxname\ and 
RDCS J1252.9$-$2927 (Holden \etal\ 2005a), star-forming galaxy
properties in \rxname\ (Homeier \etal\ 2005), the luminosity
function of \msname\ (Goto \etal\ 2005), and weak lensing in
\rxname, \msname, and RDCS J1252.9$-$2927 
(Jee \etal\ 2005a,b; Lombardi \etal\ 2005). 
Similar programs combining ground-based and HST/ACS data 
on distant cluster samples are also being carried out by 
other groups (J{\o}rgensen \etal\ 2005; Kodama \etal\ 2005).
A complementary effort within the ACS GTO program seeks to
detect and study proto-clusters around high-redshift
($z{>}2$) radio galaxies (Miley \etal\ 2004; Overzier \etal\ 2006),
at an epoch when most of the stars in our $z\sim1$ clusters were likely forming.

In the present work, we study the properties of confirmed members of
the massive X-ray selected clusters \rxname\ and \msname.  \rxname\
was found independently by a number of X-ray surveys and its properties
have been studied with ROSAT, BeppoSAX, XMM-Newton, and Chandra
(Ebeling \etal\ 2000; Della Ceca \etal\ 2000; Maughan \etal\ 2003).
It exhibits a complex X-ray morphology, including two main components
with temperatures $6.7^{+1.2}_{-1.0}$ and $8.7^{+2.4}_{-1.8}$ keV
(I.~Balestra \etal, in preparation), separated by $\sim\,$1\farcm6,
that appear to be in the early stages of merging
(e.g., Girardi \etal\ 2005).
The mean redshift of the entire system is $z=0.837$ based 
on $\sim\,$100 members (Demarco \etal\ 2005), or 
$z=0.834$ for just the early-type galaxies (see below).
\msname, the most X-ray luminous cluster in the {\it Einstein}
Extended Medium Sensitivity Survey (Gioia \etal\ 1990;
Gioia \& Luppino 1994), has also been the target of numerous
X-ray studies and has a similarly complex morphology
(Donahue \etal\ 1998; Jeltema \etal\ 2001), although in its
case the merger is more advanced, and the total
mass is roughly twice that of \rxname. While estimates
of the temperature of \msname\ have varied widely (see Gioia \etal\ 2004
for an X-ray review of this cluster), the most recent determination
using the latest {\it Chandra} calibration gives $8.9^{+1.0}_{-0.8}$ keV
(Jee \etal\ 2005b).  Its mean redshift is $z=0.831$ based
on $\sim\,$150 members (van Dokkum \etal\ 2000;
K.-V.~Tran \etal, in preparation).

The following section presents our observations of these two
clusters and describes the basic imaging reductions.
Section~\ref{sec:measure} describes our galaxy fits,
quantifies deviations from the fits, and discusses the
galaxy photometry and the
size--surface brightness relations in these clusters.
Section~\ref{sec:cmrs} presents the color-magnitude and
color--color diagrams and investigates the dependence of the
color-magnitude relation on cluster radius and local density.
Section~\ref{sec:discussion} considers these results in the
broader context of cluster galaxy evolution, and the final
section provides a summary.
Throughout this paper we adopt a cosmology with
$(h,\Omega_m,\Omega_{\Lambda}) = (0.7,0.3,0.7)$, giving a lookback time of
7.0 Gyr, an age for the universe of 6.5 Gyr, and an angular scale
of 7.6 kpc~arcsec$^{-1}$ at $z{\,=\,}0.83$.

\begin{figure*}\epsscale{1.1}
\plotone{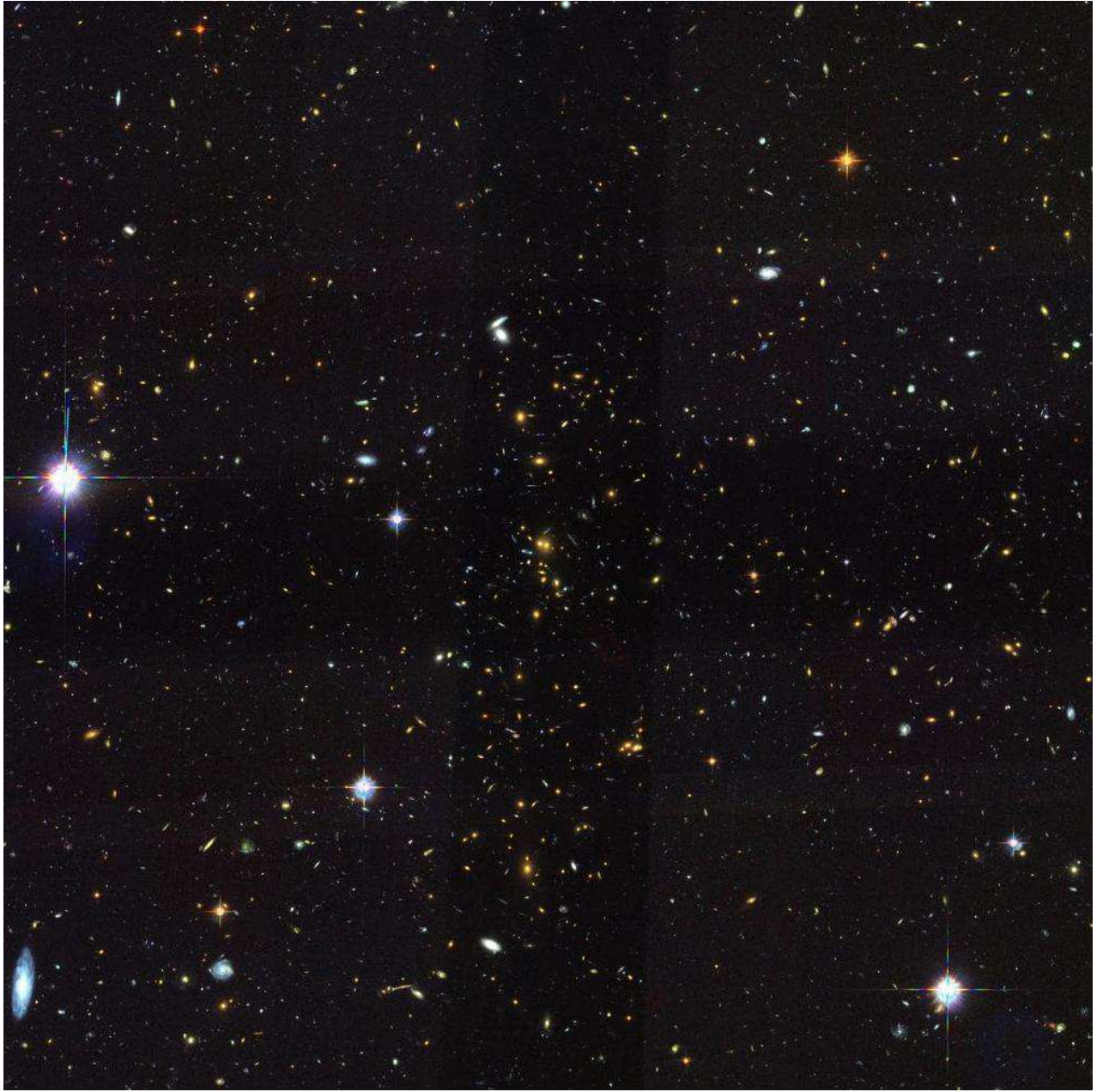}
\caption{ACS/WFC F625W/F775W/F850LP color composite image
showing 5\arcmin\ of the \clname\ mosaic in the observed orientation.
\label{fig:0152im}}
\end{figure*}

\begin{figure*}\epsscale{1.1}
\plotone{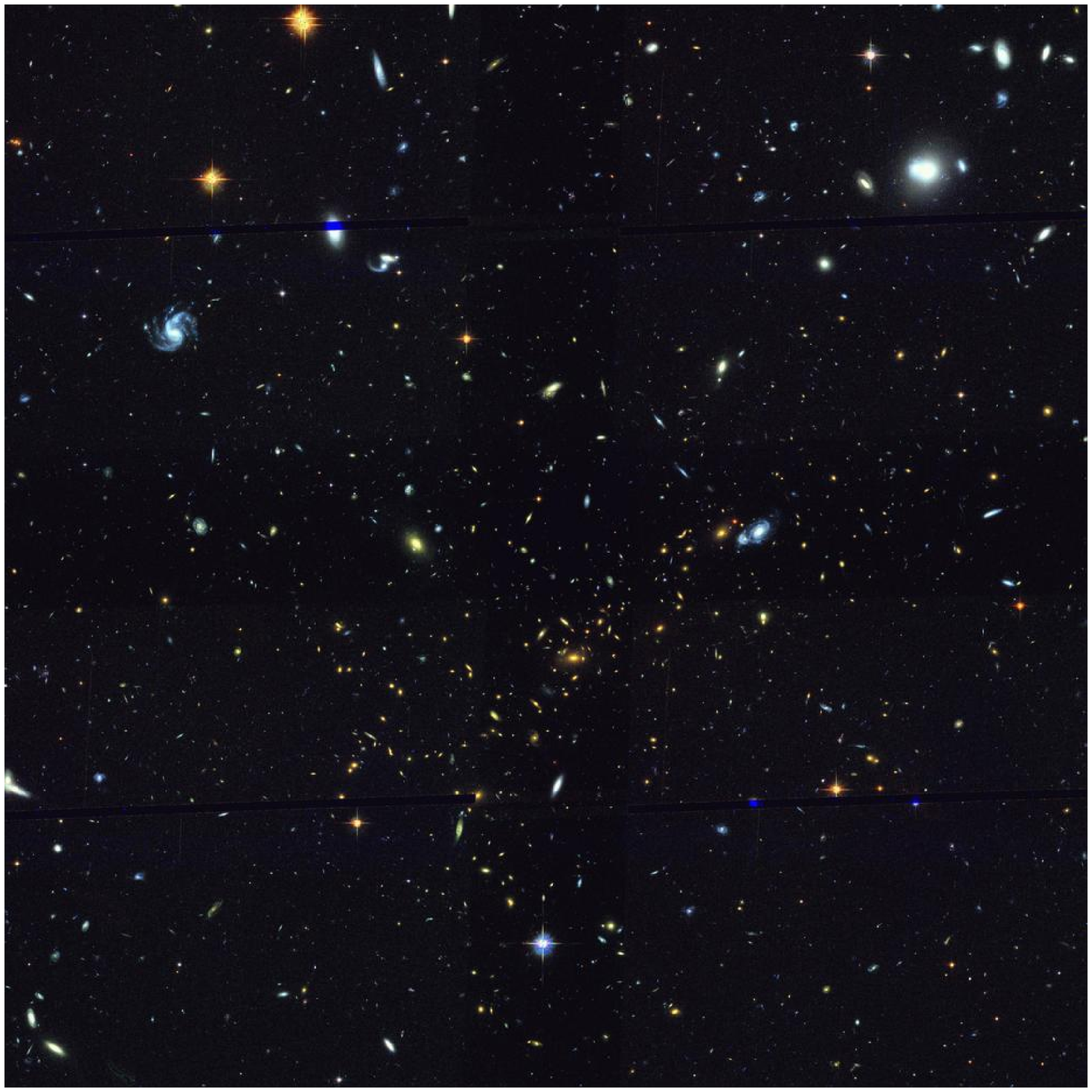}
\caption{ACS/WFC F606W/F775W/F850LP color composite image
showing 5\arcmin\ of the \msname\ mosaic field
in the observed orientation.
\label{fig:1054im}}
\end{figure*}

\section{Observations}

\subsection{HST/ACS Data}

The clusters \clname\ and \msname\ were observed with the ACS/WFC using a
$2{\times}2$ overlapping mosaic pattern.  The usable area of each mosaic
was about 5\farcm8 on a side.  \clname\ was observed in 2002~November in
the F625W, F775W, F850LP bandpasses for 2 orbits apiece at each of the
four mosaic positions.  \msname\ was observed in 2002 December in the
F775W and F850LP bandpasses for two orbits apiece, and in 2004
January/February in the F606W bandpass for one orbit, at each mosaic
position. The total orbit expenditure was therefore 24 for \clname\ and
20 for \msname.  The exposure times per pointing for \clname\ were
4750, 4800, 4750 sec in F625W, F775W, F850LP, respectively, and the
\msname\ exposure times were 2025, 4340, and 4440 sec in 
F606W, F775W, and F850LP, respectively.   
The magnitudes in the F606W, F625W, F775W, and F850LP ACS bandpasses are
sometimes for convenience referred to as \vacs, \racs, \iacs, and \zacs,
respectively.

The images were bias-subtracted and flat-fielded with the standard 
ACS calibration pipeline (Mack \etal\ 2003).
We further processed the images using the ACS GTO {Apsis} pipeline
(Blakeslee \etal\ 2003b) to measure exposure offsets and rotations
and then ``Drizzle'' (Fruchter \& Hook 2002) the data onto a 
geometrically rectified output frame.  The observations were processed
in this way both as large mosaics, and each pointing
separately, in order to assess measurement uncertainties.  The same
basic imaging data have also been used for a number of other studies
by our team (Holden \etal\ 2005a; Homeier \etal\ 2005; Jee \etal\
2005a,b; and Postman \etal\ 2005). As in the ACS/WFC photometric study
of the RDCS J1252.9$-$2927 (Blakeslee \etal\ 2003a), we use an output
pixel scale of 0\farcs05 pix$^{-1}$ and the Lanczos3 interpolation
kernel for Drizzle, which reduces the
noise correlations of adjacent pixels (see the discussion
by Mei \etal\ 2005).  Figures~\ref{fig:0152im} and \ref{fig:1054im}
show the color-composite ACS images of the two clusters.

We use the AB photometric system calibrated
with the ACS/WFC zero points from Sirianni \etal\ (2005).
The magnitudes are corrected for Galactic extinction according
to Schlegel \etal\ (1998).  The corrections for \clname\ 
are 0.038, 0.029, 0.021 mag in F625W, F775W, F850LP, respectively,
and for \msname, they are 0.098, 0.070, 0.052 mag in F606W,
F775W, F850LP, respectively.
The morphological classifications for our sample galaxies are from
Postman \etal\ (2005), who classified all galaxies in these ACS fields
down to an AB magnitude limit of $\iacs < 23.5$.  Postman \etal\
(2005) illustrate the different morphological types with many example
galaxies from these ACS images.

\subsection{Spectroscopic and Morphological Data}

The galaxies in our sample include all confirmed cluster members
within the area of our ACS/WFC images from the spectroscopic surveys of
\clname\ by Demarco \etal\ (2005) and of \msname\ by 
van Dokkum \etal\ (2000) and Tran \etal\ (2005a),
with the VLT and Keck observatories.
Those works provide full details on the spectroscopic observations
and selection; here we briefly summarize.
The \clname\ spectroscopic survey used photometric redshifts based
on $BVRIJK_s$ color data to select likely cluster members down to
a limiting magnitude of $R<24$.  The \msname\ survey targeted an
$I$-band selected sample of limiting magnitude $I<22.7$.  
In both cases, the completeness drops precipitously near the
magnitude limits.  When we intercompare the results of our
photometric analyses of these clusters below, we do so 
over common F775W magnitude ranges where the samples are
mainly complete.  We have used all redshifts regardless of
their quality flags in the original works, since they are
likely to be members even if their spectra are low quality.
Thus, our samples may be more inclusive than those adopted
for the cluster dynamical analyses in the works quoted above.
The sample galaxies have visual morphological classifications
from Postman \etal\ (2005) based on these ACS data.

\section{Structural and Photometric Measurements}
\label{sec:measure}

Our general procedure for measuring the galaxy photometry follows
van Dokkum et al.\ (2000) and Blakeslee \etal\ (2003a).  We select
galaxies in the \clname\ and \msname\ fields with measured redshifts,
fit their 2-D light profiles,
remove the effect of differential blurring from the
point spread functions (PSF) in the individual bands,
measure the colors of each galaxy within its circular effective radius 
in each of the separate ACS mosaic pointings in which it is
found, then average the measurements and use the scatter to
assess the errors.  In the following sections
we discuss these steps in more detail.

\subsection{Profile Fitting}

We fitted the 2-d surface brightness profiles of galaxies
in each field to S\'ersic (1968) models using
the program GALFIT (Peng \etal\ 2002).  For the purpose
of these fits, we selected galaxies over a broad range in
spectroscopic redshift $0.6<z<0.9$, yielding 149 and 182 galaxies
with redshifts in the \rxname\ and \msname\ fields, respectively.
The broad interval represents the range over which
the F775W bandpass provides a reasonable match to rest-frame~$B$,
and allows us to increases the fraction of late-type 
field galaxies at similar redshifts for purposes of comparison.
Each galaxy was fitted to a PSF-convolved S\'ersic model with the position,
orientation, major-axis effective radius $R_e$, and ellipticity as free
parameters, and with the index $n$ constrained such that $0< n\leq4$.
The upper bound is introduced because larger values of $n$
usually do not improve the fit much, but 
the covariance between $n$ and $R_e$ can
lead to an overestimate of $R_e$ for large $n$.
Neighboring galaxies with $\iacs < 25$ AB were fitted 
simultaneously in order to avoid biasing the fits,
and fainter objects were masked.  We used empirical
PSFs constructed from high signal-to-noise stars in
archival ACS/WFC images.  We also experimented with analytic
representations for position-dependent PSFs in these fields
(Jee \etal\ 2005a).  While these describe the shape of the
core of the PSF very well for optimally measuring 
ellipticities of faint sources, they do not adequately
model the scattering of light to large radii (the PSF halo --
see Sirianni \etal\ 2005).  As a result, the fitted $R_e$
values are over-estimated by 15--20\%; we were able to 
reproduce this effect by truncating our empirical PSFs.
Thus, we do not use the variable model PSFs for this purpose.
For similar reasons, we chose not to use PSFs generated
from TinyTIM (Krist \& Hook 2001).

The S\'ersic profiles were fitted in both the F775W and F850LP bands,
both with and without simultaneous fitting of the local sky. When not
fitting the sky, we used the SExtractor (Bertin \& Arnouts 1996) sky
map to remove the background.  In general, we obtained good agreement
in $R_e$ with the various approaches, except in a few cases where the
covariance between the sky and galaxy profile resulted in an
anomalously large $R_e$ and negative sky. These cases were generally
clear from the large uncertainties in the fitted values, and we chose
instead the results from the fits using the fixed local sky;
otherwise, we used the results with the sky fitting.

\subsubsection{Effective radii}

\begin{figure}\epsscale{1.1}
\plottwo{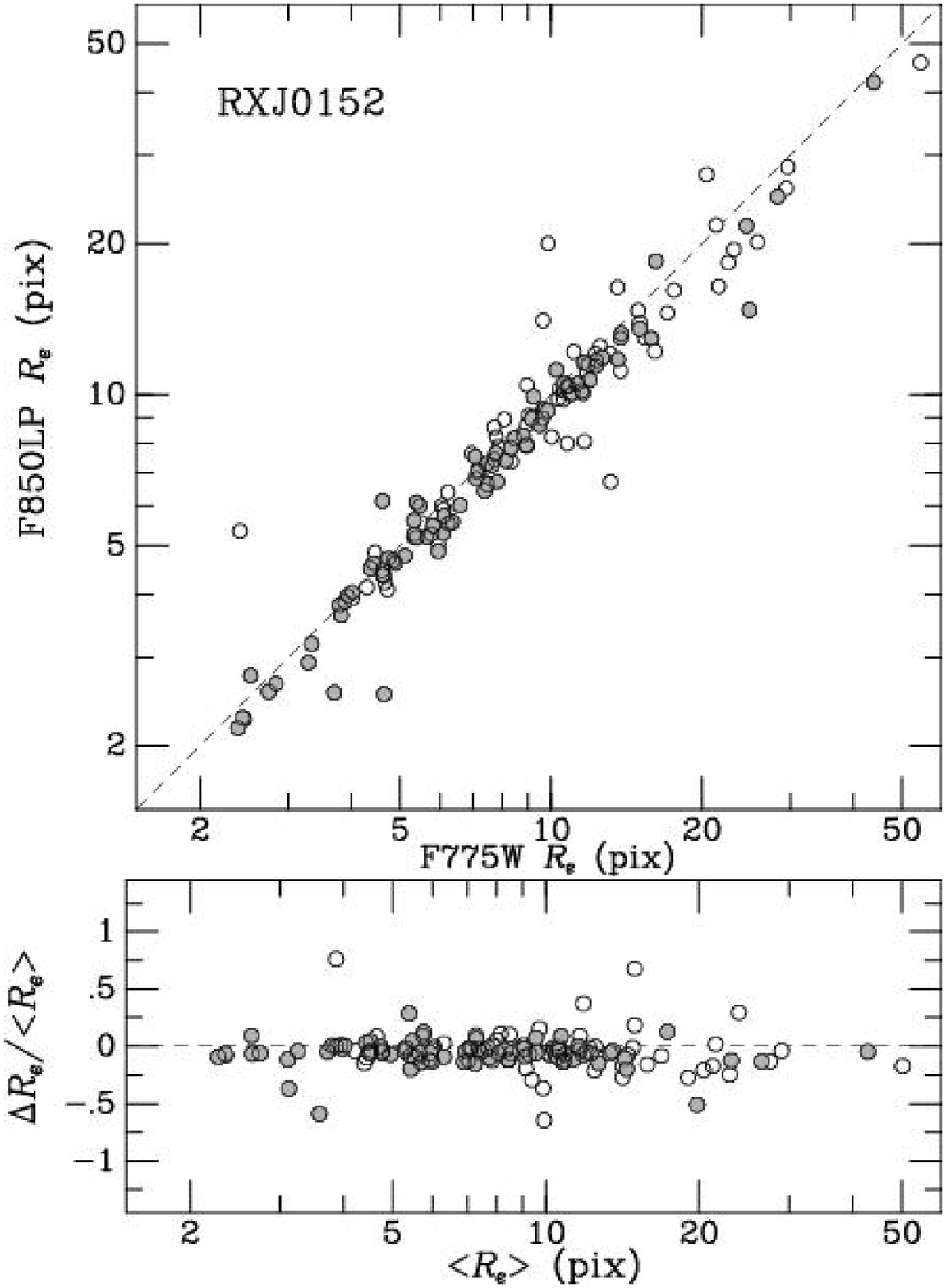}{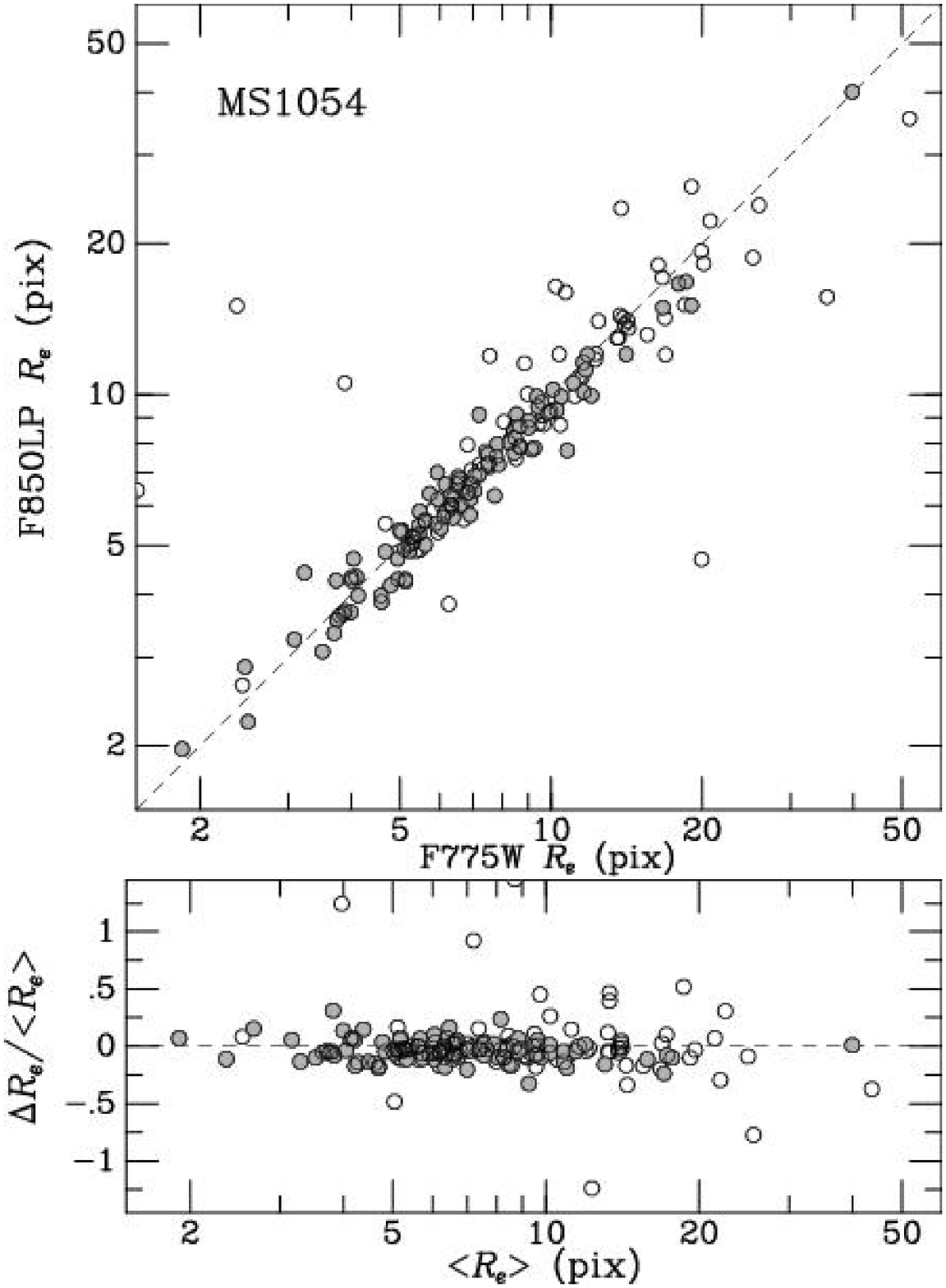}
\caption{Comparison of the fitted $R_e$ values from the
F850LP and F775W images for \rxname\ (left) and \msname\ (right).
These results are from fits with the sky previously subtracted,
rather than fitted simultaneously.
Shaded and open symbols indicate early- and late-type galaxies,
respectively. The dashed lines indicate unity and are not fits
to the data.  The lower panels show the difference in the $R_e$
values for the two bands divided by the average.  The robust
biweight estimator gives scatters of 9.1\% and 7.0\% for
the full sample and early-types, respectively, in the left
panel, and 9.7\% and 8.7\% for full and early-type samples
in the right panel.
\label{fig:Recomp}}
\end{figure}

Figure~\ref{fig:Recomp} compares the fitted (major-axis) $R_e$ 
values from GALFIT for the two different bands in which the fits
were done for the two clusters.  The scatter in
the differences based on the robust biweight scatter estimator (Beers \etal\
1990) is similar for the two fields, being about 9.5\% for all galaxy
types and 8\% for just the early-types, suggesting an internal error
in the $R_e$ determination of $\sim\,$7\% in each band ($\sim\,$6\%
for the early-types).  A few late-type
galaxies with irregular and bandpass-dependent structure are
significant outliers in each cluster.  There is also a
small offset between the results from the two bands, with the 
$R_e$ values being in the median 4\% (\msname) and 5\% (\rxname)
larger in F775W than in F850LP.  These offsets are thus about half as
large as the scatters.   While this could result from imperfectly 
accounting for the PSF (with the template F850LP PSF being too broad
relative to the F775W one), because the offset does not depend strongly
on $R_e$, it may simply reflect the mean color gradient in
the sample galaxies. With the exception of nuclear starbursts, the
outer parts of galaxies tend to be bluer,
and therefore $R_e$ is slightly
larger in bluer bands; McIntosh \etal\ (2004) have used this
same ratio as a crude measure of color gradients in local
cluster and field samples.  But the effect here is small: 
if the effective radii for an $R^{1/4}$--law galaxy
differ by 4--5\%  in two bandpasses as a result of a color gradient,
then the color within $R_e$ will differ by 0.02--0.03 mag
from the color integrated to infinity. However, small changes 
in the adopted $R_e$ (e.g., choosing the $R_e$ value 
from one band or the other) will negligibly change the color
measured within~$R_e$.


We adopt the F775W-band fit results because these images have
the highest signal-to-noise and this band approximates rest-frame
$B$ for the cluster galaxies.  The reduced $\chi^2_\nu$ values reported
by GALFIT using the Apsis error arrays are generally good:
the median and average are $\sim\,$0.82 and $\sim\,$1.0 for both 
clusters (possibly indicating a slight overestimation of 
the errors) with biweight scatters of $\sim0.15$.  All of the
galaxies in the \rxname\ field have $\chi^2_\nu<1.9$ except the two
cluster AGNs (Demarco \etal\ 2005) which yield $\chi^2_\nu\sim10$.
There is only galaxy in the \msname\ field with a similarly
large $\chi^2$, a highly irregular late-type galaxy 
that appears to be undergoing a gas-rich merger, although it
is not one of the outliers in Figure~\ref{fig:Recomp}.

\subsubsection{S\'ersic $n$ parameters}

\begin{figure*}\epsscale{0.9}
\plottwo{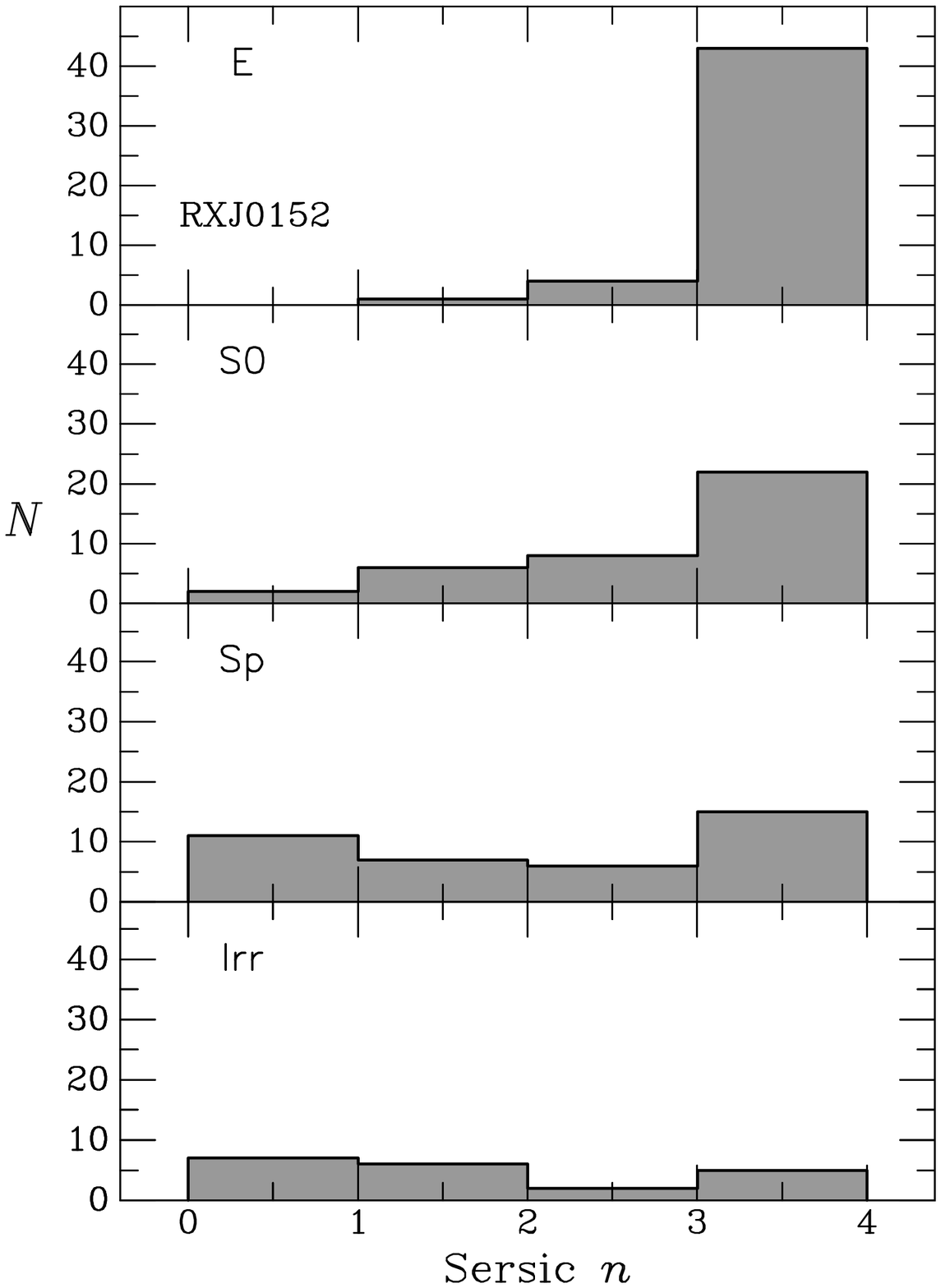}{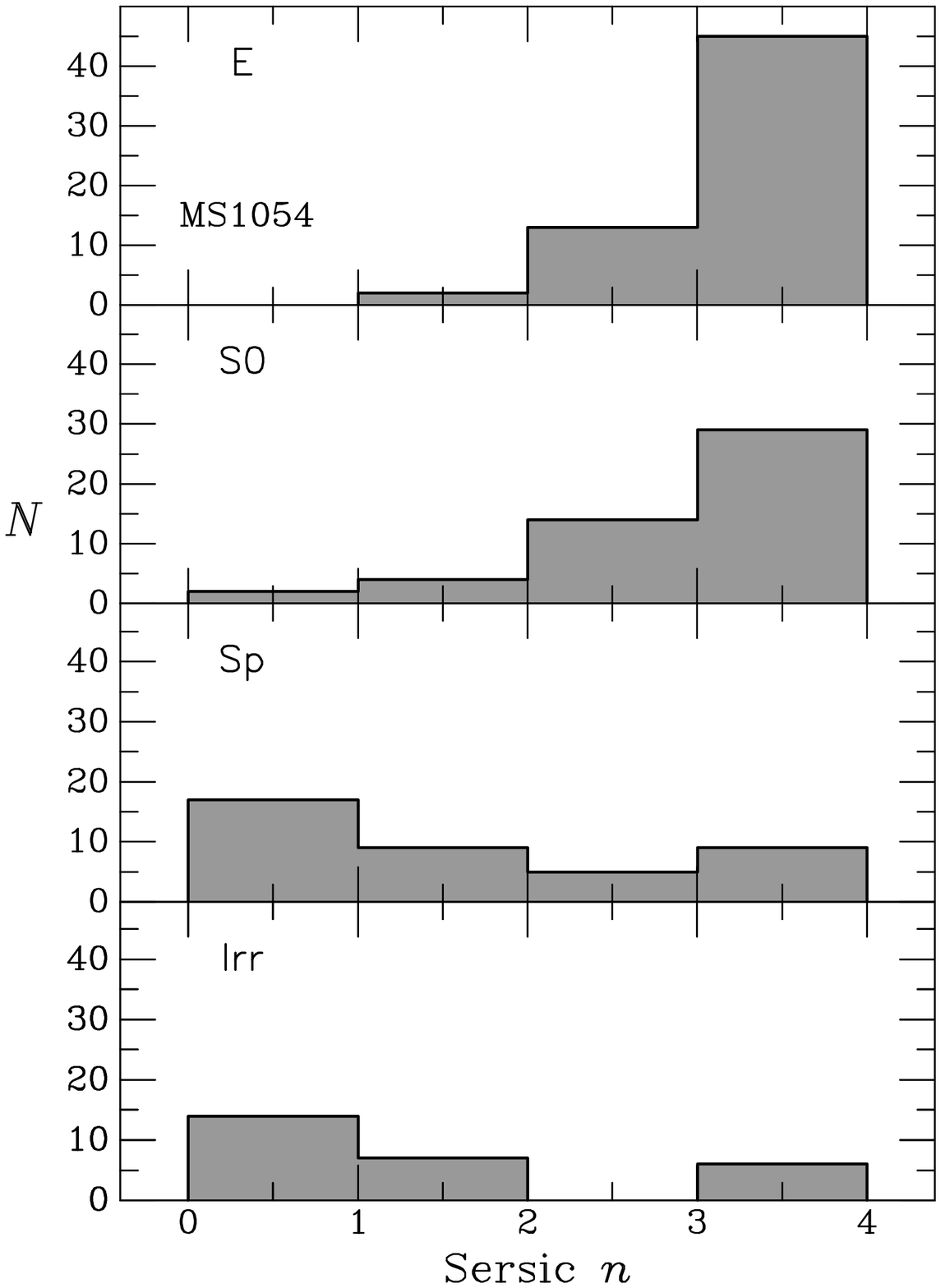}
\caption{Histograms of the S\'ersic index $n$ for different
galaxy morphological types in the \rxname\ (left) and \msname\
(right) fields.
\label{fig:nsersic}}
\end{figure*}

Of the fitted galaxies, 97\% in each field have reliable morphological
classifications from Postman \etal\ (2005) (we omit the two AGNs, as
their light is dominated by the non-thermal emission).
Histograms of the S\'ersic $n$ parameters for the different 
galaxy morphological types are shown in Figure~\ref{fig:nsersic}.
As expected, the elliptical and lenticular (S0) galaxies preferentially
have larger $n$ parameters.  Among these early-type galaxies, 
82\% (81\%) have $n{\,>\,}2.5$ and 92\% (93\%) have
$n{\,>\,}2$ in \rxname\ (\msname); considering just the
ellipticals, these percentages become 
96\% (92\%) with $n{\,>\,}2.5$ and 
98\% (97\%) with $n{\,>\,}2$.

\begin{figure}
\plotone{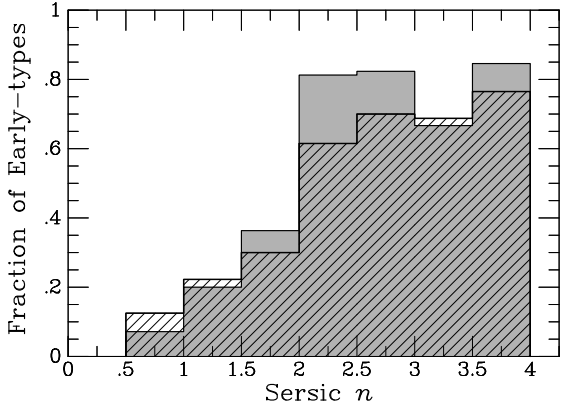}
\caption{Fraction of early-type galaxies as a function the
S\'ersic index $n$ for the fitted galaxies in the
\msname\ (grey histogram) and \rxname\ (hatched histogram) fields.
\label{fig:nfrac}}
\end{figure}

A more interesting question is how good is the $n$
parameter at selecting out galaxy types. Therefore, we show
in Figure~\ref{fig:nfrac} the fraction of early-types as
a function of $n$.  
In \rxname, 18\% of the galaxies (7 of 38) with $n<2$ are
early-type, while 74\% (78 of 106) with $n>2$ are early-type;
in \msname, 15\% (8 of 53) with $n<2$ and 82\% (101 of 123) with
$n>2$ are early-type.
At the lowest values of $n\lta1$, the galaxies
are about 90\% late-type (late-types are only about 40\% of the
sample in both fields), while at the highest values they are
about 80\% early-type.  The significant fraction of late-type
galaxies with large $n$ values occurs mainly because those fits
are driven by one bright compact clump near the center, with an 
additional skirt of material that is fitted by an extended outer profile.
For such galaxies, additional information concerning
deviations from the smooth model, parameters measuring
flocculence and/or asymmetry, would be an important
additional discriminant, as we discuss below.

\subsubsection{Profile bumpiness}
\label{sec:bumpy}

Several approaches exist for quantifying irregular structure
within galaxies, including galaxy asymmetry measurements
(e.g., Abraham et al.\ 1996) and relative intensity variations
on high spatial frequencies (e.g., Takamiya 1999; Conselice 2003).
In our case, we already had smooth parametric fits for all galaxies
and wanted a simple way to measure deviations from the smooth fits.
We explored several possibilities, including the reduced
$\chi^2$ and the normalized sum of the absolute values of the
residuals, but we found that a simple ratio of the standard deviation
to the mean of the light distribution provided the best
measure of the galaxy higher-order structure.
This is similar to the ratio of moments used in the surface brightness
fluctuations (SBF) method (Tonry \& Schneider 1988; Tonry \etal\
1997), although there it is the variance divided by the mean and is
measured on a very different physical scale.  Borrowing from the
SBF terminology (e.g., Blakeslee \etal\ 1997), 
we label this parameter the `bumpiness'~$B$.

To measure this parameter, we take the residuals from the
galaxy model subtraction, smooth them on a scale slightly smaller
than the PSF (we used a Gaussian of FWHM = 0\farcs085) to reduce
the shot noise, mask any region within a 2~pix radius of the galaxy 
center to avoid problems related to central deviations from the
S\'ersic profile or slight PSF mismatch, mask any other pixels 
that were masked before fitting, measure the root mean square (rms)
of the unmasked residuals within 2\,$R_e$, and divide by the mean of
the galaxy model over the same exact region.  That is, we
measure:
\begin{equation}
B = { \sqrt{ \frac{1}{N}\sum_{r_{i,j}<2R_e}\left[\, I_{i,j}
 - S(x_i,y_j|R_e,n) \right]_s^2 - \sigma_s^2}
     \over
     \frac{1}{N} \sum_{r_{i,j}<2R_e} S(x_i,y_j|R_e,n)
    \label{eq:Bdef}} \,,
\end{equation}
where the sum is over all pixels within $2R_e$,
omitting the galaxy center and masked pixels as described above;
$N$ is the number of unmasked pixels over which the sum is evaluated;
$I_{i,j}$ is the galaxy intensity of pixel $(i,j)$;
$S$ is the PSF-convolved S\'ersic model with fitted parameters $R_e$ and $n$;
the subscript $s$ in the sum indicates smoothing on a scale slightly
less than the PSF; and $\sigma_s$ is an estimate of the photometric error 
in the smoothed residuals.

\begin{figure*}\epsscale{1.1}
\plottwo{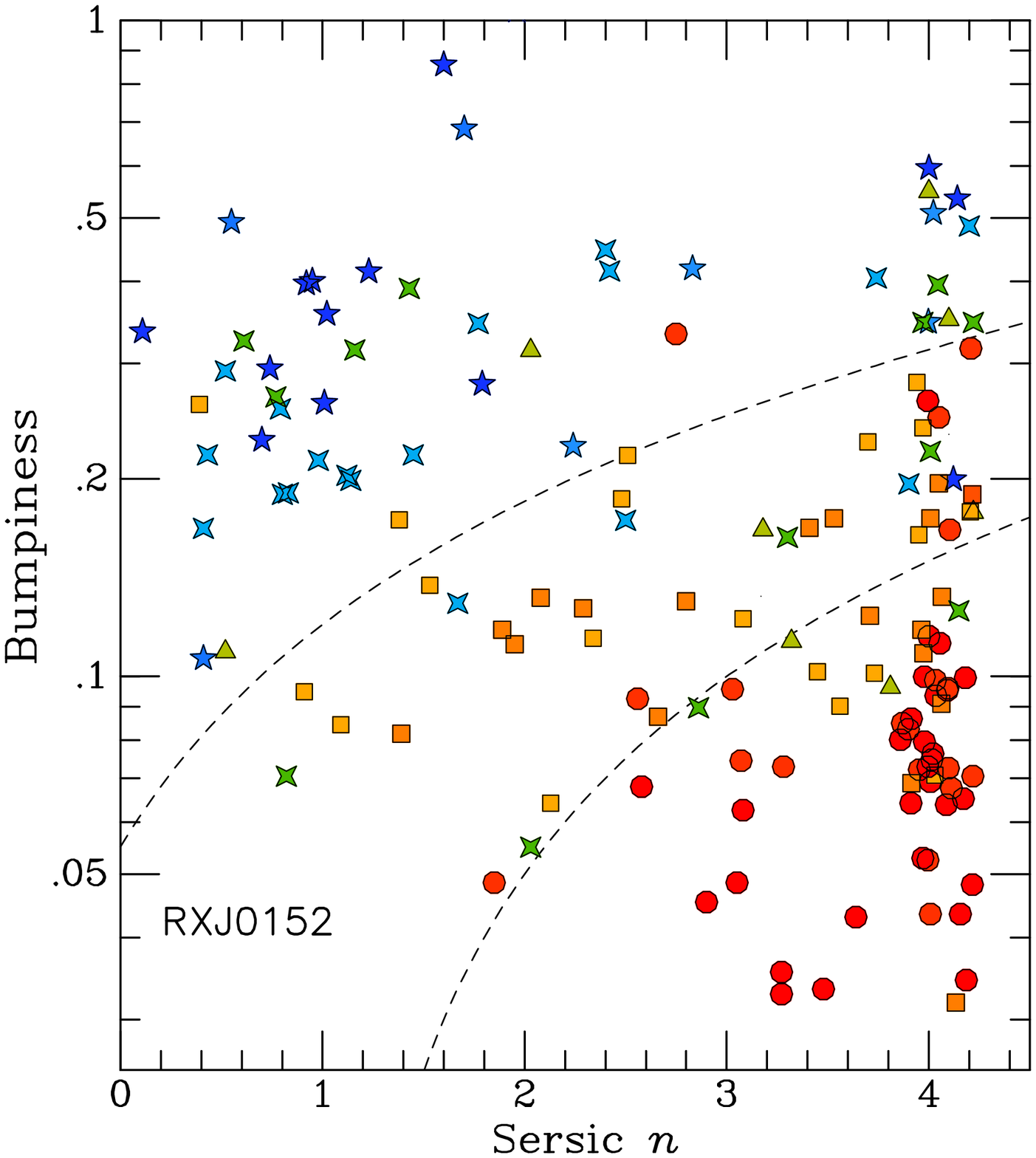}{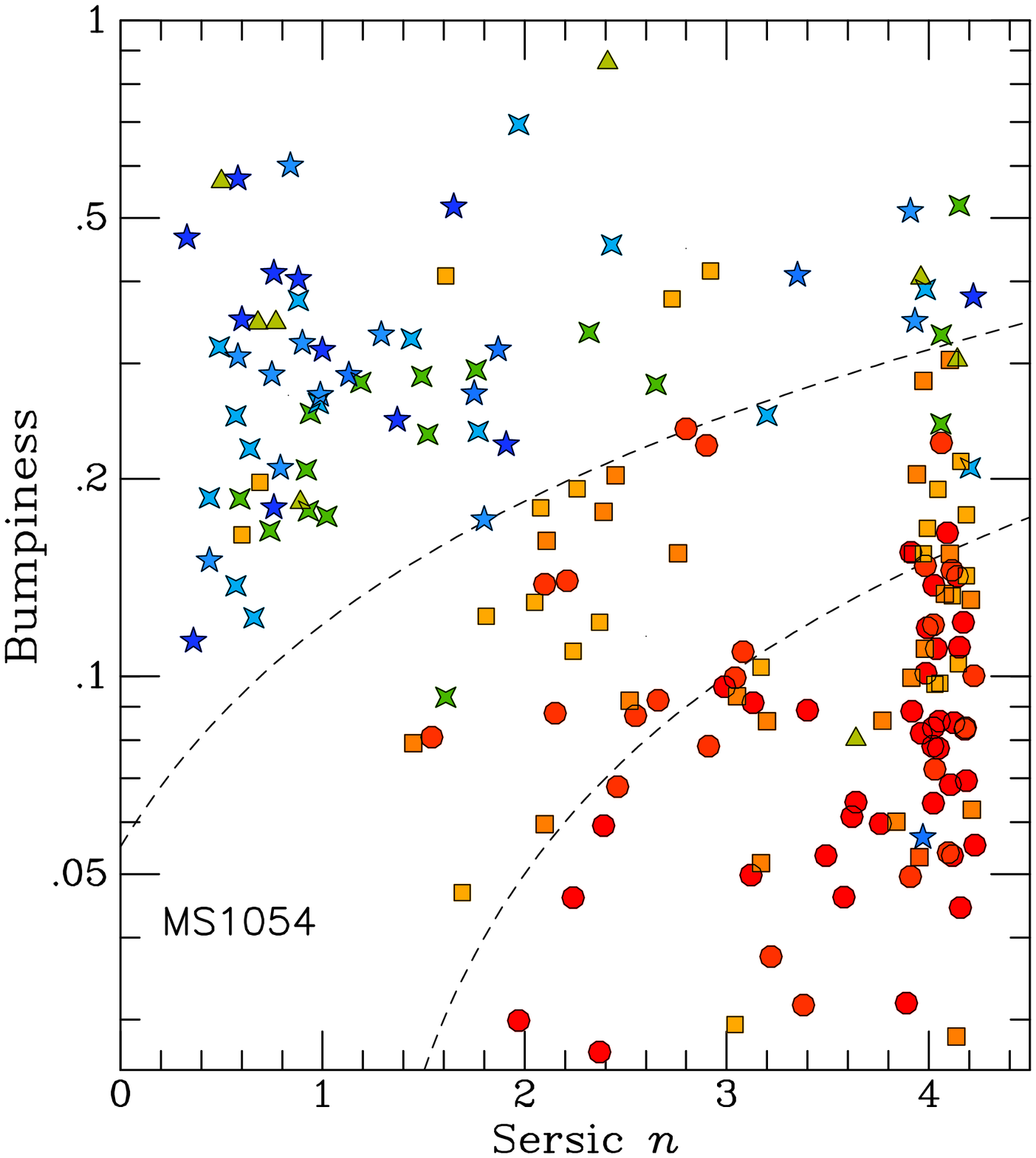}
\caption{``Bumpiness'' parameter -- ratio of the rms of residuals
after galaxy model subtraction to the mean galaxy model --
is plotted versus the $n$ index of the fitted S\'ersic model
for galaxies in the \rxname\ and \msname\ fields.  Point shapes are coded
according to broad morphological type: ellipticals (circles),
S0s (squares), S0/Sa  (triangles), 
spirals (four-point stars), and irregulars (five-point stars). 
Colors vary from red to blue according to finer morphological
$T$-type; ellipticals of types $-5$ and $-4$ are colored red; 
orange squares represent earlier--type S0s than yellow squares;
the yellow--green triangles are $T{\,=\,}0$ transition objects;
green--blue is used for early spirals, and bluer colors
are for progressively later--type galaxies. 
The upper dashed curve in each panel is the linear relation
$B = 0.065\,(n{+}0.8)$, 
which approximately splits the sample between
early-- and late--type galaxies. The lower curves are
 $B = 0.05\,(n{-}1)$, 
a suggested division between S0s and ellipticals.
%
\label{fig:bumpy}}
\end{figure*}

\begin{figure*}\epsscale{1.1}
\plottwo{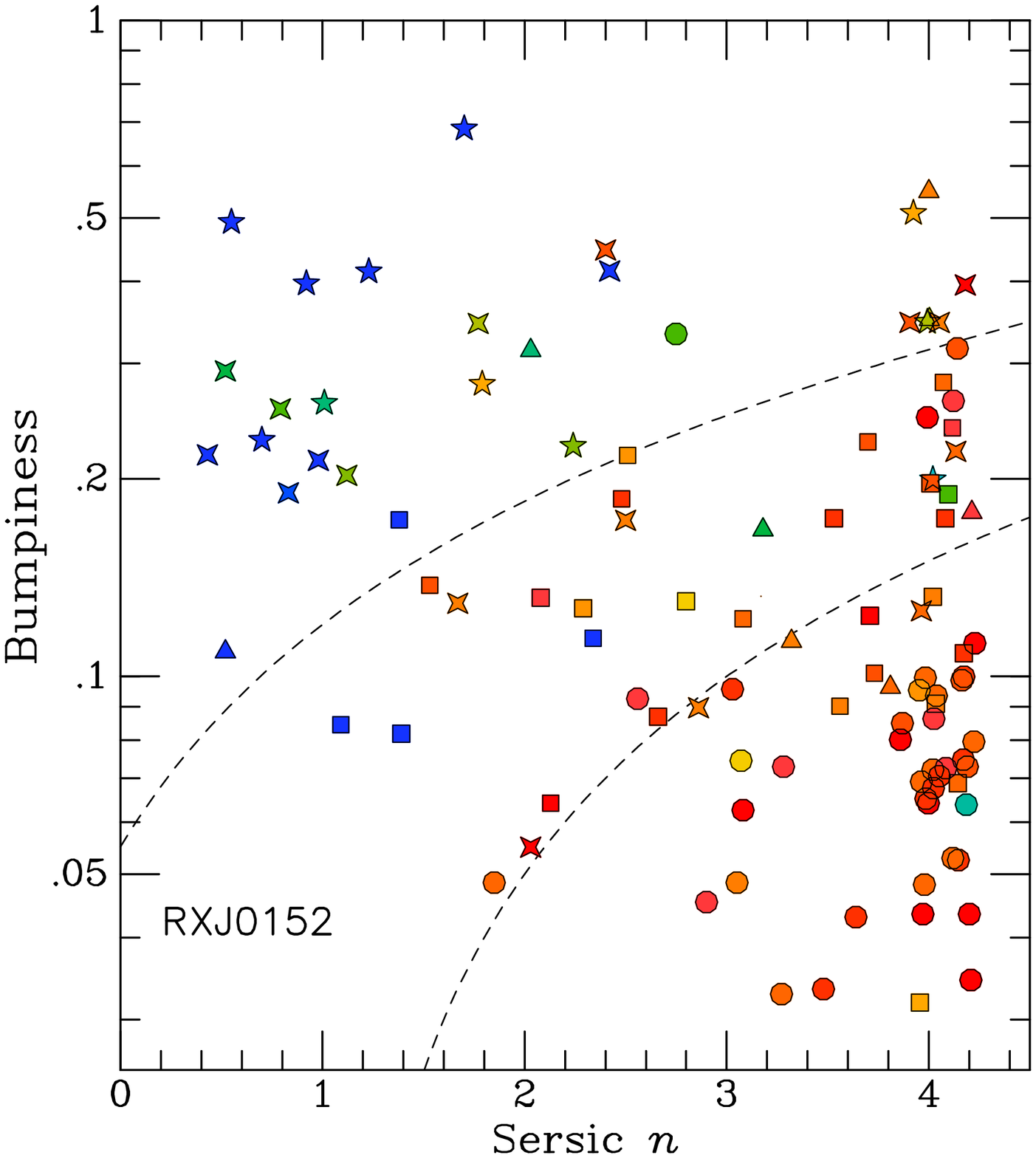}{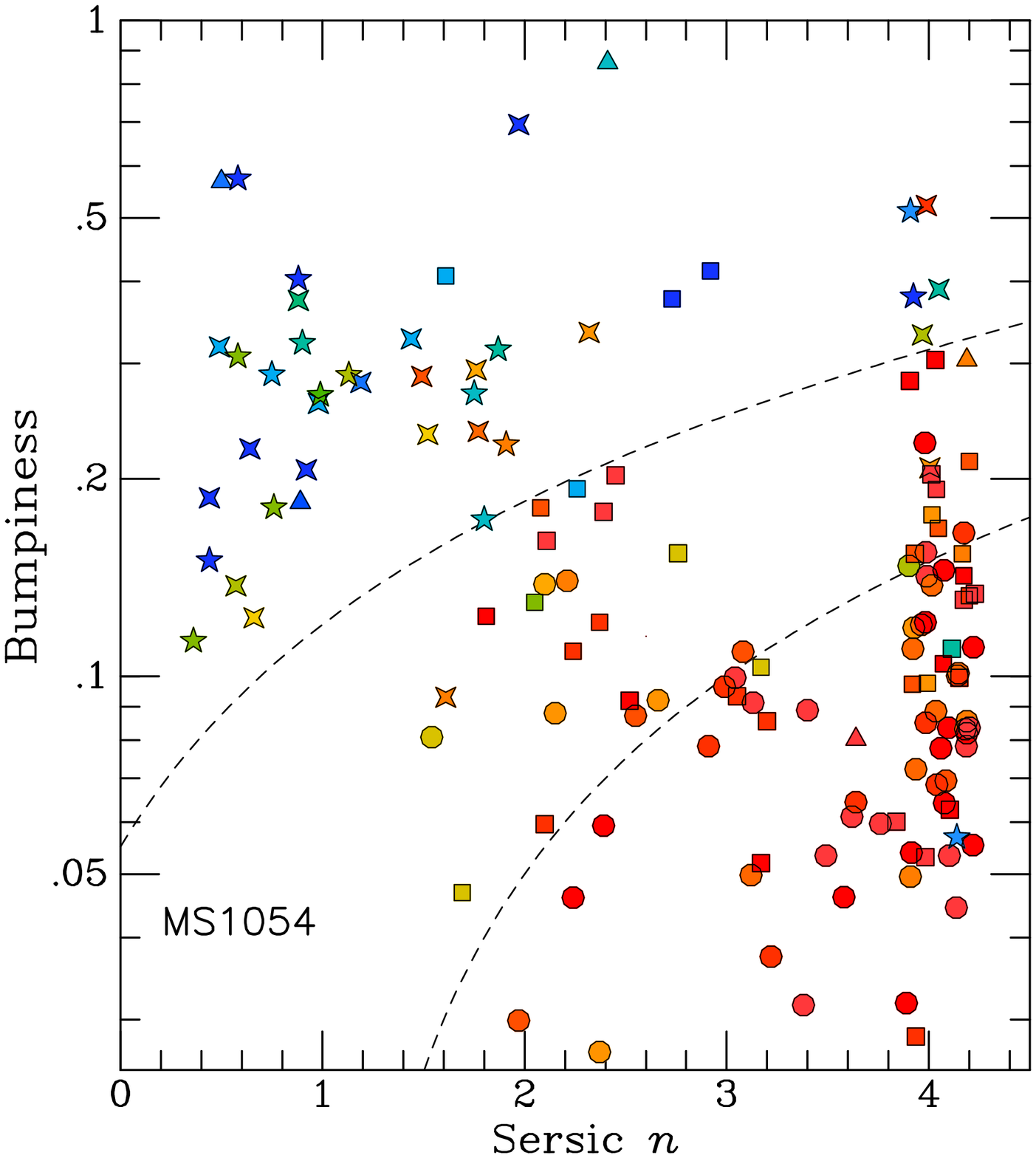}
\caption{Bumpiness versus S\'ersic $n$ parameter with point color
coded according to photometric color.
This plot is similar to Figure~\ref{fig:bumpy}, but here the
colors of the symbols are coded according to the galaxy \rzcolor\ 
(\rxname) or \vzcolor\ (\msname) color.
We use SExtractor isophotal colors here in order to avoid
an explicit dependence of the photometric aperture used for
the color measurement on the profile fitting used to determine
the bumpiness and $n$ parameters.  Further, we restrict the
redshift range of the galaxies to $z=0.80$--0.87, so that
the colors accurately reflect stellar population,
rather than redshift, variations.  The symbol color varies 
from deep blue for galaxies 2$\,\sigma$ bluer than the median color
to deep red for the photometrically reddest galaxies.
\label{fig:bumpy_cc}}
\end{figure*}

Figure~\ref{fig:bumpy} shows $B$ plotted against S\'ersic $n$ index
for galaxies in the \rxname\ and \msname\ samples.
These parameters can be used to segregate galaxies by broad morphological type;
we find the following demarcation lines work fairly well:
$B_{\rm l/e} = 0.065\,(n + 0.85)$ (separating early-- and late--types);
$B_{\rm S0/E} = 0.05\,(n - 1)$ (separating S0s and Es).
While we make no claim for universality, we find that these
lines work equally well for both of these massive $z\approx0.83$
clusters  observed in very similar manners with the ACS/WFC.  
Figure~\ref{fig:bumpy_cc} is similar to the previous figure,
but restricts the redshifts to be in the range of the clusters, and
now code the point type according to the galaxy photometric color.
It is remarkable that several of the visually classified early-type
galaxies that lie near or above the late--/early--type demarcation
line are found to have blue colors, even though the analysis is done 
in a single band (\iacs\ $\sim$ rest-frame $B$) without regard to color.

We quantify the accuracy of this automatic ``\bn\ classification'' by
calculating what fraction of the \bn\ classes have the same ``by-eye''
classification from Postman \etal\ (2005), and conversely, what fraction
of the by-eye classes were ``recovered'' by the \bn\ classes.
For the broadest separation between early (E+S0) and late types, 
we find that 91\% of the galaxies classified as early-type with the
\bn\ relation are also classified as early-type by~eye,
while 96\% of the by-eye early-types are also classified as
early-type with the \bn\ relation.  
On the other hand, 92\% of the \bn\ late-type galaxies
are classified as late-type by-eye, while 80\% of the by-eye
late-types are classified as late-type with \bn.
Thus, there is some tendency to classify galaxies as earlier
type with the chosen \bn\ segregation lines.
For the more difficult separation between elliptical (E) and
lenticular (S0), 61\% of the \bn\ S0s are classified as S0 by~eye,
while 58\% of the by-eye S0s are classified as S0 with \bn.
Furthermore, 72\% of the \bn\ Es are classified as E by~eye,
while 83\% of the by-eye Es are classified as E with \bn.
Thus, not surprisingly, the agreement for the middle S0 class is 
lowest, but still a respectable $\sim\,$60\%.

\begin{figure*}\epsscale{1.1}
\plottwo{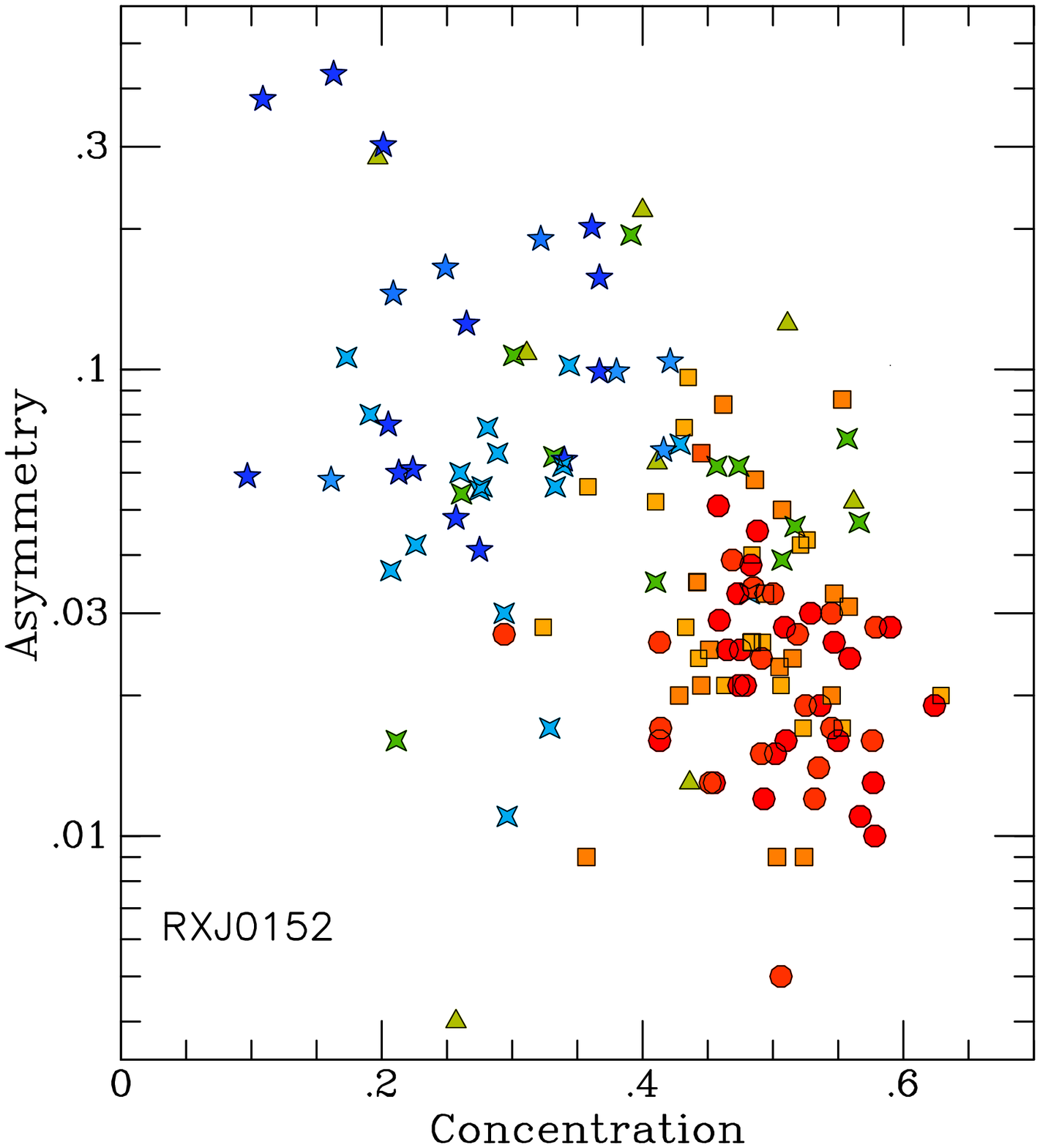}{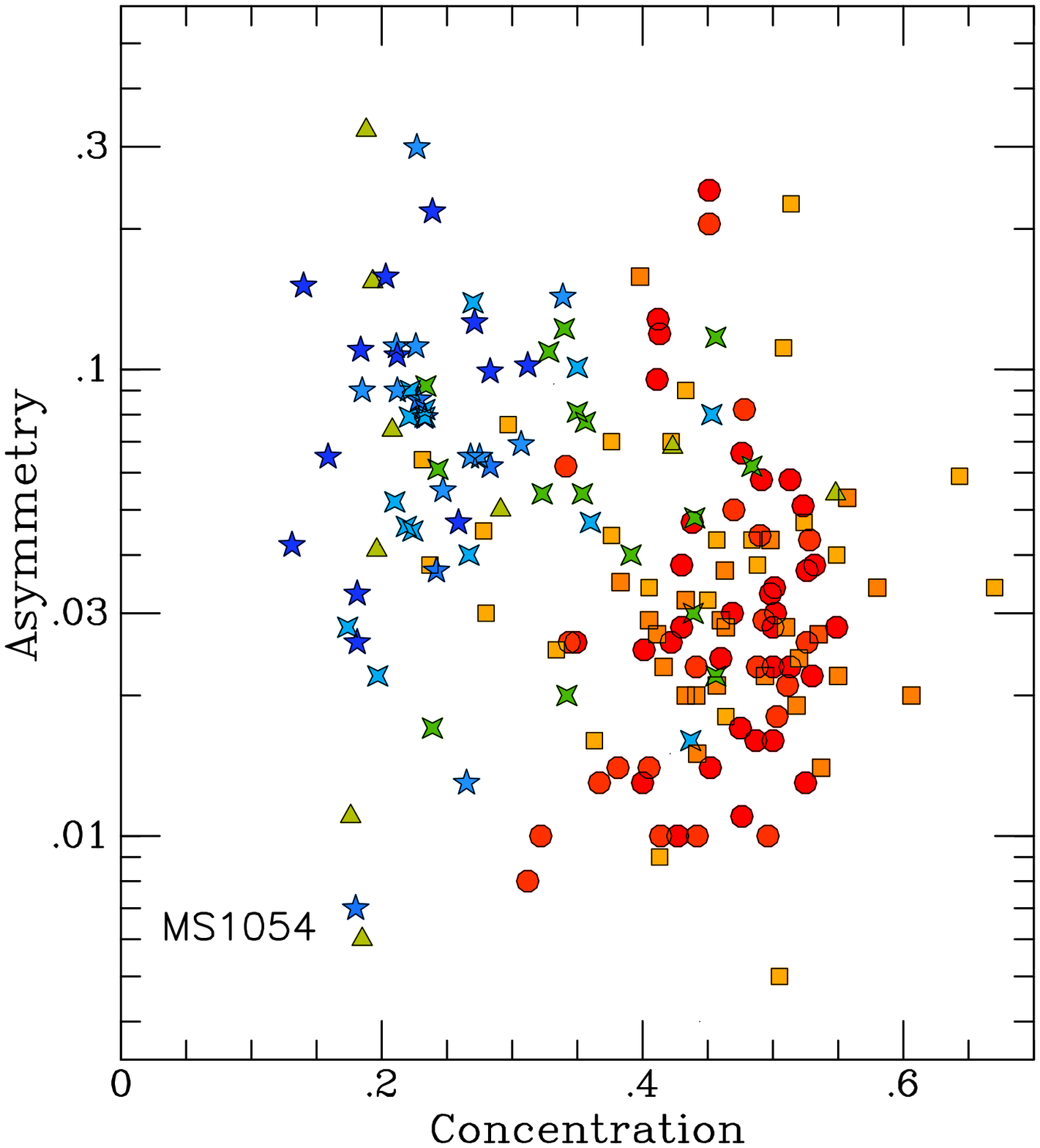}
\caption{Asymmetry $A$ and concentration $C$ parameters for
galaxies in \rxname\ and \msname. 
Symbols are coded as in Figure~\ref{fig:bumpy}.
\label{fig:ac_diag}}
\end{figure*}

\begin{figure*}\epsscale{1.1}
\plottwo{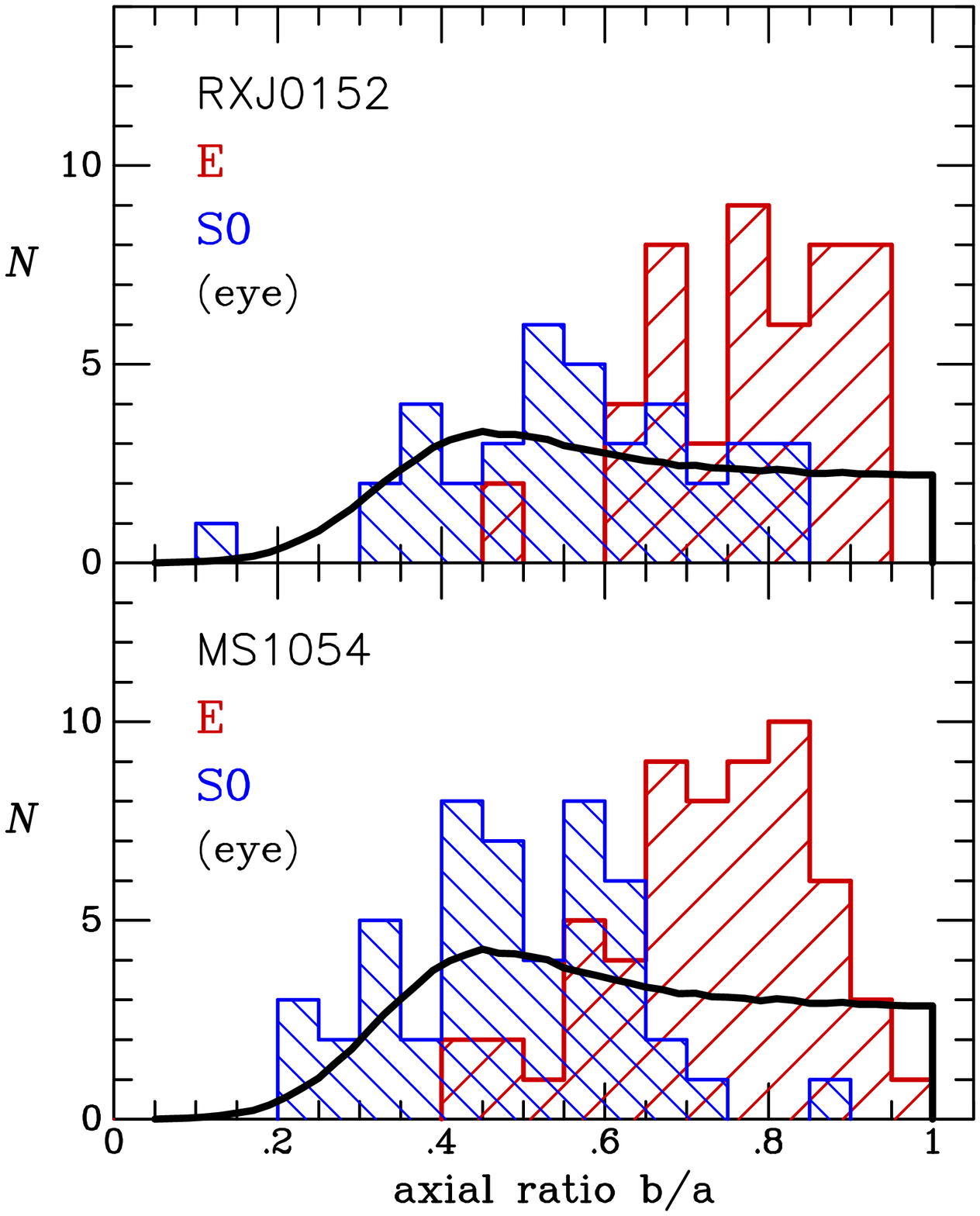}{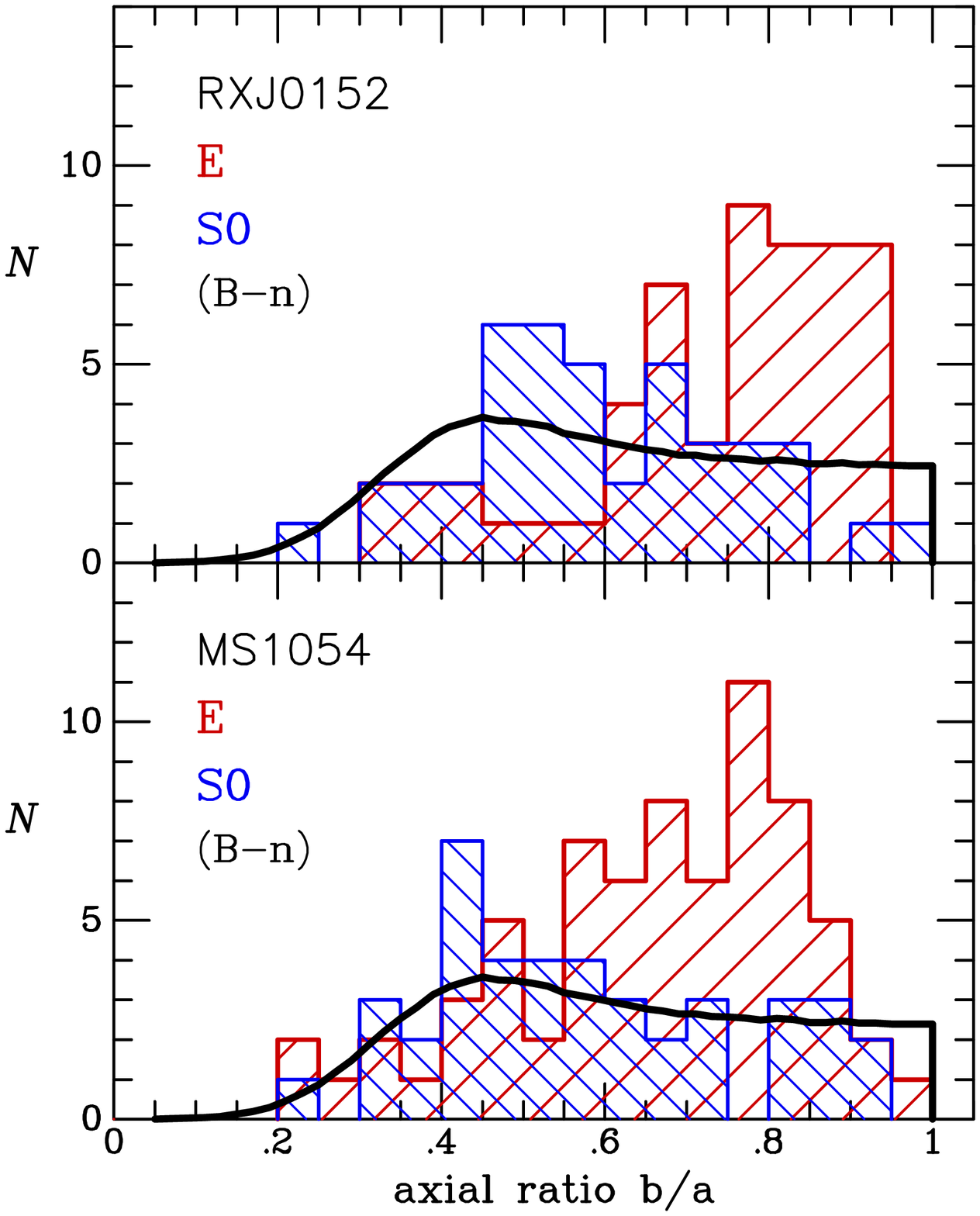}
\caption{Histograms of galaxy axial ratios determined from 
GALFIT for S0s (blue histograms with narrower hatching) and
ellipticals (red histograms with broader hatching) classified
by eye (left) and from the bumpiness--$n$ classifications
described in the text and preceding figures.  The thick black
curves show the expected distribution from a ``random disk'' model
of the S0s: oblate ellipsoids of mean intrinsic axial ratio 0.35
and Gaussian dispersion 0.1 viewed at random orientations.
\label{fig:axhistos}}
\end{figure*}


Figure~\ref{fig:ac_diag} shows for comparison
plots of the asymmetry $A$ versus concentration index $C$ (e.g., Abraham et
al.\ 1996) for galaxies in the \rxname\ and \msname\ fields,
measured using PyCA (Menanteau \etal\ 2006). 
This combination of parameters also effectively separates early--
and late--type galaxies, but there is no clear separation
of the S0 population from the ellipticals.  The most similar
automatic classification method in the literature to the
\bn\ approach given here was presented by Im \etal\ (2002).
They fitted their sample galaxies using GIM2D (Simard \etal\ 2002)
and found that plotting the bulge-to-total ratio $B/T$ versus a
parameter called $R$ (the sum of the symmetric and anti-symmetric
summed absolute values of the fit residuals, normalized by the total
galaxy light) provided good separation between early-- and late--type
galaxies but not S0s and ellipticals.

Figure~\ref{fig:axhistos} compares the histograms of axial ratios $b/a$
for E and S0 galaxies classified from the by-eye and the \bn\ analyses.
J{\o}rgensen \& Franx (1994) found a strong deficit of round S0 galaxies
in the nearby Coma cluster and concluded that 
projection effects played an important role in the visual classifications,
such that face-on disk galaxies are preferentially classified as E.
The ellipticity distribution of the Coma S0 galaxies 
could not be satisfactorily matched with any simple distribution of
intrinsic ellipticities viewed at random orientations 
(calculated as in Sandage et al.\ 1970).
Their best-fitting model of the intrinsic S0 ellipticities was a Gaussian 
distribution with mean 0.65 (mean axial ratio 0.35) and dispersion 0.1,
but it had a probability of only 11\% based on a K-S test
(partly because of this, they concluded that the fainter E and S0 galaxies
form a single class with the individual classifications being determined
by bulge-to-disk ratio and observed orientation).
We include the expected distribution from J{\o}rgensen \& Franx's
best-fitting S0-alone ellipticity model in Figure~\ref{fig:axhistos};
it predicts a mean observed axial ratio of 0.61.  This mean will drop
if the disks are more intrinsically flattened, but the frequency
distribution should remain flat as it approaches $b/a\sim1$; 
a dip at high values indicates the objects are significantly
triaxial (or prolate, e.g., van den Bergh 1994), or that there
is an orientation bias.

For these clusters, the by-eye and \bn\ classifications yield
similar relative numbers of elliptical and S0 galaxies, and we find
mean observed axial ratios for the visually classified S0s of 0.56 in
\rxname\ and 0.48 in \msname, while for the \bn\ classified S0s, the
mean ratios are 0.59 in \rxname\ and 0.57 in \msname.  These are
fairly small differences; however, the distributions at high values of
$b/a$ are more uniform for the \bn\ S0s, especially in \msname. 
%
%
Based on a K-S test, the probabilities that the by-eye classified
S0s have the same axial ratio distribution as the randomly oriented
disk model are 4\% and 0\% for \rxname\ and \msname, respectively;
these probabilities become 19\% and 35\%, respectively, for the \bn\ S0s.
That is, the ellipticity distributions of these galaxies are now reasonably
consistent with oblate spheroids viewed at random orientations.

The $B$ parameter can distinguish S0s because these
galaxies have both strong bulge and disk components, meaning
that the ellipticity and radial slope of the isophotes change with radius,
causing deviations from the fitted simple 2-D S\'ersic models.
There is likely still some orientation bias, since the ellipticity
will change less for more face-on systems, but the lower mean $n$ value
of S0s regardless of orientation,
and the bumpiness caused by the change in slope, make this method
relatively robust.  While we consider these results promising, 
we will generally stick to the by-eye classifications for
consistency with past works and because of the extensive tests performed
by Postman \etal\ (2005), though we do compare the color--magnitude
diagrams of the \bn\ classified types to the visually
classified ones in \S\ref{ssec:cmds}.

\subsection{Size--Surface Brightness Relation}

Holden \etal\ (2005a) study the evolution 
of the early-type galaxy size--surface brightness relation (Kormendy 1977)
in three rich clusters out to redshift $z=1.24$, including \rxname.
They find an evolution in the rest-frame $B$-band surface
brightness with redshift of $(-1.13\pm0.15){\,\times\,}z$ mag,
similar to the value obtained from the fundamental plane
(Holden \etal\ 2005b).   This result for the evolution
depends on the universality of the size--surface brightness
relation at a given redshift.  

\begin{figure}
\plotone{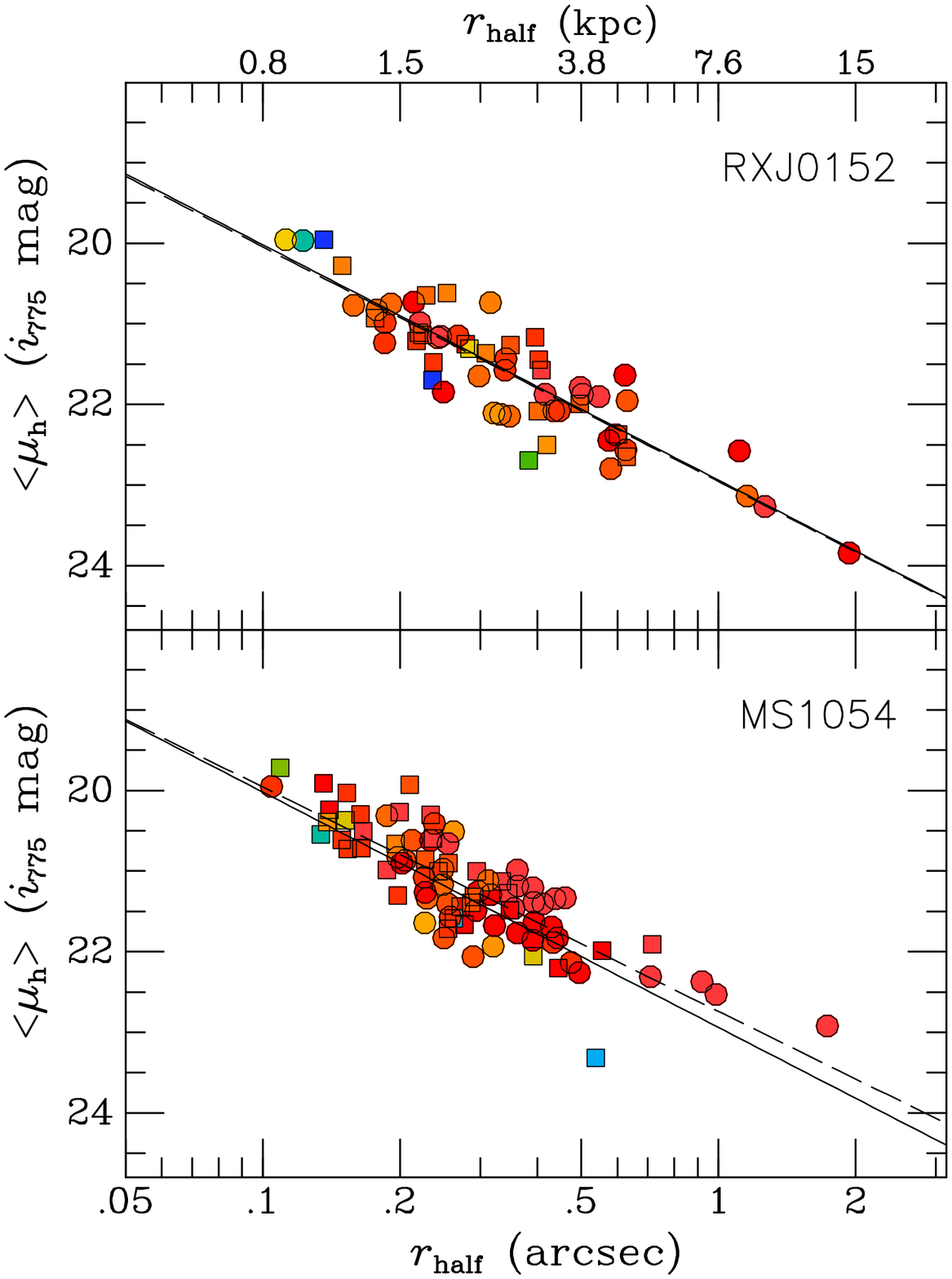}
\caption{Size--surface brightness relations in \rxname\
and \msname.  Circular effective radius $r_h$ is indicated along
the horizontal axis in arcsec at bottom and in kpc at top.
Symbol color is coded according to galaxy
photometric color as in Figures~\ref{fig:bumpy_cc}.
Solid lines show the relation for \rxname\ from Holden \etal\ (2005), 
while dashed lines are least-squares fits to the plotted points.
The slopes and scatters are the same within the errors,
and there is a small offset of $\sim\,$0.1~mag.
\label{fig:sizeSB}}
\end{figure}

In Figure~\ref{fig:sizeSB} we compare the relations obtained from our
galaxy surface brightness modeling for \rxname\ and \msname\ in the
\iacs\ band.
Following Holden \etal\ (2005a), we use the equivalent circular
effective radius $r_h \equiv R_e\sqrt{(b/a)}$, where $b/a$ is the fitted
mean axial ratio, and we restrict the sample to galaxies with
$r_h > 0\farcs1$ and $\iacs < 23$.  The solid lines in the
two panels show the relation for \rxname\ from Holden \etal:
$\langle \mu_h(\iacs)\rangle = 21.41 + 2.92\,\log_{10}(r_h/0\farcs3)$,
while the dashed lines show the fitted linear relations for the
two samples.  For \rxname, our result (using the same basic
imaging data as Holden \etal\ 2005a, but different fits) is nearly identical:
$\langle \mu_h(\iacs)\rangle = (21.43\pm0.05) + (2.91\pm0.15)\,\log_{10}(r_h/0\farcs3)$,
where we quote bootstrap errors.
For \msname, we obtain:
$\langle \mu_h(\iacs)\rangle = (21.29\pm0.04) + (2.82\pm0.23)\,\log_{10}(r_h/0\farcs3)$.
The biweight scatters for these fits are
$0.37\pm0.04$ and $0.38\pm0.04$ mag for \rxname\ and \msname, respectively.
Thus, the slopes and scatters are identical within the errors, and there
is an offset of 0.14 mag, or 2.2\,$\sigma$, in the zero~point
(the offset is the same if we force identical slopes).  The offset
is therefore marginally significant, but consistent with the scatter
in the evolution found by Holden \etal\ (2005a).

There is only one 3-$\sigma$ outlier in the two panels,
the blue S0 which appears anomalously faint by 3.5\,$\sigma$
 in the plot of the \msname\ 
size--surface brightness relation.  It may seem surprising that a blue
galaxy would have an anomalously faint surface brightness, since the blue
color would likely indicate recent star formation that would tend to 
brighten the galaxy.  However, this galaxy (ID \#24) is composed of a
bright nucleus (or knot of star formation) offset from the center of
an extended low-surface brightness disk.  The fitted S\'ersic parameter is 1.6,
or nearly exponential, and the \bn\ parameters indicate it is late-type. 
The bluer F606W band (approximately rest-frame U) shows irregular,
or possibly spiral, structure in the low-surface brightness disk.
We conclude that this is a late-type galaxy; however, leaving it out
of the fit does not significantly change any of the above fit parameters.

\subsection{Galaxy Photometry}\label{ssec:gxyphot}
%
Before measuring the galaxy colors, we apply the CLEAN algorithm
(H{\" o}gbom 1974)
to `postage stamp' images of the program galaxies using the same
routine as Blakeslee \etal\ (2003a) in order to remove the differential
blurring effects of the PSF in the different bands.  This is particularly
important for colors involving the F850LP bandpass because of the
long-wavelength halo in the ACS CCDs  (Sirianni \etal\ 1998, 2005).
For instance, the point-source aperture correction to the 
($\racs{-}\zacs$) color is 0.15 mag for a radius of 0\farcs15,
dropping to 0.07 mag at 0\farcs25
and 0.03 mag at 0\farcs5.
We use CLEAN because it is straightforward to enforce flux conservation by
adding the residuals from the method back to the derived CLEAN maps.  But
first the CLEAN maps are smoothed with a $3{\times}3$ Gaussian in order to
reduce the pixel-to-pixel noise that results from this algorithm.
%
In order to  minimize the effects of color  gradients, we then measure
the  galaxy colors  within  one equivalent  circular effective  radius
$r_h = R_e\sqrt{b/a}$, where $R_e$ is the major-axis effective radius
and $b/a$ is the mean axial ratio, both as determined by GALFIT.

We measure the galaxy colors in this way in each of the four separate
ACS/WFC pointings in which they are found, and then average the colors
for galaxies imaged in multiple pointings.  The scatter in the
multiple measurements is then used to estimate the random error in
the colors.  This error for single-pointing observations is typically 
around 0.02 mag for galaxies of magnitude $\iacs=22$ and $\sim0.03$ mag
at $\iacs=23$.  These errors are larger than in our study of 
RDCS J1252.9$-$2927 because we have fewer orbits per pointing here.
We also found some ($\sim\,$0.02 mag) systematic variation in
the galaxy photometry with position in the image, 
apparently resulting from imperfect flat fielding.  We were
only partially
successful in removing this effect; details are provided in
Appendix~\ref{app:pgrad}.

\subsection{Redshift Corrections}
The galaxies in both \rxname\ and \msname\ have large velocity 
ranges, and this adds additional scatter to their observed colors.
The main effect comes from the 4000~\AA\ break moving through the
F775W bandpass at redshifts above $\sim\,$0.8, causing 
the \ricolor\ and \vicolor\ colors to become bluer and 
\izcolor\ to get redder.  The cluster velocity dispersions 
can add more than 0.02 mag of scatter in these colors for
early-type galaxy spectra, while the broader baseline
\rzcolor\ and \vzcolor\ colors are not very sensitive to 
the redshift variation within these
clusters.  We fitted the early-type galaxy color variations 
over the 0.80--0.87 redshift range of these cluster and determined 
the following corrections:
$\Delta\vicolor = 1.0{\,\times\,}\Delta z$,
$\Delta\ricolor = 1.2{\,\times\,}\Delta z$, and
$\Delta\izcolor = -1.5{\,\times\,}\Delta z$,
where $\Delta z$ is the offset in redshift with respect to
the cluster mean.  The \vzcolor\ and \rzcolor\ color corrections,
which are the appropriate sums of the other corrections,
are smaller because the 4000 \AA\ break lies between the two bandpasses.  
We apply the corrections only for the early-types, 
as their intrinsic color scatter is small, and thus the corrections
can be significant.  Late-type galaxies
would have smaller corrections because their spectra are less
steep through the F775W band, and the effect is generally
negligible compared to the intrinsic color scatter.
We adopt $\langle z\rangle{\,=\,}0.834\pm0.001$ 
as the mean systematic redshift for \rxname\  
because it is the biweight mean of the 68 confirmed
early-type members (the median is 0.833), 
although the median and biweight mean of the full $\sim\,$100 galaxy
sample is 0.836 (and the straight average is 0.837).
For \msname\ we use  $\langle z\rangle{\,=\,}0.831$,
which is the median and mean of both the full and early-type samples.
The difference in \iacs\ magnitude between these two mean cluster
redshifts for an early-type galaxy spectrum is $\lta\,$0.02 mag
(from distance modulus and $k$-correction);
the difference in \izcolor\ would be $\lta\,$0.007 mag, while
the difference in \rzcolor\ would be $<\,$0.001~mag.

Tables~\ref{tab:dat0152} and \ref{tab:dat1054} present all
of our measurements from both the structural (S\'ersic fits and
bumpiness measurements) and photometric analyses.  The galaxy
coordinates for the \msname\ members have been omitted from Table~\ref{tab:dat1054}
because the analysis of the spectroscopic sample is ongoing; however, 
our ID numbers will be cross-referenced when the membership
catalogue and coordinates are published (K.-V.~Tran \etal, in preparation).

\section{Color-Magnitude and Color--color Relations}
\label{sec:cmrs}

We use the colors as measured in the preceding section and adopt total
magnitudes from the SExtractor (Bertin \& Arnouts 1996) ``\magauto''
parameter (with a Kron factor of 2.5)
to study the galaxy color--color and color-magnitude
relations in \rxname\ and \msname.  The SExtractor \magauto\ correlates
well with the GALFIT total magnitudes for early-type galaxies,
with a biweight scatter in the differences of 0.14 mag for \rxname\
and 0.12 mag for \msname\ over a 4~mag range. But there is an offset of
$0.20\pm0.01$ mag in both cases, in the sense that \magauto\ misses
$\sim\,$20\% of the light.  This is consistent with the conclusions of
\txitxo\ \etal\ (2004), based on extensive testing of SExtractor
using simulated galaxy images.
However, \magauto\ is more robust than the GALFIT magnitudes for
late-type galaxies, which often are not well approximated by S\'ersic profiles.
We therefore adopt the \magauto\ values but with an overall
`aperture correction' of $-$0.20~mag.  The random error in the total
magnitude does not add significantly to the scatter in the
color-magnitude relation because the slope is so shallow.

\begin{figure*}
\plotone{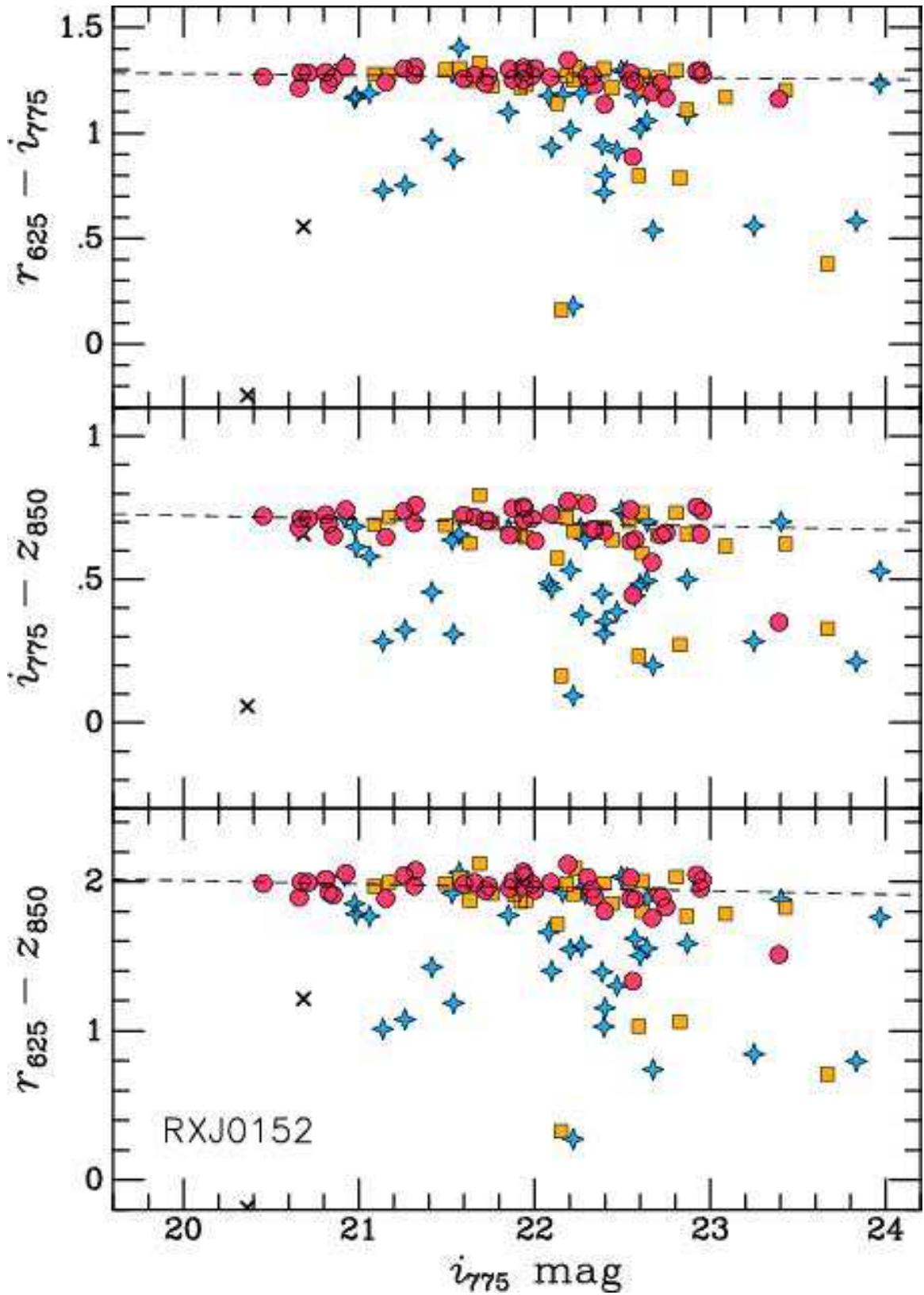}\epsscale{0.9}
\caption{Color--magnitude relations in \rxname\
for visually-classified ellipticals (red circles), S0s (orange squares),
and late-type galaxies (blue stars); two 
spectroscopically identified AGNe are represented by crosses.
The dashed lines indicate the fitted relations for the elliptical
galaxies given in the text.
\label{fig:cmrs0152}}
\end{figure*}

\begin{figure*}
\plotone{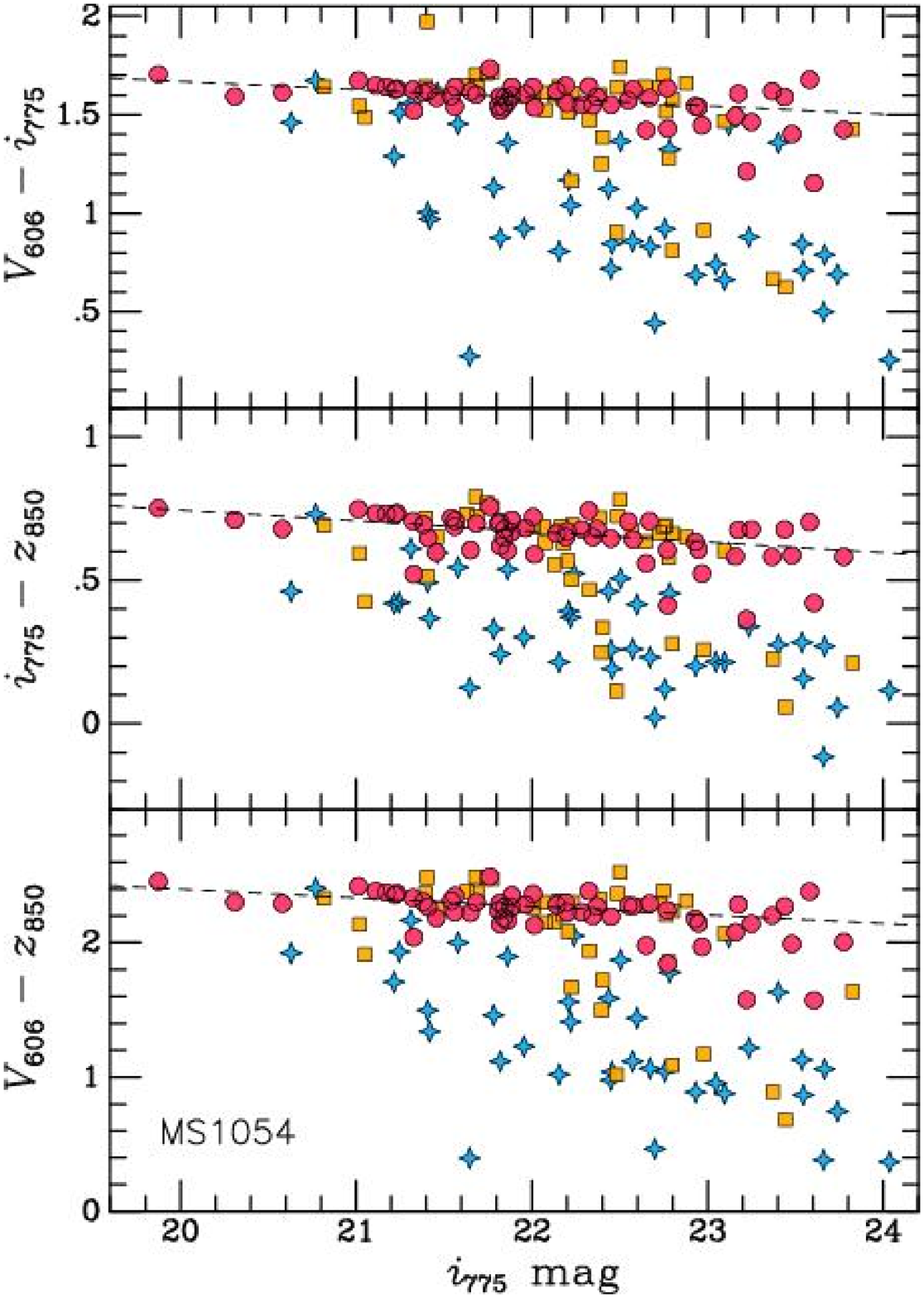}
\caption{Color--magnitude relations in \msname\
for visually classified ellipticals (red circles), S0s (orange squares),
and late-type galaxies (blue stars).  The dashed lines indicate the 
fitted relations for the elliptical galaxies given in the text.
\label{fig:cmrs1054}}
\end{figure*}

\subsection{Color-Magnitude Diagrams}
\label{ssec:cmds}

We adopt a value of $M_B^* = -21.2$ AB mag for the fiducial magnitude of
the Schechter (1976) function in the Johnson~$B$ band at $z=0.83$,
based on the Norberg \etal\ (2002) galaxy luminosity function
with $B \approx b_J + 0.25$
and $-0.95$ mag of luminosity evolution (van Dokkum \& Stanford 2003;
Holden \etal\ 2005b).  This corresponds to an observed magnitude of
$i_{775}^{*} = 22.0\pm0.1$ AB for an early-type spectrum, and agrees well
with the best-fit value obtained by Goto \etal\ (2005) 
in fitting the galaxy luminosity function of \msname.

Figures~\ref{fig:cmrs0152} and~\ref{fig:cmrs1054} show the color--magnitude
diagrams for the confirmed members of both clusters.  The different visually
classified morphological types are represented as circles (ellipticals), S0s
(orange squares), and stars (late-type galaxies).  All the colors are shown
as a function of \iacs\ magnitude to facilitate intercomparison of the plots.
Remarkably, there are no blue early-type galaxies in \rxname\ brighter than
$M^*$ ($\iacs < 22$).  Fainter than this, there are several blue S0 galaxies,
but only two significantly blue ellipticals (5$\,\sigma$ or more bluer
than the ``red sequence,'' the main locus of early-type galaxies).
The situation is similar in \msname: there are a few early-type galaxies
that scatter towards the blue at $\iacs < 22$, but the very blue S0s
are fainter than $\iacs = 22.2$, and there are no very blue ellipticals
brighter than $\iacs =23$.  In both clusters, late-type galaxies 
at magnitudes $\iacs>21$ scatter far to the blue.  Thus, the star-formation
is occurring in mainly faint, low-mass systems.

\subsubsection{Scatters}

To assess the color
scatter and offsets of the different galaxy subsamples within each cluster,
we first fit linear relations to the color-magnitude data for all the
elliptical members, down to a magnitude limit of $\iacs = 23.0$, where the
samples are largely complete.  For the purposes of this fit, we use simple
least squares with iterative 3-$\sigma$ clipping of the outliers, as in
Blakeslee \etal\ (2003a).  Van Dokkum \etal\ (2000) used a linear fit that
minimized the biweight scatter estimator, but found the result differed
negligibly from the clipped least squares fit.  
For \rxname, we obtain the following linear relations:
\begin{eqnarray}
%
\ricolor \,&=&\,  (1.267 \pm 0.008) - (0.007\pm0.011)\,(\iacs - 22)\; \nonumber\\
\izcolor \,&=&\,  (0.698 \pm 0.008) - (0.012\pm0.011)\,(\iacs - 22)~~~~~ \label{eq:cmrs0152}\\
\rzcolor \,&=&\,  (1.961 \pm 0.015) - (0.023\pm0.019)\,(\iacs - 22) \;. \nonumber
\end{eqnarray}
For \msname, we obtain:
\begin{eqnarray}
\vicolor \,&=&\,  (1.586 \pm 0.009)  - (0.042\pm 0.014)\,(\iacs - 22)\; \nonumber\\  
\izcolor \,&=&\,  (0.670 \pm 0.007)  - (0.037\pm 0.011)\,(\iacs - 22)~~~~~ \label{eq:cmrs1054}\\
\vzcolor \,&=&\,  (2.270 \pm 0.015)  - (0.064\pm 0.023)\,(\iacs - 22) \;. \nonumber
\end{eqnarray}
We then calculate the offsets and scatters with respect to these
relations using the biweight location and scatter estimators (Beers
\etal\ 1990) for various galaxy subsamples within each cluster.
The subsamples include groups based on morphology and magnitude alone;
morphology, magnitude, and radial position; and morphology, magnitude,
and local mass density.
Table~\ref{tab:stats0152} and Table~\ref{tab:stats1054} present the
results of these calculations for \rxname\ and \msname, respectively.
All uncertainties are estimated from bootstrap resampling.  The
intrinsic scatters $\sigma_{\rm int}$ for the subsamples are estimated
by correcting for the measurement errors described in
\S\ref{ssec:gxyphot}.

Tables~\ref{tab:stats0152} and~\ref{tab:stats1054} reveal several
interesting results on the galaxy populations in these two clusters.
First, as is typical, the elliptical galaxies have 
the smallest color scatters.  In \clname, the S0s exhibit color 
scatters roughly 50\% larger, while in \msname, the difference is a 
factor of two.  However, in neither cluster is there a significant
offset in the biweight mean locations of the ellipticals and S0s
brighter than $\iacs=23$.  The late-type confirmed members in
\msname\ scatter more to the blue than those in \rxname, but this 
is likely the result of the selection process in \rxname, 
where very blue members would be less likely to be included in sample
(see Demarco \etal\ 2005).

The tables also show results for subsamples limited in radius and
local surface density $\Sigma$, expressed in units of the critical density
$\Sigma_c = {c^2 \over 4\pi G D_L \langle\beta\rangle}$,
where $D_L$ is the angular diameter distance of the cluster lens,
and $\langle\beta\rangle$ is the mean ratio of the angular diameter distances 
between lens and source and between observer and source.
In Jee et al.\ (2005a,b), $\langle\beta\rangle$ is determined to be 0.282 and 0.290, 
for CL0152 and MS1054, respectively; thus, $\Sigma=1$ corresponds to
a surface mass density of $\sim\,$3650 $M_{\odot}\,$pc$^{-2}$ (the mass-sheet
degeneracy was lifted based on parametric fitting results).
For the radius in \rxname, since there are two similarly massive
concentrations merging, we use the harmonic mean of the radii from
the two centers, as given by Demarco \etal\ (2005); the harmonic
mean is constant on elongated contours that roughly follow the
constant density contours.  The local
densities are taken from the weak lensing mass maps of 
Jee \etal\ (2005a,b).  We find that for \rxname, the scatter is
essentially unchanged for either the elliptical (E) or early-type (E+S0)
subsamples when only galaxies within a mean radius of 1\farcm1 (0.5 Mpc)
are used.  However, there are marginally significant reductions in 
the scatter when only galaxies in locations with $\Sigma > 0.1$ are used.
Since the azimuthally averaged $\Sigma$ reaches 0.1 at a radius
near 1\farcm1 (Fig.\,16 of Jee \etal\ 2005a),
these two different types of selection yield about
the same number of objects, and the lower scatter in the density-selected
subsample reinforces the conclusion that the local density is 
responsible for the galaxy transformations. 

On the other hand, for
\msname, the reduction in color scatter is about the same for the 
$R<1\farcm1$ and $\Sigma>0.1$ samples.  But the mass distribution is
rounder in this cluster, making radius a better tracer of surface density 
(outside the complex central region), as well as more extended, so the mean
surface density reaches 0.1 at a larger radius of nearly two arcminutes.
The $R<1\farcm1$ radial cut effectively
occurs at a mean surface density $\Sigma\approx0.16$,
and includes $\sim\,$30\% fewer galaxies. Thus, we again conclude
that density is more important than radius, and will examine these
effects further in the following sections.  

We have also computed the offsets and scatters for samples using
a brighter magnitude cut of $\iacs<22.5$ and find that the scatters
are also reduced using this cut, particularly in \msname, for which
the $\vacs$ data are only a single orbit and shallower than the rest
of the images.  For \clname, the reduction in scatter is less for
this magnitude cut than for the density cut.

Finally, we note that in \S\ref{sec:bumpy}, we discussed an alternative set of
morphological classifications based on the bumpiness $B$ and S\'ersic index
$n$ parameters.  We find that the fitted linear relations and scatter
measurements using these \bn\ classifications are virtually identical to those
found above using the more standard visual ones, as expected from the good
correlation between the two types of classification.  This is highly promising
for future large datasets lacking visual classes.

\begin{figure*}\epsscale{0.88}
\plottwo{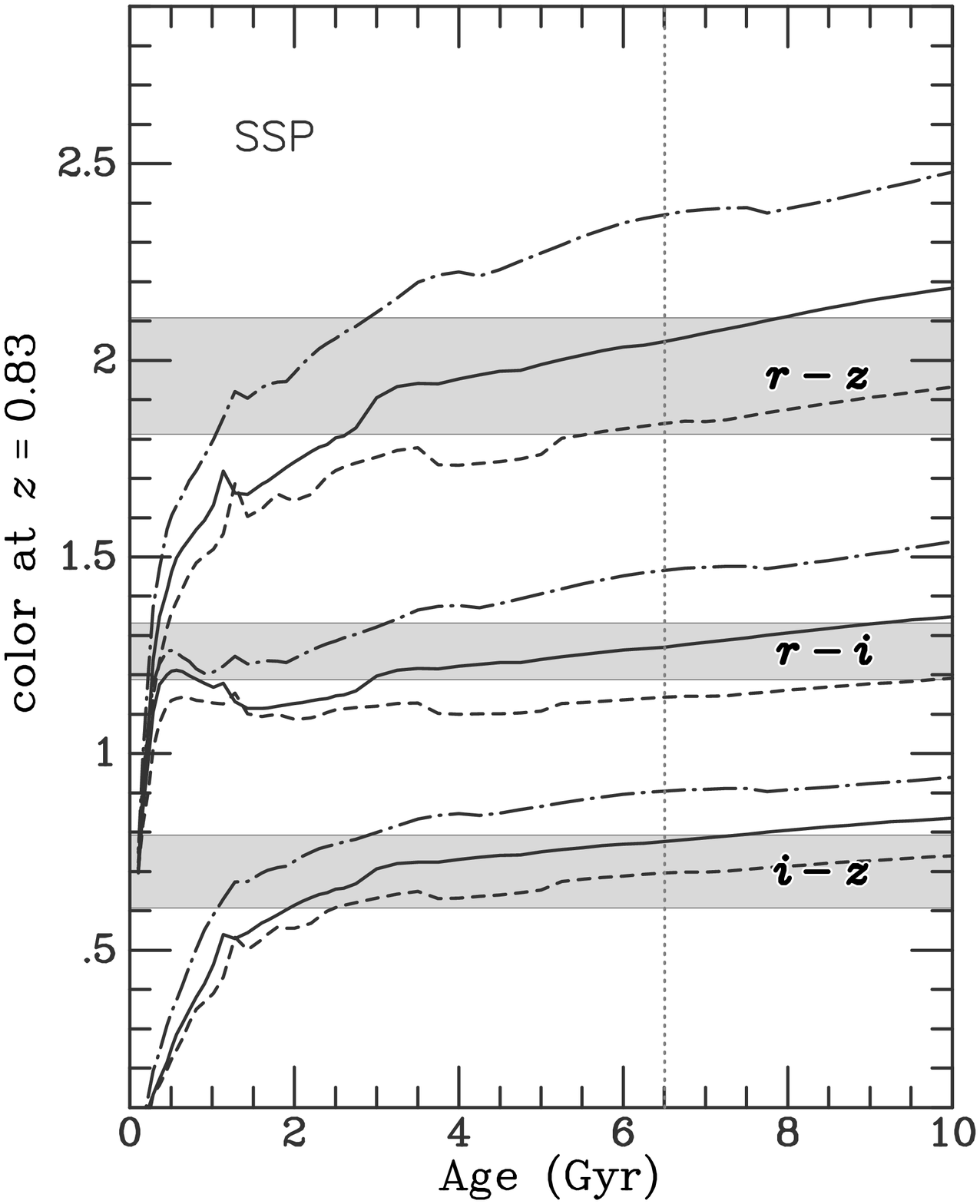}{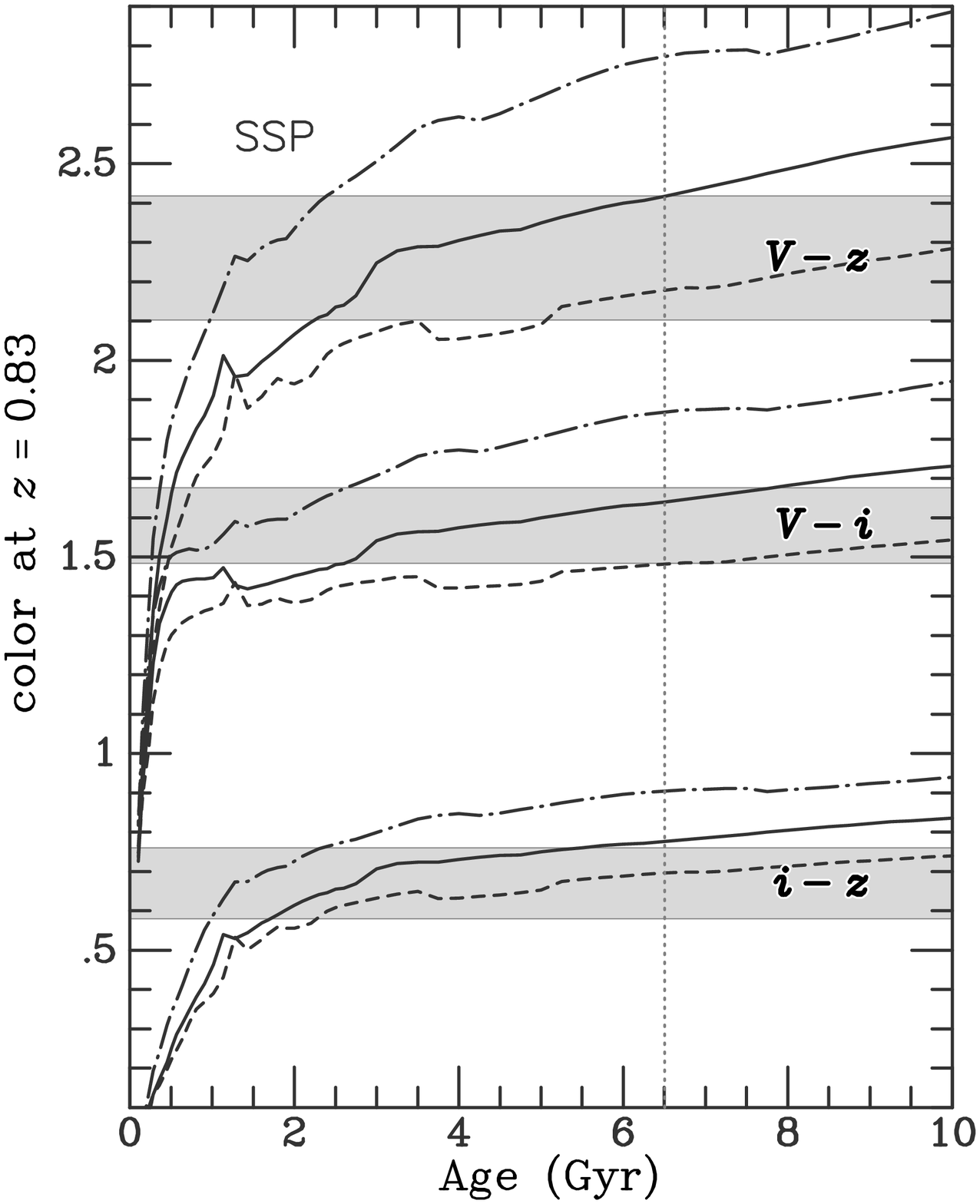}
\caption{%
Predicted evolution in the observed colors for Bruzual \& Charlot (2003)
single-burst stellar population (SSP) models of metallicity
$Z = 0.008$ (dashed line),
$Z = 0.02$ (solid line), and $Z = 0.05$ (dot-dashed line).
The shaded areas delineate the 2--$\sigma$ loci of the
elliptical galaxies in \rxname\ (left) and
\msname\ (right).  The vertical dotted line indicates
the approximate age of the universe at this redshift.
\label{fig:evol_ssp}}
\end{figure*}

\begin{figure*}\epsscale{0.88}
\plottwo{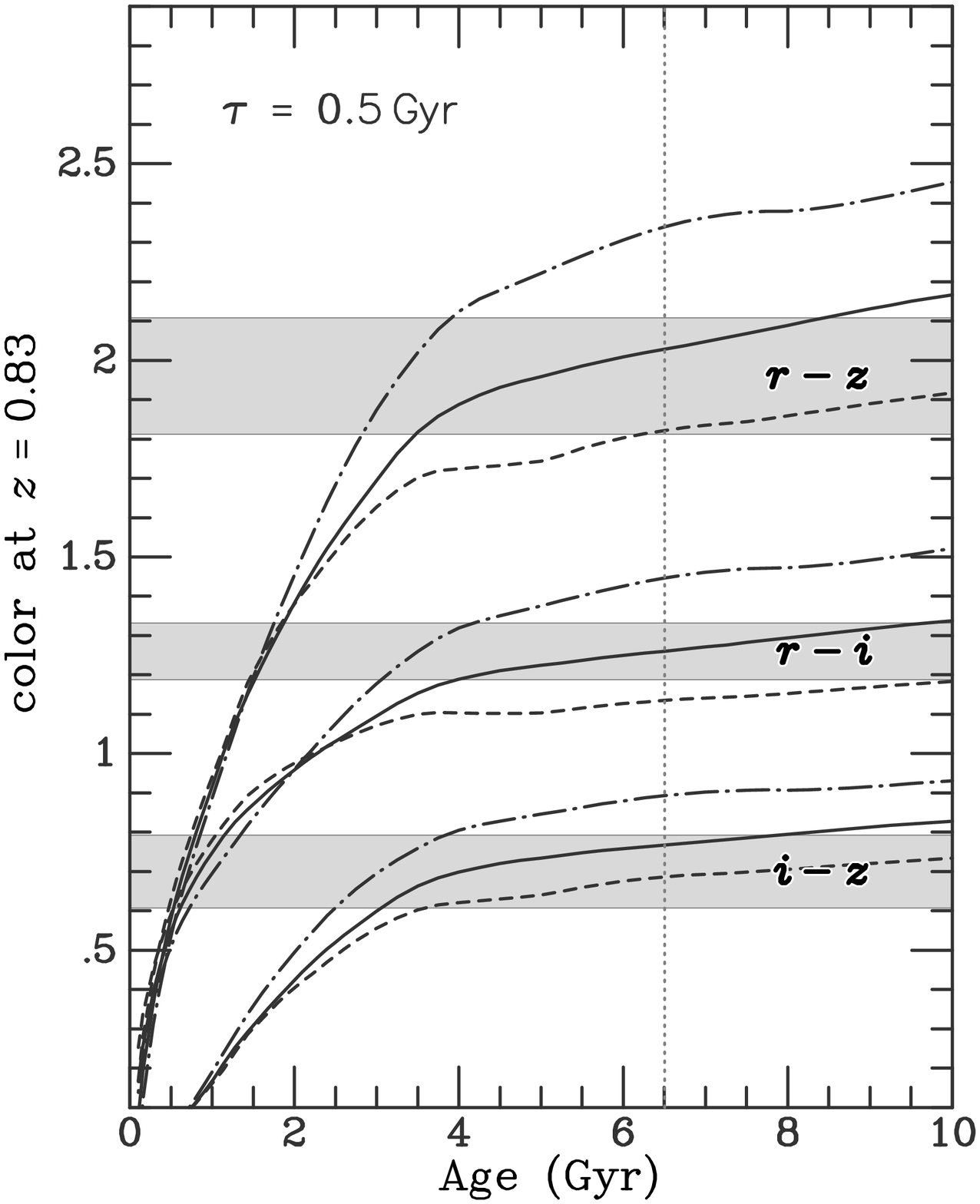}{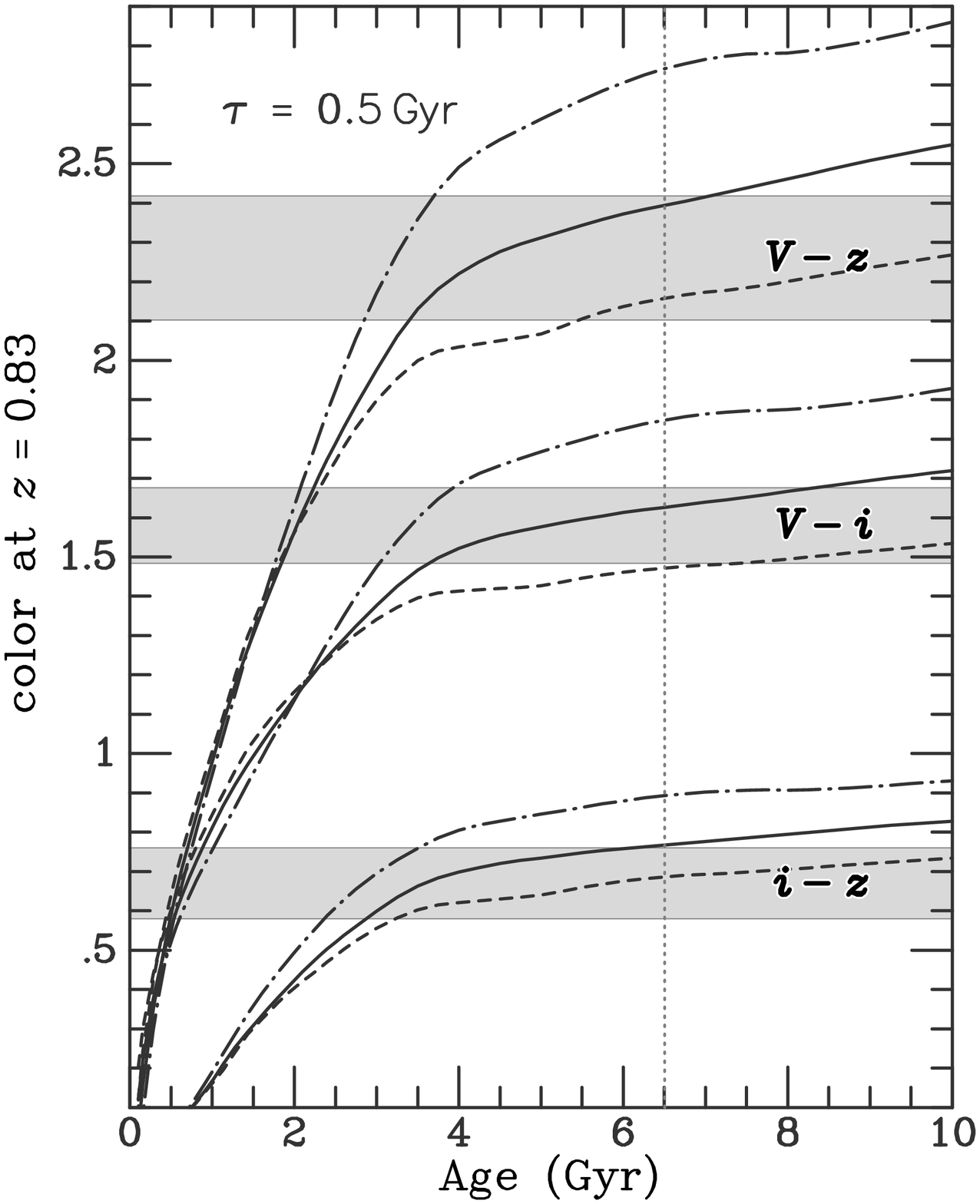}
\caption{%
Predicted evolution in the observed colors for Bruzual \& Charlot (2003)
exponentially declining star formation models of decay constant $\tau = 0.5$ Gyr
(no cutoff time), and metallicities
$Z = 0.008$ (dashed line),
$Z = 0.02$ (solid line), and $Z = 0.05$ (dot-dashed line).
The shaded areas delineate the 2--$\sigma$ loci of the
elliptical galaxies in \rxname\ (left) and
\msname\ (right).  The vertical dotted line indicates
the approximate age of the universe at this redshift.
Comparison of these models to the SSP ones in
Figure~\ref{fig:evol_ssp} illustrates how small changes in star-formation
history can affect the color evolution.
While these are still not true chemo-evolutionary models,
they are probably more realistic
in predicting smoother evolution.
\label{fig:evol_tau}}
\end{figure*}

\subsubsection{Color Evolution}

In order to assess which of the measured colors is most ``useful,'' we
show in Figures~\ref{fig:evol_ssp} and~\ref{fig:evol_tau} the predictions for
our measured color-indices at $z\approx0.83$ for single-burst stellar population
(SSP) and exponentially declining star formation models (``$\tau$ models'')
with time constant $\tau{\,=\,}0.5$~Gyr.
Neither type of model is very realistic, since they both neglect
chemical evolution. However, the $\tau$ models evolve more smoothly because
they mix stars at a range of ages, and thus avoid the artificial
``bumps'' seen in Figure~\ref{fig:evol_ssp} when stars of a single
specific spectral type suddenly come to dominate the galaxy spectrum.
We see that solar-metallicity populations of
ages $\sim$ 4--5 Gyr, or higher metallicity models of ages 1.5--3.5 Gyr,
can approximately reproduce the galaxy colors (the well-known age--metallicity
degeneracy in broad-band colors, e.g., Worthey 1994).  We return to this 
below in the presentation of the color--color diagrams.
We also note that the broadest baseline colors, \rzcolor\ for \rxname\
and \vzcolor\ for \msname, are the most sensitive to age and
metallicity changes, while being the least sensitive
to photometric errors and the small changes in redshift (since the
4000\,\AA\ break lies in the \iacs\ band).  We therefore concentrate on these 
colors when examining trends with radius and density below.

\begin{figure}\epsscale{1}
\plotone{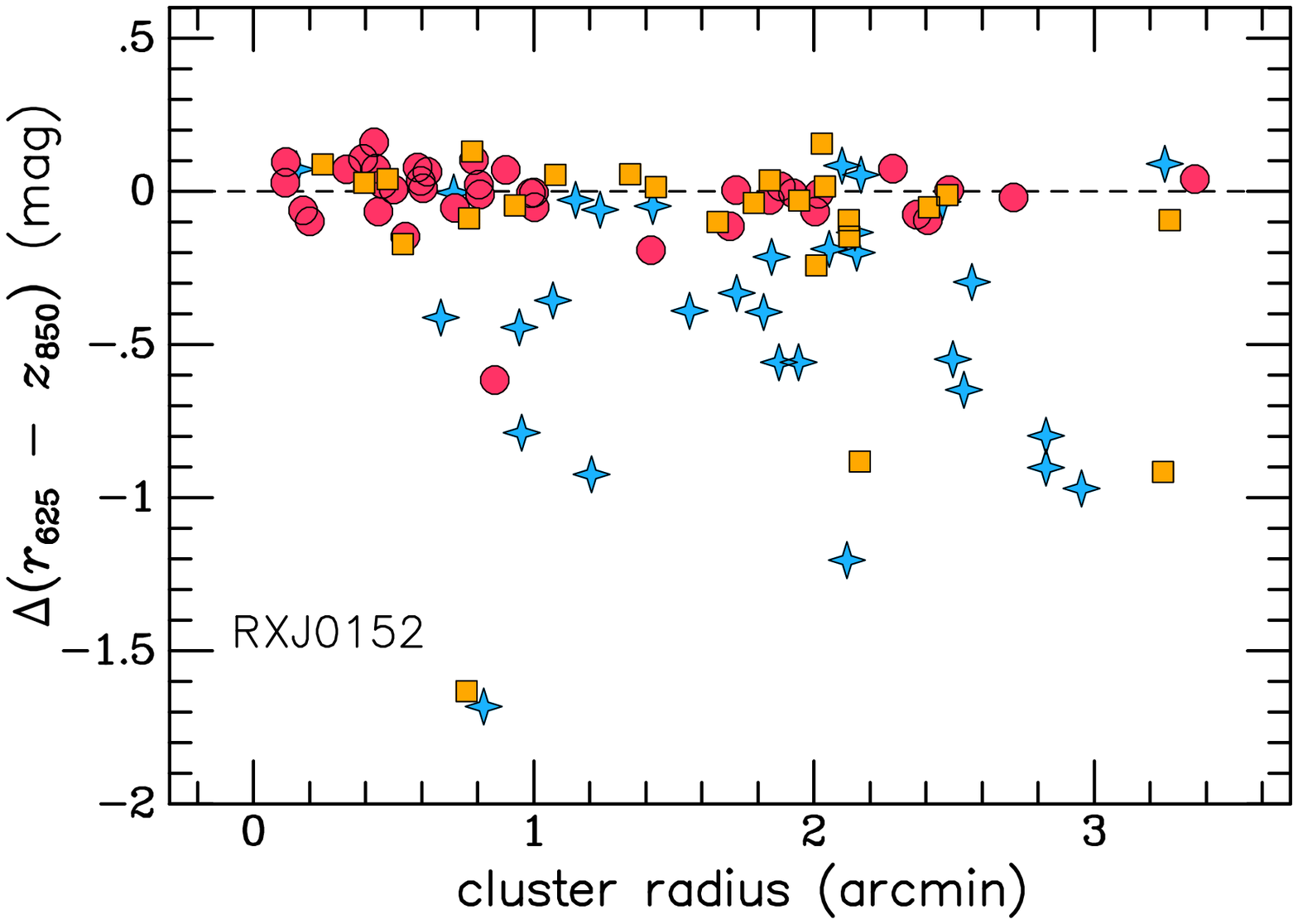}
\plotone{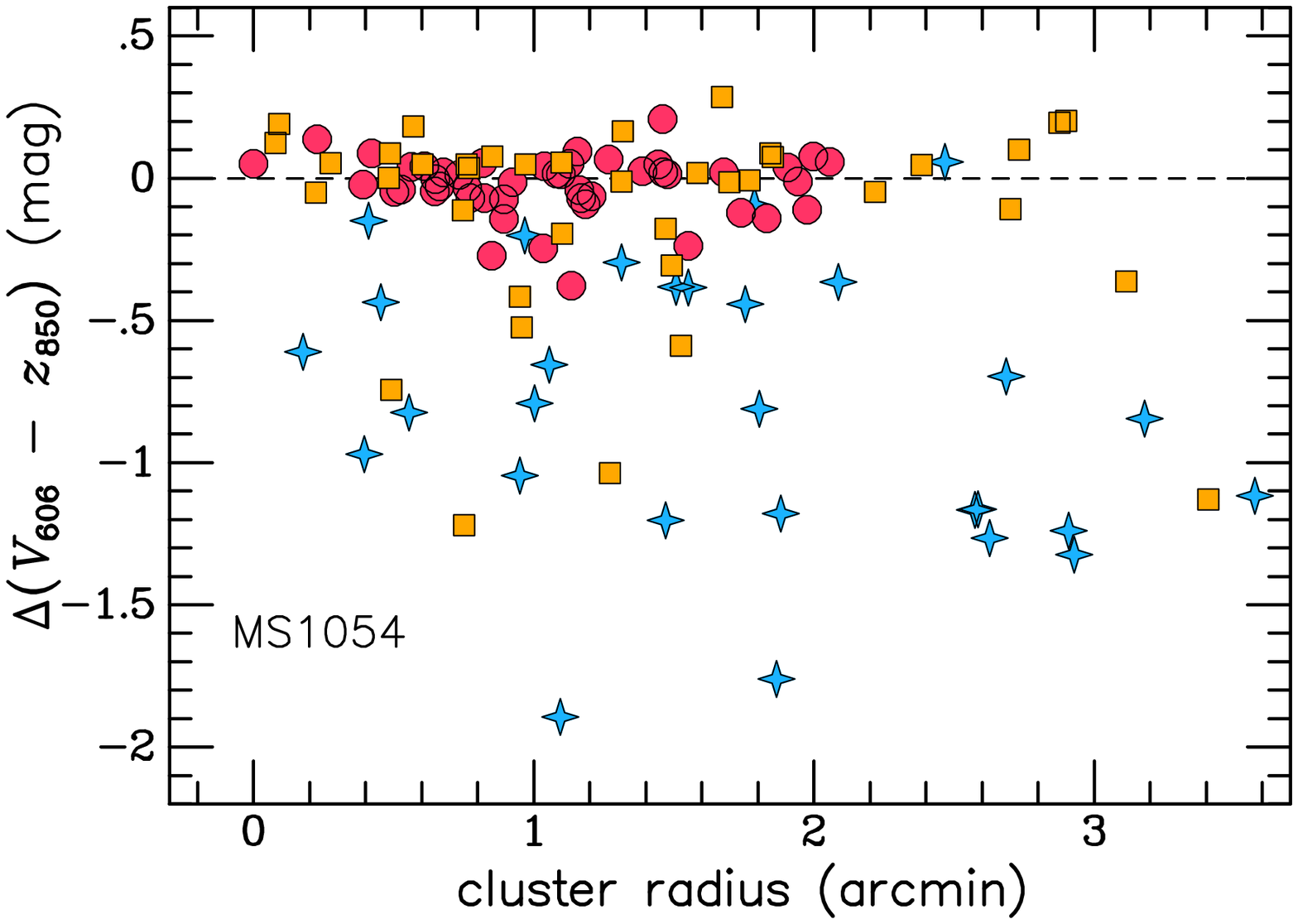}
\caption{Residuals with respect to the elliptical galaxy
color--magnitude relation in \rxname\ (top) and \msname\ (bottom)
as a function of radial distance from the cluster centers.  
Symbols are as in Fig.\,\ref{fig:cmrs0152}.
Since there are two major, distinct mass centers in \rxname, 
we use the harmonic mean of the radii from the two centers, defined as
in Demarco \etal\ (2005), in this case.  For \msname, we use
the location of the centrally dominant cD galaxy. 
\label{fig:crads}}
\end{figure}

\begin{figure}\epsscale{1}
\plotone{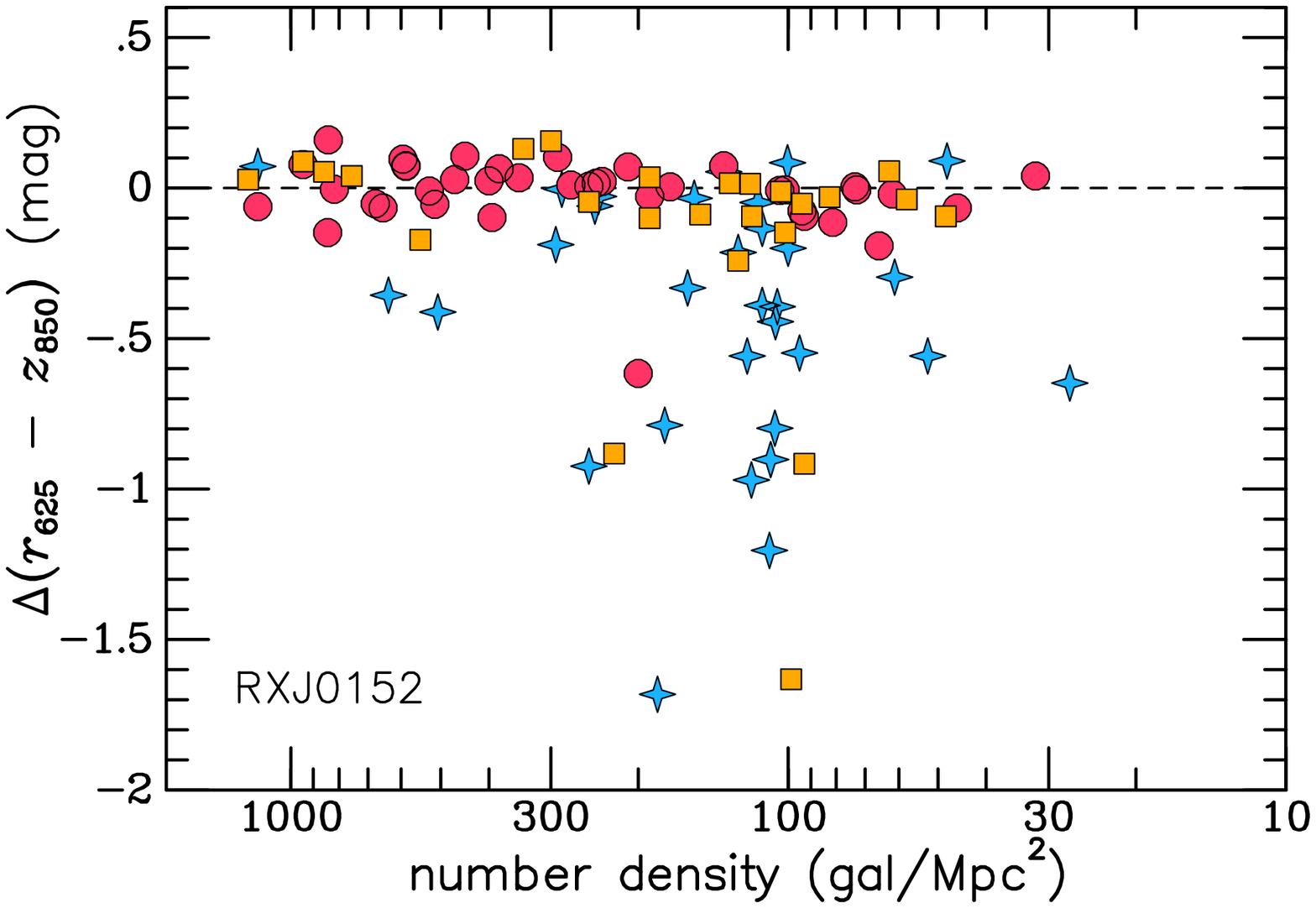}
\plotone{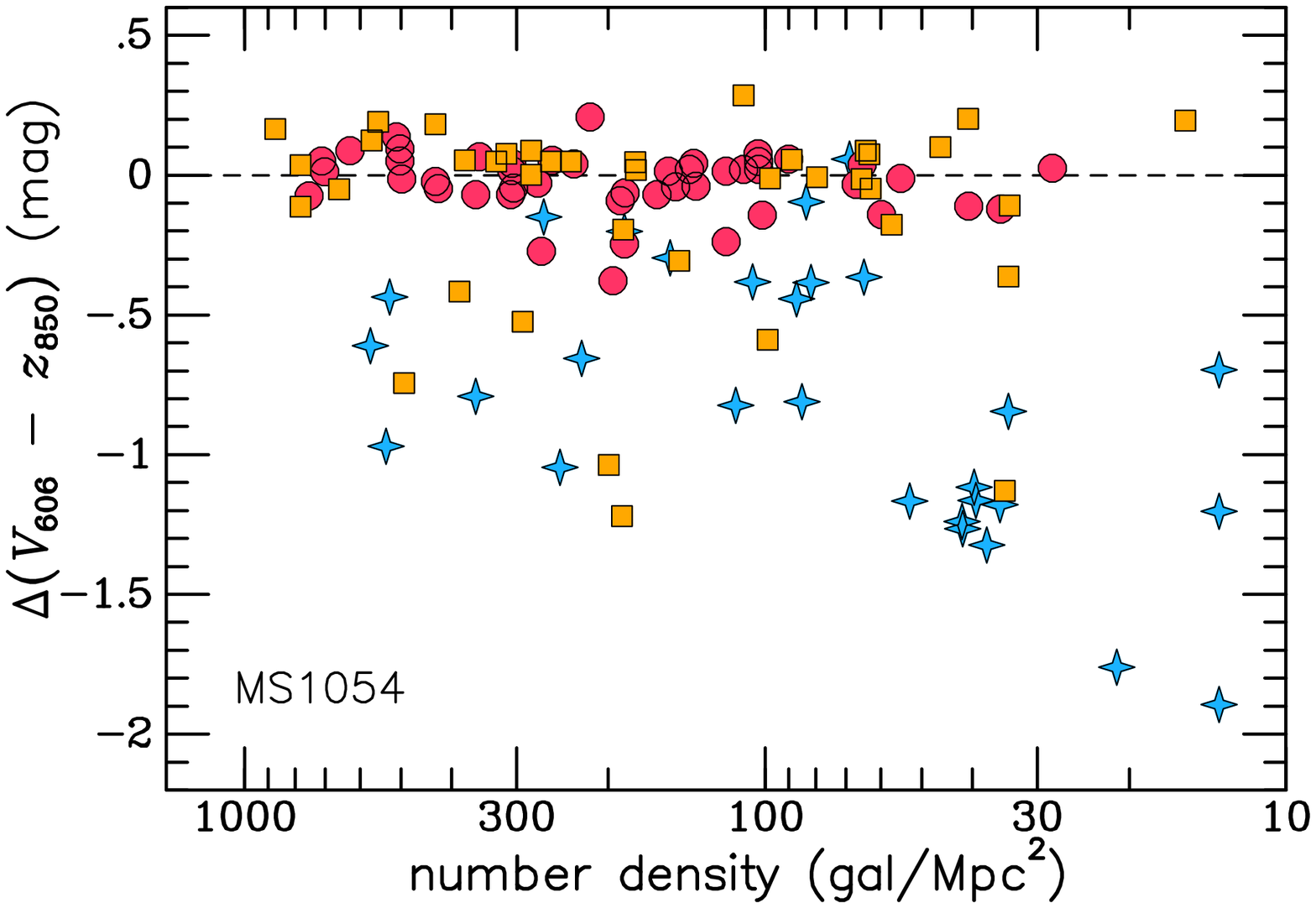}
\caption{Residuals with respect to the elliptical galaxy
color--magnitude relation in \rxname\ (top) and \msname\ (bottom)
as a function of the local number density of cluster galaxies,
from Postman \etal\ (2005).  Symbols are as in Fig.\,\ref{fig:cmrs0152}.
\label{fig:cNdens}}
\end{figure}

\begin{figure}\epsscale{1}
\plotone{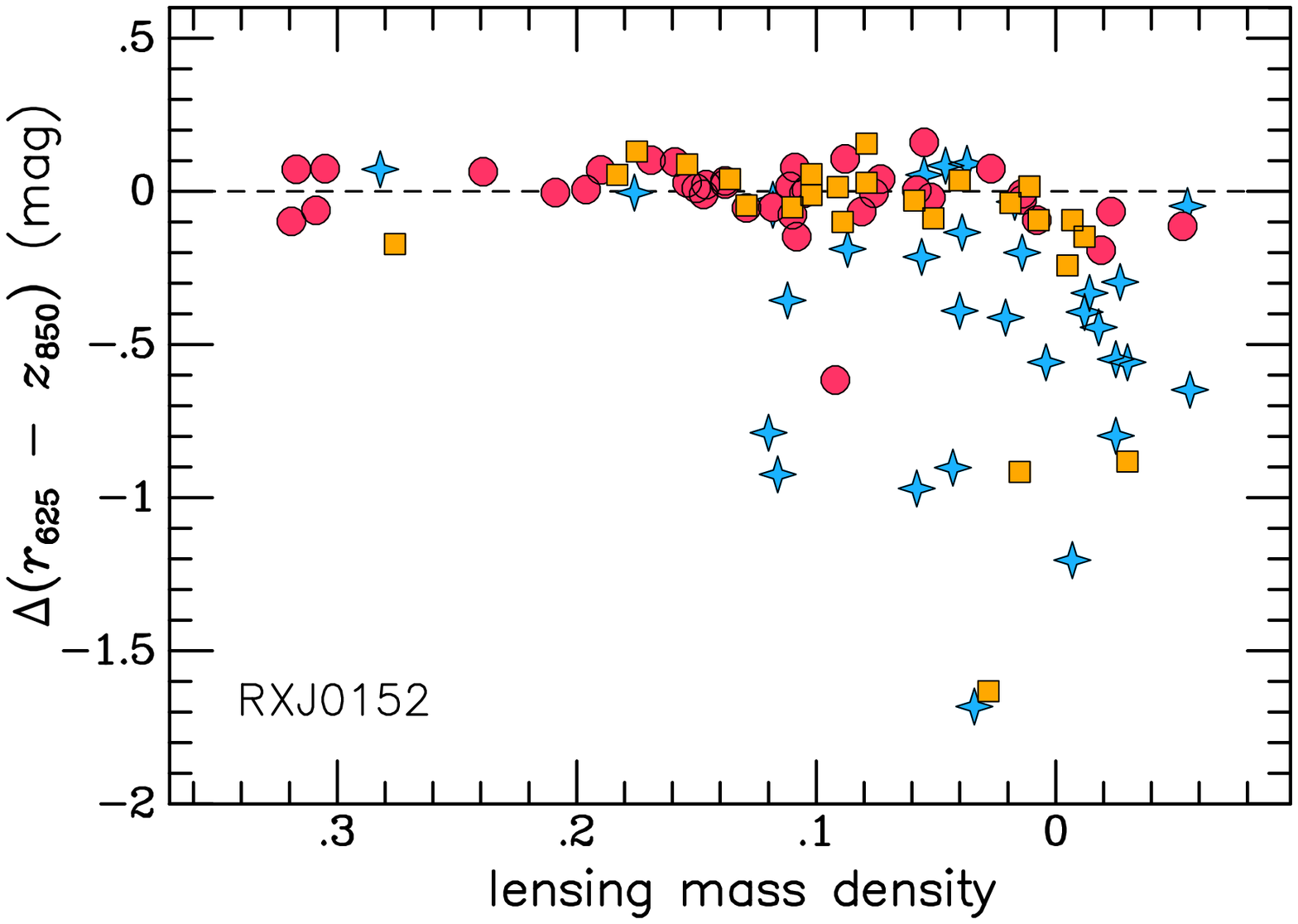}
\plotone{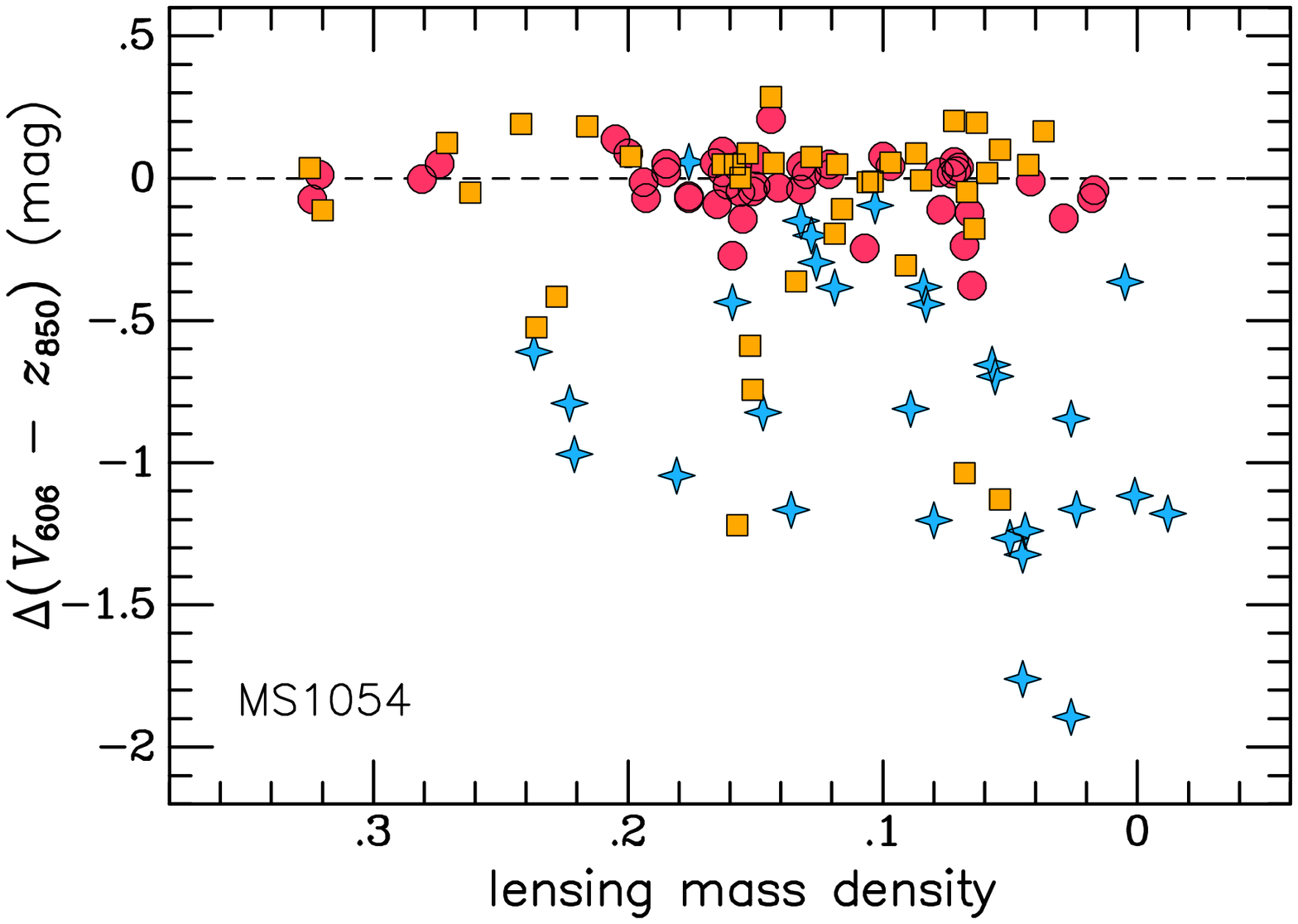}
\caption{Residuals with respect to the elliptical galaxy
color--magnitude relation in \rxname\ (top) and \msname\ (bottom)
as a function of the dimensionless surface mass density $\Sigma$ 
from weak lensing (Jee \etal\ 2005a,b).
Symbols are as in Fig.\,\ref{fig:cmrs0152}.
\label{fig:cMdens}}
\end{figure}

\subsection{Dependence on Cluster Radius and Density}

Figure~\ref{fig:crads} shows the deviations of the galaxy colors
from the elliptical galaxy color-magnitude relations, given by
Eqs.~\ref{eq:cmrs0152} and~\ref{eq:cmrs1054}, as a function of radius.
We use the broadest baseline colors for the reasons given immediately above
and plot only galaxies with $\iacs < 23$, the 
limit used for the color-magnitude fits.
We tested the significance of the correlations between 
color residuals and cluster radius, for different galaxy classes and 
the population as a whole, using both Spearman rank ordering
and Pearson's~$r$.   Table~\ref{tab:rescorrels}
shows these estimated significance levels, as well as those
for the number- and mass-density correlations discussed below.
In both clusters, the overall galaxy populations (all types)
become significantly bluer with radius. This results from the well-known
morphology--density relation (Dressler \etal\ 1980) and the fact that
late-type galaxies are bluer.  There is no strong evidence for
correlations in the color residuals for individual galaxy classes with radius,
although there is some correlation for the combined early-type (E+S0)
galaxies in \rxname.

The correlations become stronger when plotted against measures
of the local density. Figure~\ref{fig:cNdens} shows the same color
residuals, but now plotted as a function of galaxy number density
as calculated by Postman \etal\ (2005).
As evident from Table~\ref{tab:rescorrels}, the colors for the full 
galaxy sample in both clusters show strong correlations with number
density based on both measures of significance.
This again mainly reflects the segregation between the spatially 
concentrated red early-type and dispersed blue late-type galaxies.  
However, now we also find significant color--density
correlations for the early-type galaxies in \rxname\
and for the late-type galaxies in \msname,
but not other subsamples within these clusters.
Figure~\ref{fig:cMdens} plots the color residuals against 
surface mass density $\Sigma$ from the weak lensing maps of Jee \etal\ (2005a,b).
Once again, the correlations for the full samples are highly significant.
Furthermore, in \rxname, there is at least some evidence for 
correlations for each of the individual subsamples, although 
the evidence is weak ($\sim1.5\sigma$) for the late-type subsample.
Conversely, in \msname, we find the strongest evidence for
correlation in the late-type galaxy subsample, weak evidence for
ellipticals, and no correlation for the S0s,
which show considerable color scatter in \msname.

While the significance levels are worth computing, it is
more enlightening to observe the basic features of
Figures~\ref{fig:cNdens} and~\ref{fig:cMdens}.  In \rxname, it is
striking that there are no blue cluster members (brighter than
$\iacs{=}23$) at number densities $\gta 300$ galaxies~Mpc$^{-2}$ or
mass densities $\Sigma \gta 0.12$ ($\sim\,$440 $M_{\odot}\,$pc$^2$).
Similarly, the emission line galaxies in this cluster are found outside 
the regions of highest number density and X-ray emission 
(Demarco \etal\ 2005; J{\o}rgensen \etal\ 2005; Homeier \etal\ 2005).
Tanaka \etal\ (2005) also find that the colors
of galaxies go from predominantly blue to predominantly red above
some local number density in \rxname\ and other clusters.
Using the ACS GTO cluster sample, Homeier \etal\ (2006) find a
difference in the mean colors of late-type field and cluster
galaxies, consistent with the color--density trends that
we find for the late-type cluster galaxies.

In terms of surface mass density from weak lensing,
Gray \etal\ (2004), using  COMBO-17 data, observed a similar threshold
for the presence of blue galaxies in the A901/A902 supercluster.
However, they found that the abrupt transformation from star-forming
to quiescent galaxies occurred mainly at faint luminosities,
one magnitude below $M^*$; the trend still remained
for brighter galaxies, but was weaker and more gradual.  Our sample
only reaches to a limit of $M^* + 1$, yet we find the transition
to be clear and abrupt in \rxname.  This suggests that the time scale
for shutting off star-formation in galaxies being accreted from the field
is short (much less than the $\sim\,$1~Gyr crossing time).
Moreover, all the galaxies
(both early- and late-type) at the lowest mass densities $\Sigma\lta0.02$
in this cluster lie below the linear fit the red sequence.
This may also be an age effect: a consequence of star-formation 
occurring earlier in larger dark matter halos, coupled with the mass 
segregation of dark matter halos within the cluster.

Similar, but less striking, trends are present in \msname, with the
transition between purely quiescent and predominantly star-forming 
galaxies occurring more gradually between $\Sigma\sim0.2$ and $\Sigma\sim0.1$
Fainter galaxies with $\iacs>23$ are primarily blue and late-type,
in regions with $\Sigma\lta0.1$, so adding these makes this
threshold appear stronger.
The ellipticals in the lowest density regions, $\Sigma<.1$,
again lie mainly below the locus of the red sequence.
But the ACS weak lensing map for \msname\ does not reach surface densities
as low as in the outer parts of the \rxname\ field.

\begin{figure}\epsscale{1.1}
\plotone{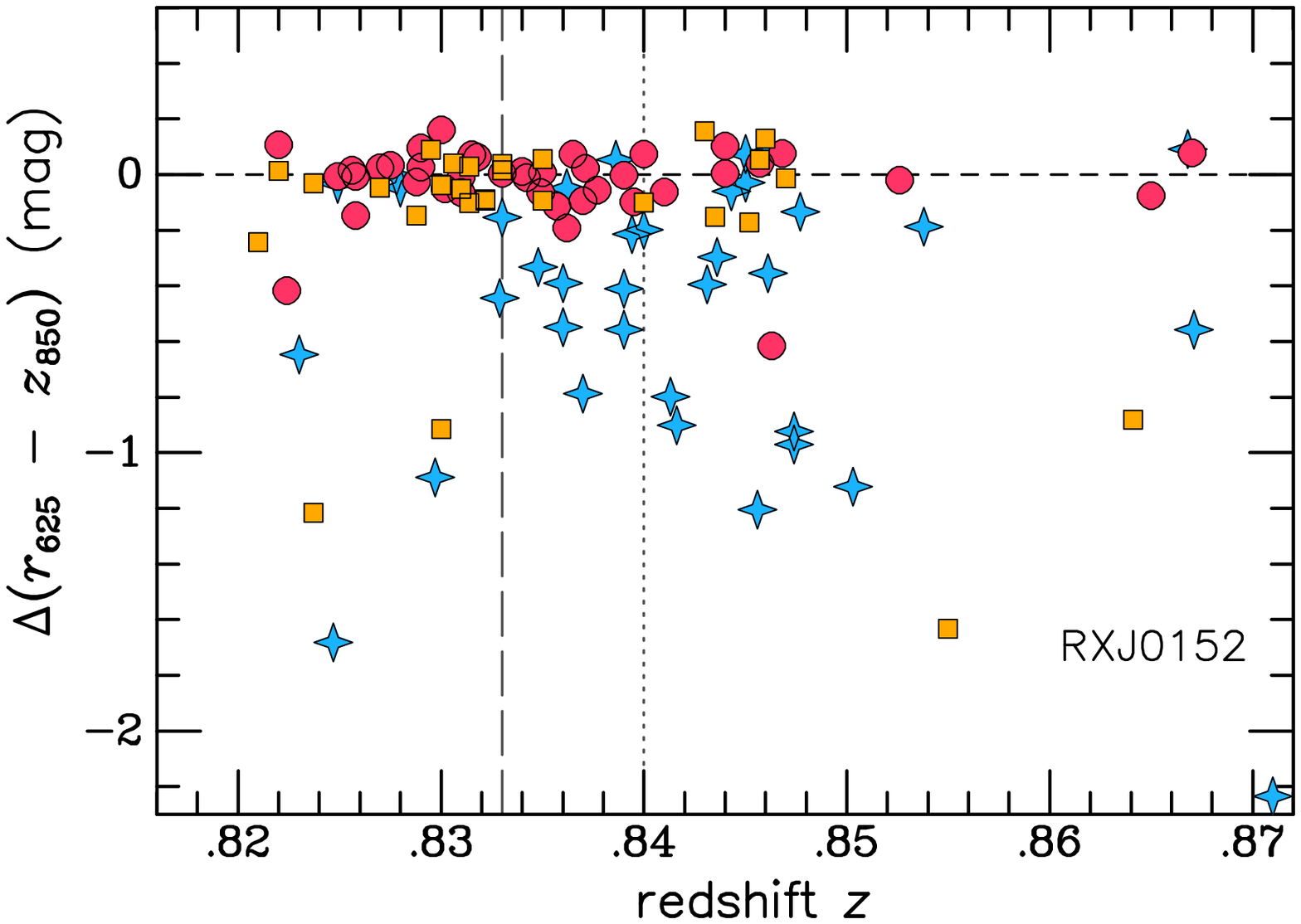}
\plotone{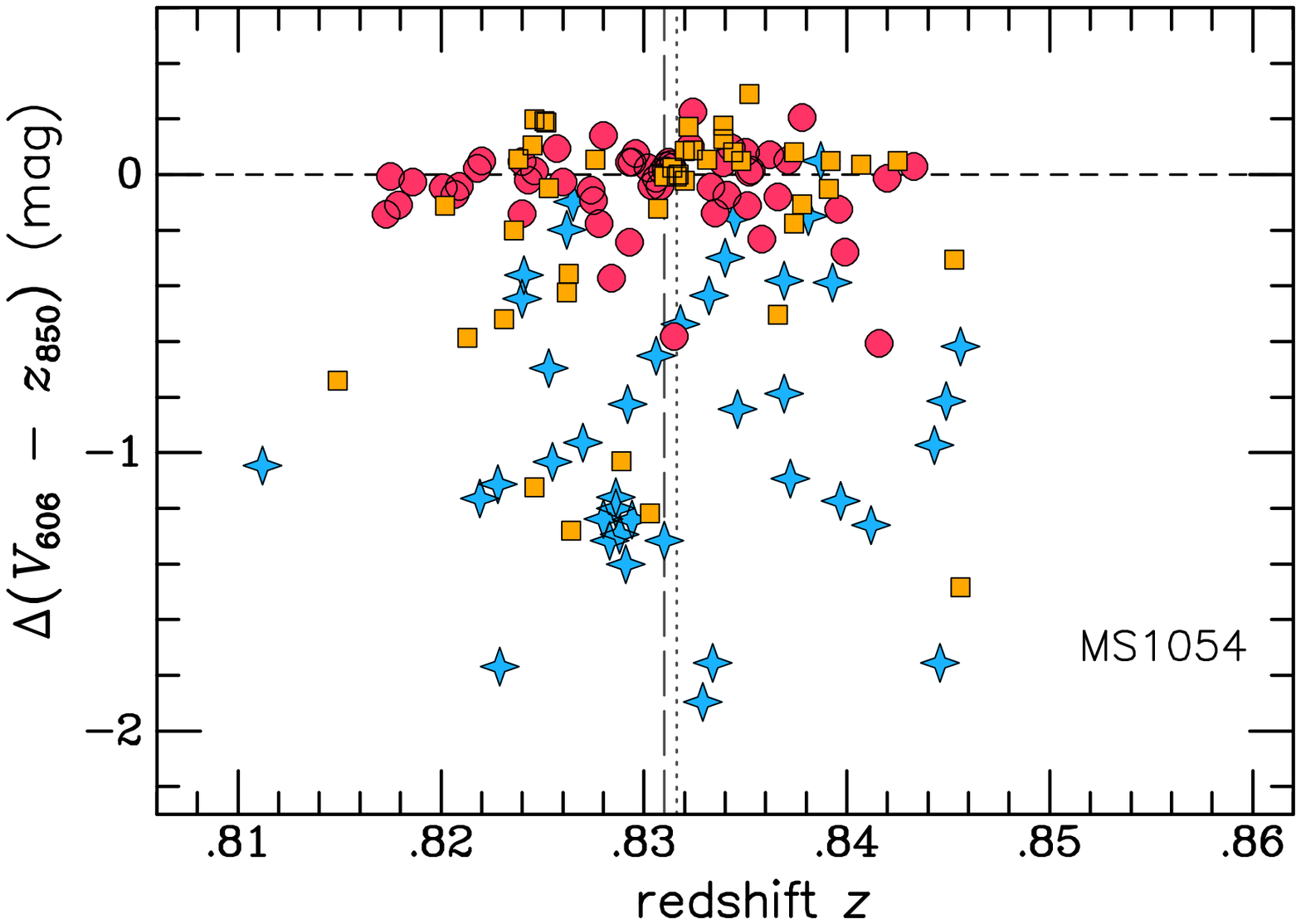}
\caption{Residuals with respect to the elliptical galaxy
color--magnitude relation in \rxname\ (top) and \msname\ (bottom)
as a function of redshift~$z$. Symbols are as in Fig.\,\ref{fig:cmrs0152}.
The same $\Delta z$ range is used
for both clusters in order to illustrated how much more dispersed
\rxname\ is in redshift space.  The dashed vertical lines mark the
median redshifts of the early-type galaxies in each cluster,
while the dotted vertical lines mark the median redshifts of the
late-type samples.
\label{fig:cvels}}
\end{figure}

\subsection{Effects of Cluster Substructure}

Both \rxname\ and \msname\ are structurally complex, having multiple
subclumps in the galaxy, mass, and X-ray maps (see references in \S1).
\rxname\ also presents a complex velocity histogram with multiple distinct
peaks, indicating the cluster is far from dynamical equilibrium (Demarco
\etal\ 2005; Girardi \etal\ 2005). 
The velocity distribution appears more regular in \msname\
(van Dokkum \etal\ 2000).  We wish to examine the galaxy colors as a function
of velocity (redshift) and look for differences relating to the cluster
substructure.  

Figure~\ref{fig:cvels} shows the \rzcolor\ and \vzcolor\
deviations from the elliptical galaxy relations in Eqs.~\ref{eq:cmrs0152}
and~\ref{eq:cmrs1054} as a function of galaxy redshift.
The size of the redshift interval is the same in both panels;
clearly, \rxname\ is more dispersed and multimodal.
Adopting the nomenclature from Demarco \etal\ (2005),
the southern clump is centered at $z=0.830$, the northern clump
at $z=0.838$, and the much smaller eastern clump near $z\approx0.845$.
There is also a more diffuse swath of galaxies, mainly off
to the west, with $z\approx0.866$.  Despite this complex 
velocity structure, we find no clear trends in the mean
colors of the early-type galaxies as a function of redshift.
We note that we have applied small corrections to the galaxy
colors to correct for bandpasses differences with redshift,
but these are negligible for \rzcolor\ at $z{\,\approx\,}0.83$
because the 4000\,\AA\ break lies between the two bandpasses,
in the \iacs\ band.

\begin{figure}\epsscale{1.1}
\plotone{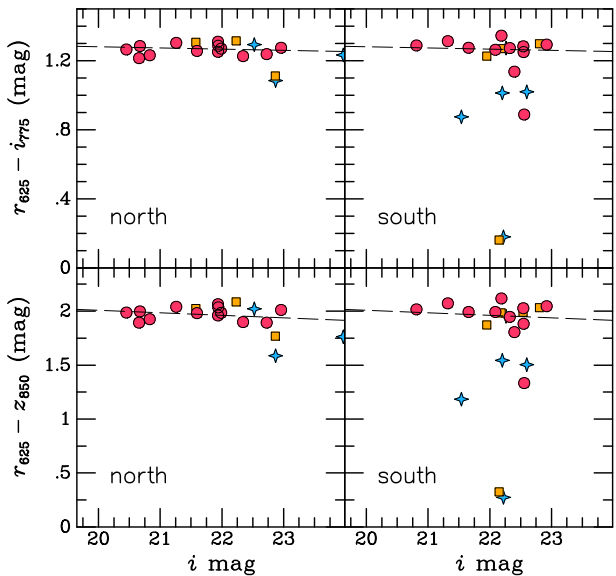}
\caption{Color-magnitude relations for the ``Northern'' and ``Southern''
clumps of \rxname, as defined by Demarco \etal\ (2005). Dashed lines
indicate the overall relations for the cluster ellipticals.
Symbols are as in Fig.\,\ref{fig:cmrs0152}.
\label{fig:cmr_clumps}}
\end{figure}

However, there is an overall trend of galaxy color becoming
bluer with velocity in \rxname.  This is caused by the presence
of a significant velocity offset between the early- and late-type
members of the cluster.  We find biweight mean redshifts of
$0.8337\pm0.0012$  and $0.83970\pm0.0017$ for the 
early- and late-types, respectively (the medians are 0.833 and 0.840).
This is a 3\,$\sigma$ difference, corresponding to about 1000~\kms,
and may indicate a large association, or sheet, of late-type
galaxies falling into \rxname\ from the ``front.''  It is unlikely
that these galaxies would have passed through the cluster core,
since the crossing time is nearly 1~Gyr, long enough to effect
the transformation to redder, earlier-type galaxies.
In this regard, it is interesting that these blue, late-type,
infalling galaxies are overwhelmingly projected onto the
low-density regions (cf.\ Fig.~\ref{fig:cMdens}), which may
simply indicate the ones infalling through the cluster center
are quickly transformed to early-type galaxies.
Alternatively, they could be a background grouping that 
simply has a larger Hubble velocity and has not yet ``turned around''
to begin infalling onto the main cluster -- the distance offset
would be $\lta10$ Mpc for our adopted cosmology.  But in this case there
is no reason why they would appear to avoid the highest density regions.
We note that there is no significant velocity offset 
between the early- and late-type galaxies in \msname\
(lower panel of Figure~\ref{fig:cvels}), in line with the
evidence that this is a more dynamically evolved cluster.

In the previous section we examined trends in the colors with
radial position in the cluster and local mass density.
It is difficult to study color trends for different mass
clumps because of the overlapping spatial positions
of galaxies in the different clumps and/or small number statistics.
For instance, in \msname, while there are several distinct 
peaks in the X-ray and mass distributions, they are not well
separated (and the X-ray and mass peaks are not always cospatial),
so there is no good way to assign galaxies to the different clumps.
However, the structure is more clear in \rxname, where there are
two massive well-separated clumps in the early stages of merging,
the northern and southern clumps defined by Demarco \etal\ (2005)
based on the X-ray enhancements.
These are distinct in both the X-ray and weak lensing maps,
and we can use the circular regions given by Demarco \etal\
to assign galaxies to them (the eastern clump and other concentrations
found in the mass map are much smaller and have relatively few galaxies).

Figure~\ref{fig:cmr_clumps} shows the \ricolor\ and \rzcolor\
color--magnitude diagrams for the two major \rxname\ clumps, along with the
mean linear relations given above.  There are 19 galaxies in our sample in
the northern clump (18 with $\iacs<23$) 
and 21 in the southern clump. Both clumps contain
16 early-type galaxies, with 13 (11) of those in the northern (southern)
clump being ellipticals.  It is immediately apparent that the
northern clump contains almost entirely red galaxies (with one of
the late-types being moderately blue), while $\sim\,$30\% of
the galaxies in the southern clump have blue colors (including two
of the early-types).  Moreover, there are twice as many galaxies brighter
than $i_{775}^*=22$ in the northern clump.  These features suggest
that the northern clump is a more evolved system, consistent with
its higher lensing mass (Jee \etal\ 2005b).

We calculated the offsets and scatters of the early-type galaxies in the two
clumps with respect to the mean color-magnitude relation.  The zero-point
offsets in \rzcolor\ are $0.009 \pm 0.022$ and $0.031 \pm 0.022$ mag (biweight
mean location with bootstrap errors), for the northern and southern clumps,
respectively.  The two zero points agree with the overall relation to
within 1.4\,$\sigma$ and with each other to less than $1\,\sigma$.
The biweight scatters in \rzcolor\ are $0.077\pm0.015$ and
$0.086\pm0.026$, respectively; for comparison, the straight rms scatters would be 
$0.079\pm0.014$ and $0.442\pm0.195$, showing the effect of the two early-type
outliers in the southern clump.  Thus, there is no significant
difference between the main part of the early-type populations in
the two clumps, but the southern clump is more ``contaminated'' by
blue galaxies, suggesting that it is the sight of more current infall.

\begin{figure*}\epsscale{0.71}
\plotone{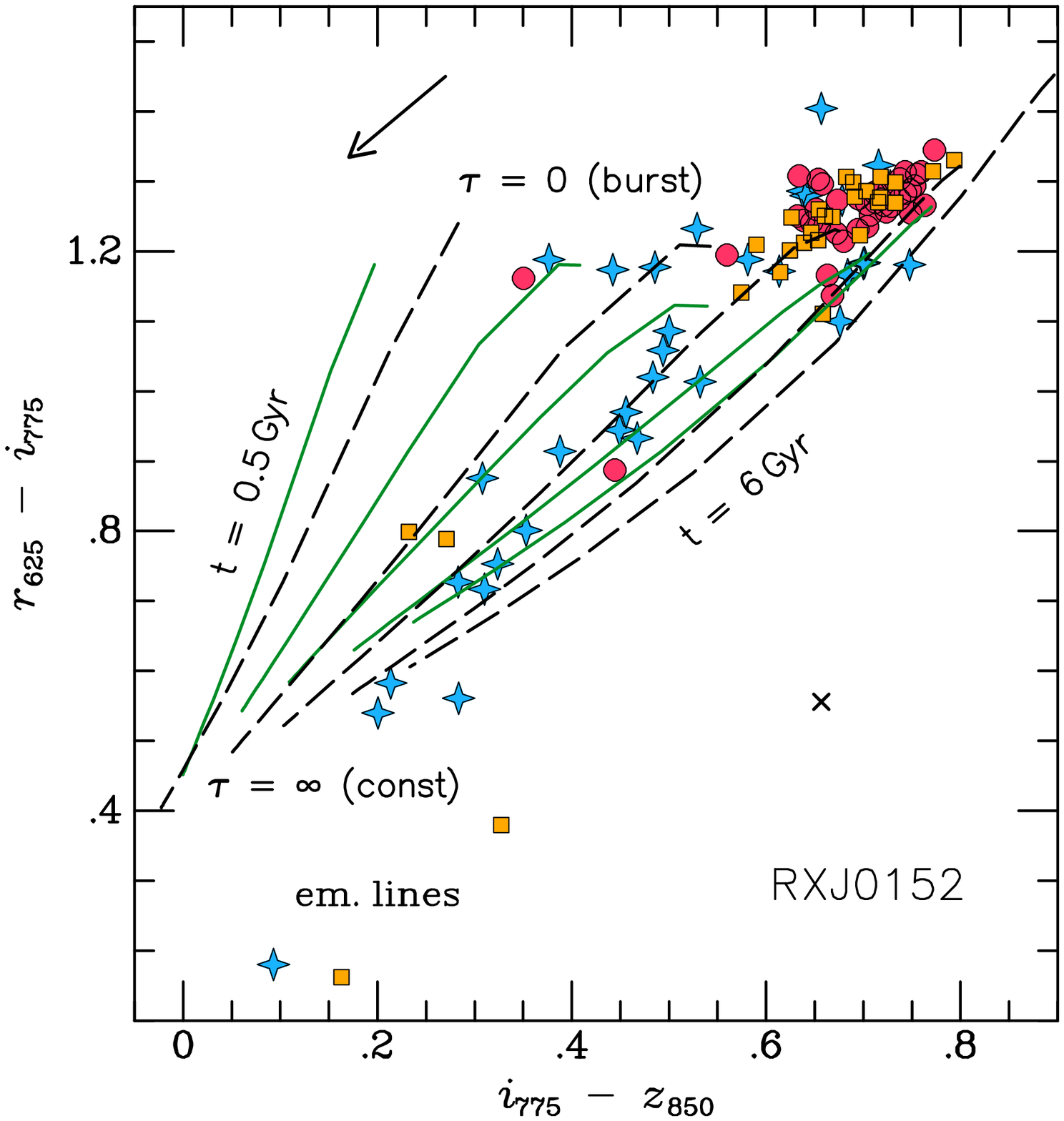}
\vspace{0.1cm}
\plotone{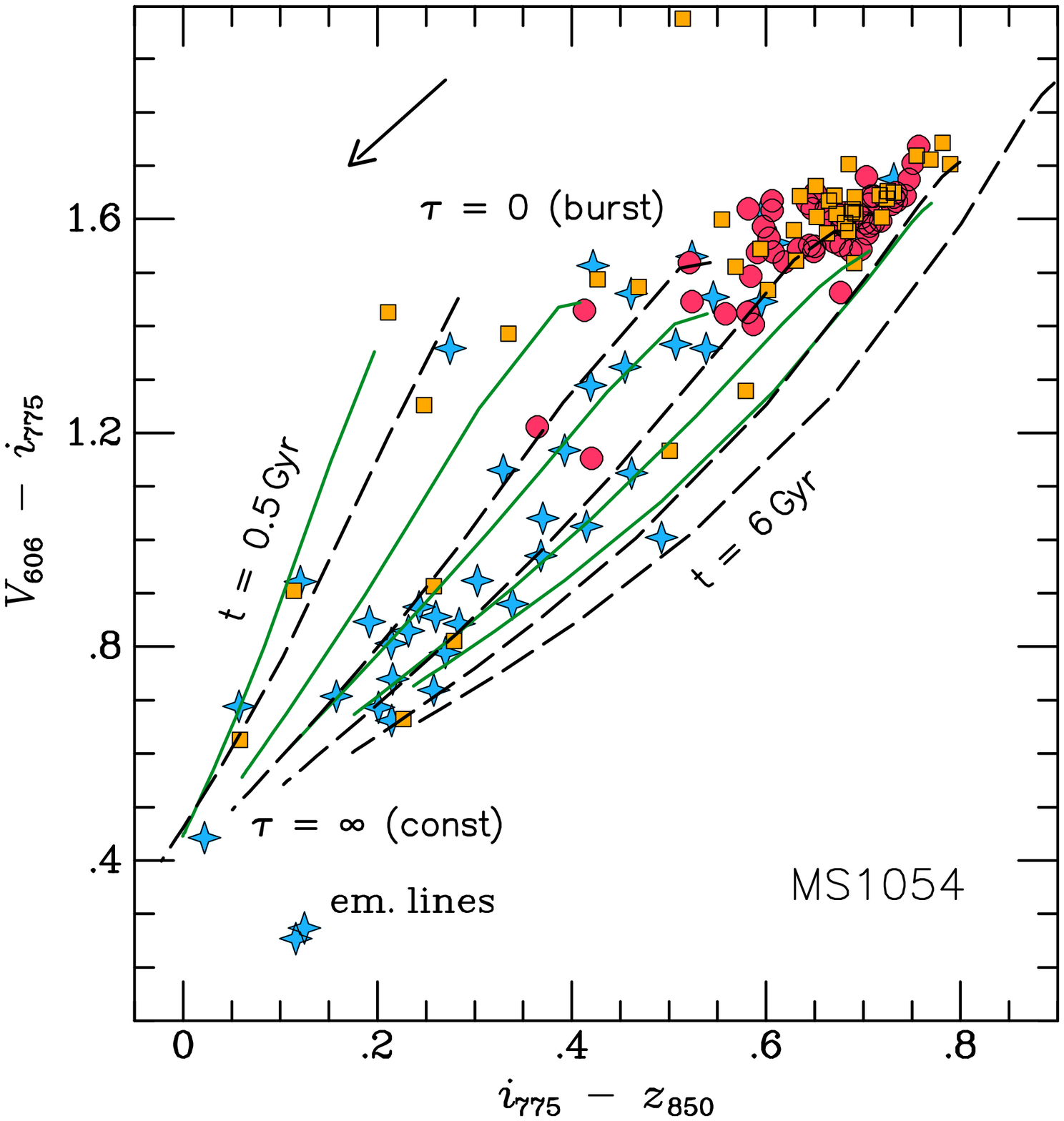}
\caption{Color--color diagrams for \rxname\ and \msname.
Symbols are as in Figure~\ref{fig:cmrs0152}.
The curves show Bruzual \& Charlot (2003) population models of varying
exponential star-formation time~scales from
constant ($\tau\approx\infty$) to single-burst ($\tau\approx0$),
varying ages since the onset of star~formation 
$t= 0.5$, 1, 1.5, 3, and 6 Gyr,
and two metallicities: solar ($Z{\,=\,}0.02$, green solid lines); 
2.5$\times$ solar ($Z{\,=\,}0.05$, black dashed lines).
The arrows near the tops of each panel show the correction vector for 
$E(B{-}V) = 0.15$ mag of internal reddening by dust.
\label{fig:clr_clr_diag}}
\end{figure*}

\subsection{Color--color Diagrams and Model Comparison}

We now examine the relation between the two color indices defined by our 3-band 
datasets. Figure~\ref{fig:clr_clr_diag} shows the
color--color diagrams for \rxname\ and \msname.
Elliptical, S0, and late-type galaxies are represented by
the same symbols as in the preceding color-magnitude
relation figures. We also show tracks from Bruzual \& Charlot (2003; BC2003) 
stellar population models of solar and 2.5$\times$ solar metallicity, 
ages ranging from 0.5 to 6~Gyr (upper limit set by age of the universe),
and star-formation time-scale ranging from $\tau\approx0$ 
(single-burst SSP models) to $\tau\approx\infty$ (constant formation models). 
The metal-rich
models ($Z{=}0.05$) approach the locus of the red sequence galaxies for 
an age of about 2~Gyr. The solar metallicity models cannot simultaneously
match the two colors for the red sequence galaxies; however, this 
result is sensitive to the shape of the F775W bandpass definition,
since the early-type galaxy spectra have a steep slope through this bandpass.
Given that these are not true chemo-evolutionary models, that they use simple
scaled solar-abundance ratios, and the uncertainties in galaxy SED
modeling and the bandpass definitions, we do not consider this small offset
of $\lta0.05$ mag to be highly significant.  It would be profitable to explore
this further using additional bandpasses and more realistic stellar
population modeling.

For local samples of morphologically normal galaxies, Larson \& Tinsley
(1978) showed that the $(U{-}B)$ vs $(B{-}V)$ color--color locus could
be approximated by assuming roughly uniform ages for all galaxies,
but varying star-formation histories, such that blue galaxies have formed
stars at nearly constant rates, while red
galaxies formed quickly in early bursts.  Morphologically irregular
galaxies showed much more scatter, indicative of complex star-forming
histories and recent bursts.
In Figure~\ref{fig:clr_clr_diag},
the blue, mainly late-type, galaxies tend to follow a locus at
roughly constant age, but varying star-formation time scale, 
reminiscent of Larson \& Tinsley (1978).
There are a few galaxies in each cluster that lie near the region
of young ages with $\tau\approx0$, i.e., galaxies that have undergone
more recent bursts of star formation.  There are also some galaxies in each
cluster that scatter off the ``bottom'' of the model curves; these
can be explained by the presence of strong [O\,{\sc ii}] 
$\lambda3727$ line emission in the bluest passbands (F606W or F625W).

\begin{figure}\epsscale{0.84}
\plotone{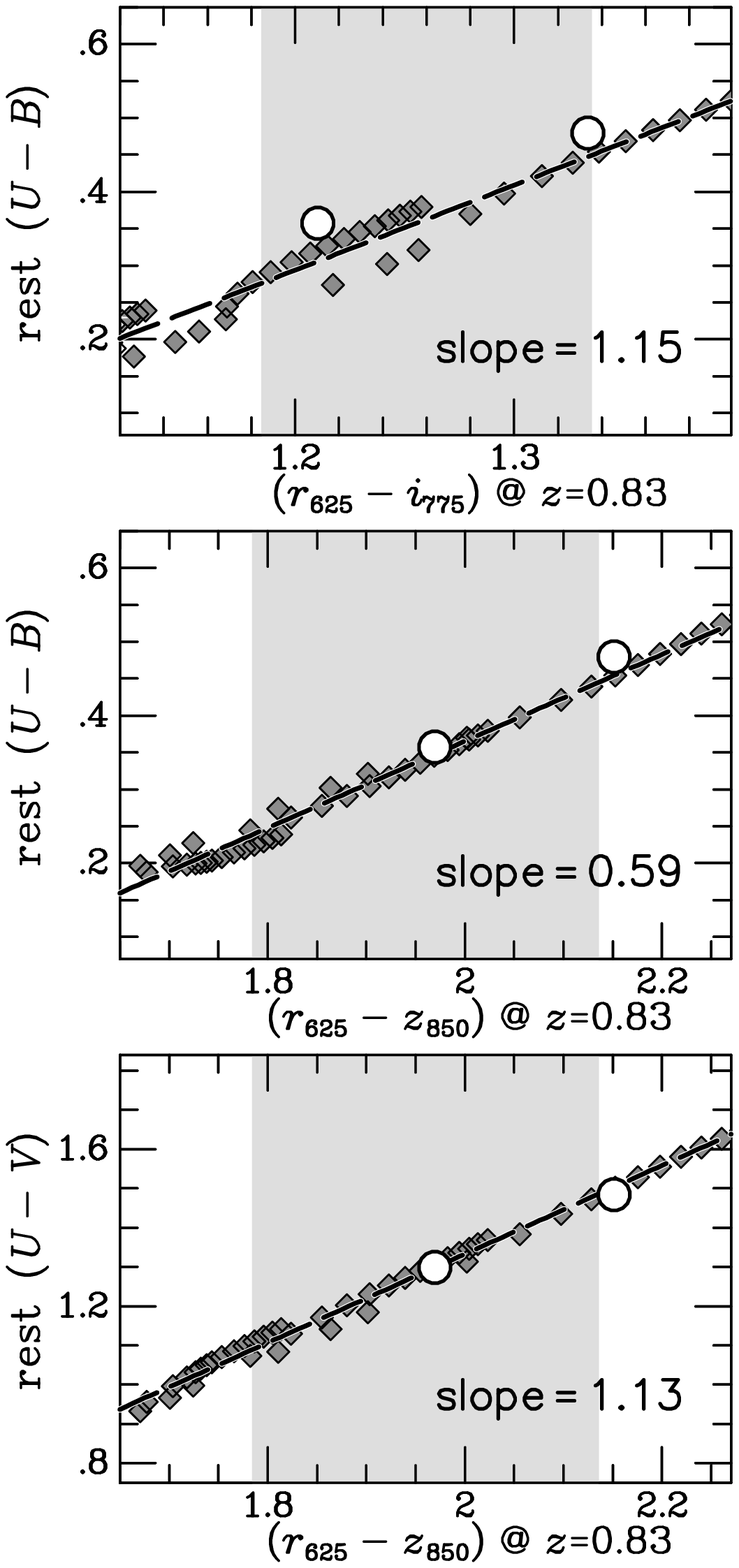}
\caption{Conversions of observed \rxname\ \ricolor\ and \rzcolor\ colors
to rest-frame \ubrest\ and \uvrest.
Small diamonds show Bruzual \& Charlot (2003) 
$\tau$ models
with metallicities $Z = 0.008$, 0.02 (solar), and 0.05, and
ages in the range 1--7 Gyr.
Large circles show the elliptical galaxy templates from
Coleman et al.\ (1980; redder color; matches local ellipticals)
and \txitxo\ et al.\ (2004; bluer color; matches distant field ellipticals).
The shaded region shows the 2.5-$\sigma$ range of the elliptical
galaxies in \rxname.
\label{fig:rzconversions}}
\end{figure}

Rudnick \etal\ (2003) found that the Larson \& Tinsley color--color
locus also matched the data for rest-frame $V$-band selected field
galaxies out to $z\approx2.8$, but they were able to reproduce this
locus using a model that aged with the redshift, had a
star-formation time scale $\tau=6$ Gyr, and invoked greater dust
extinction at high redshift.  This model is more appropriate when
studying galaxies over a large in redshift, while the 
Larson \& Tinsley approach of coeval models with variable $\tau$ 
makes more sense for a cluster of galaxies at a given redshift.

We have also calculated the reddening vectors due to internal dust
extinction at this redshift using the reddening law from O'Donnell (1994;
similar to Cardelli \etal\ 1989).  The arrows near the tops of
the figures show the direction of the reddening correction.
There are two galaxies that lie significantly above the red
galaxy loci in their respective color--color plots: a galaxy in
\rxname\ classified as spiral that is $\sim\,$0.2 mag ``too red'' in \ricolor,
and a galaxy in \msname\ classified as S0 that is $\sim\,$0.45 mag 
``too red'' in \vicolor.  Both of these appear to be nearly edge-on galaxies
with disks cutting across their bulges; the bulges appear significantly
extinguished in the bluest bands (rest-frame UV).
Both galaxies project back along the reddening vector
to SSP models of ages $\sim\,$0.5 Gyr, suggesting that they may
recently have undergone a starburst and are now heavily extincted
by dust, similar to some distant cluster galaxies found by
Poggianti \etal\ (1999).  The dust correction required for the anomalous
\rxname\ galaxy corresponds to an internal redding 
%
of $A_V\approx0.8$ mag, while the correction for the \msname\ galaxy
corresponds to $A_V\approx1.3$ mag.  These are just the most extreme
cases; given the prevalence of dust in nearby bright early-type systems
(Ferrarese \etal\ 2006), it may be that internal extinction 
causes some of the scatter in the colors of early-type galaxies at
these redshifts and should be included when performing detailed
SED-fitting analyses.

\begin{figure}\epsscale{0.84}
\plotone{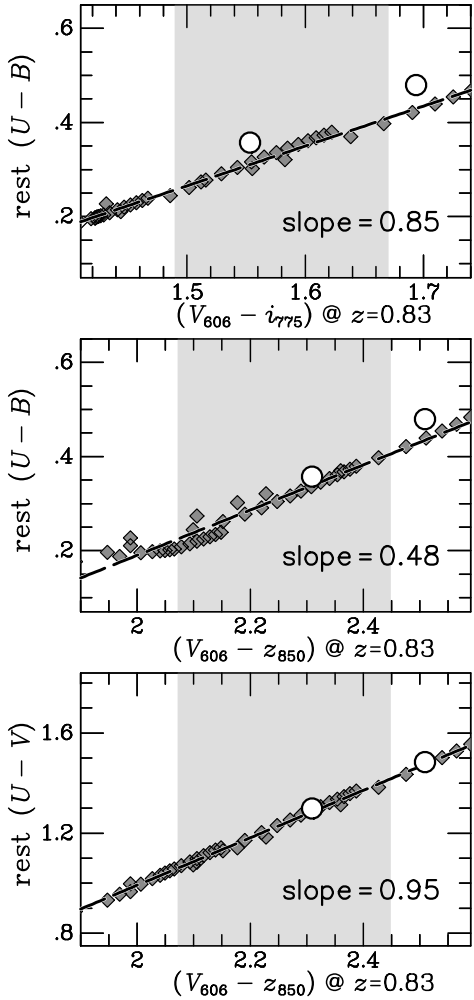}
\caption{Conversions of observed \msname\ \vicolor\ and \vzcolor\ colors
to rest-frame \ubrest\ and \uvrest.
Lines and symbols the same as in Fig.~\ref{fig:rzconversions}.
The shaded region shows the 2.5-$\sigma$ range of the elliptical
galaxies in \msname.
\label{fig:vzconversions}}
\end{figure}

\subsection{Transformation to Rest-frame Colors}

Throughout this analysis we have presented the data in the observed
bandpasses, using the natural system of the ACS/WFC instrument in order to
preserve the model-independence of the results, and have transformed model
spectra to the observed frame when needed.  But in order to compare to other
studies and redshifts, it is useful also to transform the results to
rest-frame values.  We do this by computing standard UBV rest-frame colors
for model spectra and observed ACS/WFC system colors for the same models
redshifted to the cluster redshifts, then fitting a relation to the
rest-frame colors as a function of the observed ones.
Figures~\ref{fig:rzconversions} and~\ref{fig:vzconversions}
show the results of this exercise for models with colors similar
to the cluster early-type galaxies.  At bluer colors appropriate to
late-type star-forming galaxies, the relations are no longer linear
and are less accurate due to increased scatter in the models.
Here we use the BC2003 $\tau$ models discussed above, but the
resulting transformation slopes are virtually identical if we use the SSP models.
We use the ACS and Johnson bandpass definitions distributed by
Sirianni \etal\ (2005). We caution that there are systematic
variations in the ``standard'' Johnson bandpass definitions,
particularly in the $B$ band, as Sirianni \etal\ discuss,
and using another common $B$ band definition gives 
slopes that differ by 10\%--15\%.

The fitted transformation slopes are shown in Figures~\ref{fig:rzconversions}
and~\ref{fig:vzconversions} and can be used to convert the 
measured scatters, slopes, and small offsets in
the \rxname\ and \msname\ color-magnitude relations to rest-frame
\ubrest\ and \uvrest\ (following van Dokkum \etal\ 2000,
we us the $z$~subscript on the color
to indicate rest-frame color at the observed redshift, e.g., no evolution
corrections, etc).  The transformations are most reliable when the slope
is $\sim\,1$ because this indicates the closest match between redshifted
and rest-frame bandpasses. For instance, the intrinsic \ricolor\ scatter  
$\sigma = 0.027$ mag for the full elliptical sample in \rxname\ 
transforms to $\sigma_{U{-}B} = 0.031\pm0.010$ mag, while the \rzcolor\ scatter 
converts to $\sigma_{U{-}V} = 0.081\pm0.011$ mag.  Transforming the
intrinsic \vicolor\ and \vzcolor\ scatters for \msname\ ellipticals
gives $\sigma_{U{-}B} = 0.034\pm0.009$ and $\sigma_{U{-}V} = 0.076\pm0.014$ mag.  
Thus, the color scatters for the ellipticals in these clusters
are essentially identical.
We discuss these results, and comparisons to other works,
in more detail in the following section.

\section{Cluster Galaxies at Half Hubble Time}
\label{sec:discussion}

As discussed above, the mean colors of the early-type galaxies
in these clusters are consistent with ages of 2--5 Gyr, depending
upon the metallicity and star formation histories.
We have also performed simulations of the evolution
in color scatter as function of age using the simple models 
described in Blakeslee \etal\ (2003a) and Mei \etal\ (2006).
From the stochastic burst models, we find mean ages of 
3.7~Gyr and 3.5~Gyr for the ellipticals in \rxname\
from the \ricolor\ and \rzcolor\ scatters, respectively,
and 3.5~Gyr and 3.4~Gyr for ellipticals in \msname\
from the \vicolor\ and \vzcolor\ scatters, respectively.
The mean ages increase by $\lta\,$5\% if we limit the analysis
to lower-scatter subsamples (e.g., the higher density regions
in \rxname, or the brighter samples in the two clusters; 
conversely, the inferred ages are younger where the scatter is larger.
The ages also decrease by $\sim\,$10\%
if we use the constant star-formation models instead
(to represent the other extreme).  But the data are generally
consistent with a mean age of $3.5\pm1.0$ Gyr for the 
cluster ellipticals, or a formation epoch around $z=2.2^{+1.0}_{-0.6}$.

We find that the galaxy populations in \rxname\ and \msname\ are basically
similar, although there are some notable differences, such as the
increased color scatter of the S0s in \msname, or the velocity offsets
between early- and late-type galaxies in \rxname.
However, there is a hint that the slope of the
color-magnitude relation is steeper in \msname\ than in \rxname, 
which may indicate either differing enrichment histories
(e.g., a steeper mass-metallicity relation in \msname), or a difference
in mean age with the early-type population of \msname\ being older
(Gladders \& Yee 1998).  
An earlier formation epoch for \msname\ galaxies might 
be expected in a hierarchical model because of its larger mass.
For the one color-magnitude combination in common between the
two clusters, \izcolor\ vs \iacs, the difference in slope
is $0.025\pm0.016$, or 1.6\,$\sigma$.  Converting the bluer
colors to \ubrest, we find a slope difference of $0.028\pm0.017$,
also 1.6\,$\sigma$, and if we convert the slopes for the broadest
baseline colors to \uvrest, the difference is 1.2$\,\sigma$.
These differences remain intriguing and suggestive,
but not highly significant.

This has been the first detailed study of the color-magnitude relation
in \rxname\ using \hst\ photometry.  However, van Dokkum \etal\ (2000)
made a detailed study of the relation in \msname\ using 2-band
\hst/WFPC2 photometry.  These authors reported a slope in
\ubrest\ color with $B$ magnitude of $-0.032$ and an intrinsic 
\ubrest\ scatter $\sigma = 0.024\pm0.006$ mag for 30 early-type cluster
members (they did not distinguish between elliptical and lenticular).
Our value for the slope in \vicolor\ transforms to $-0.036\pm0.012$,
in good agreement with their result.
However, based again on the \vicolor\ data, we find a scatter in \ubrest\ 
of $\sigma = 0.052\pm0.009$ mag for our early-type sample of 84 
galaxies with $\iacs<23$.  Our sample size is nearly three times larger,
and the scatter agrees better with the 
0.05~mag scatter found by van Dokkum \etal\ for their sample of
43 early-type plus merger galaxies (which they viewed as the
possible progenitor sample for modern ellipticals). 
The larger scatter we find may be due to the inclusion of fainter 
galaxies and more S0 galaxies, both of which tend to increase the scatter
(also, Postman \etal\ 2005 classified the individual components of 
early-type pairs, rather than assigning them to a ``merger'' class).
If we use only the 46 ellipticals, we find $\sigma = 0.034\pm0.008$ mag,
and if we use a magnitude limit of $\iacs<22.5$, closer to the limit of the
van Dokkum \etal\ sample, it becomes  $\sigma = 0.031\pm0.007$ mag for
37 galaxies.  Thus, the results agree better when we limit the sample
to earlier types and brighter magnitudes, which also results in a
sample size more similar to that of van~Dokkum et~al.
While this could result from an underestimate of our errors for
smaller galaxies, this is less likely because of the empirical
calibration, and it would be a natural result of ``down-sizing,''
the increase in age scatter at lower luminosity along the red sequence;
this has been discussed in the context of \msname\ by Tran \etal\ (2003).

The redshift of these two clusters marks an interesting point in the 
evolution of cosmic structure, slightly more than half the age of the universe
ago, when today's massive clusters were still in the process of assembly.
The lookback time of 7~Gyr corresponds, fortuitously, 
to the median age found for local elliptical galaxy samples based on 
stellar absorption features (e.g., Terlevich \& Forbes 2002; Worthey \&
Collobert 2003).  If this is correct, then the median local
elliptical would not be present in the samples at these high
redshifts, a phenomenon known as ``progenitor bias'' (e.g., van Dokkum
\& Franx 2001); the effect of this is to make elliptical
galaxies appear to be about half the age of the universe at
whatever epoch they are observed.  However, the giant ellipticals in
these $z=0.83$ clusters are clearly already in place: the evidence
suggests that bright cluster ellipticals mainly undergo passive
evolution at $z\lta1$ (Holden \etal\ 2005a,b) and most of the
morphological evolution occurs in the transformation of field spirals
and irregulars into cluster S0s and dwarfs (e.g., Larson \etal\ 1980;
Dressler \etal\ 1997; Postman \etal\ 2005). 
The resolution lies in correctly identifying
the descendants of the distant cluster ellipticals; the data for the
brightest ellipticals in nearby massive clusters actually indicate
ages of $\sim\,$10~Gyr or more (e.g, Poggianti \etal\ 2001; Vazdekis
\etal\ 2001), in line with the ages we measure for their likely
progenitors at $z>0.8$. The lingering star formation seen
in these clusters (e.g., Homeier \etal\ 2005) is likely only of
minor importance in terms of mass, a point which we will take up
in a forthcoming paper (B.~Holden \etal, in preparation).

On the other hand, we note that {\it passive} luminosity growth,
through galaxy mergers without associated star formation, has occurred.
\msname\ is famous for having a high fraction of red galaxy pairs
with small relative velocities (van Dokkum \etal\ 1999;
Tran \etal\ 2005a), direct evidence for passive luminosity growth.
Although we have not discussed it here, we find a very similar
fraction of red galaxy merger candidates in \clname\
(F.~Bartko \etal, in preparation), and our ACS/WFC analysis of the
X-ray selected cluster RDCS J0910+5422 at $z{\,=\,}1.1$ also shows 
a significant number of red galaxy pairs (Mei \etal\ 2006).
Thus, passive mergers appear to have been relatively common in
massive X-ray luminous clusters at this epoch.  This is not
surprising, given that the clusters themselves were still assembling
from smaller subunits in which the velocity dispersions would
have been lower.

Finally, we remark on the two galaxies  in these clusters
with very high internal extinction
(one of which was classified as early-type).
It seems likely that these are just the tip of a (dusty) iceberg.  
For instance, in a high-resolution two-band ACS/WFC study of early-type
galaxies in the Virgo cluster, Ferrarese \etal\ (2006) find that 11 of the
21 early-type galaxies with magnitude $B<12$ (corresponding to $M^*_B+1$,
the limit of our color-magnitude study) have optically detectable
dust; 4 others have diffuse dust detected in the far-infrared
(thus, $\gta\,$70\% in all).
Since the primary source of the dust
is unclear (e.g., intermittent accretion of satellite galaxies 
or continuous transpiration from the atmospheres of evolved stars;
Lauer \etal\ 2005; Ferrarese \etal\ 2006), we know very little about
the dust content of early-type galaxies at $z\sim1$. By using a
broad range of bandpasses from the rest-frame UV to the near-IR,
it should be possible to test whether dust contributes in any significant
way to the scatter in the optical colors of $z{\,\sim\,}1$ early-type galaxies,
such as the S0s that scatter towards the red in \msname.


\section{Summary and Conclusions}
\label{sec:conclusions}

We have studied the galaxy populations in \rxname\ and \msname\
using 3-band ACS/WFC mosaic imaging data combined with ground-based
spectroscopic membership information.  The samples contain 
105 and 140 confirmed members brighter than $\iacs=24$ in \rxname\ and \msname,
respectively; 94 and 114 of these are brighter than $\iacs=23$ mag
($M^*{+}1$) and have reliable (non-AGN) morphological classifications,
comprising the full samples used in our color-magnitude analyses.

We fitted S\'ersic models
to the 2-D light profiles for larger samples of galaxies 
in the redshift range  $0.6<z<0.9$ (in order to include more
late-type field galaxies, with at least similar redshifts, in the analysis).
The resulting S\'ersic indices correlate well with the visual morphological 
classifications from Postman \etal\ (2005).
We introduced a new parameter called the bumpiness $B$ which quantifies
the rms deviations from the smooth S\'ersic model fits, and used
this $B$ parameter in combination with the S\'ersic index $n$ to
group the galaxies objectively into E, S0, and late-type classes,
calibrated against the visual classifications.  The
agreement between the visual and \bn~classified morphologies is
excellent, but the latter appear to be less biased by orientation effects.
We also calculated the asymmetry $A$ and concentration $C$ parameters for
this sample of galaxies and find similarly good agreement with the visual types.
The $A$--$C$ parameters appear to be more effective in separating galaxies into
early-type, spiral, and irregular classes, while the \bn\ parameters are
more effective in distinguishing between ellipticals, S0s, and late-types.

We also used the results of our profile fitting to compare
the size--surface brightness (Kormendy) relations in \msname\
and \rxname\ in the F775W bandpass (rest-frame $\approx B$). 
The slopes and scatters of these relations for the two clusters
are the same within the errors, but there is a zero-point
offset of $0.14\pm0.064$ mag between the two.  This is
consistent with the scatter in the evolution of this relation
found by Holden \etal\ (2005a).

Following the structural analysis, we proceeded to study the galaxy
color-magnitude relations for the two clusters and correlations of
the residuals in these relations with other observables.  We concentrated
on the broadest baseline colors, as these are relatively less sensitive to 
errors and more sensitive to stellar population effects.
We found correlations of the color residuals with local density,
measured from both galaxy number density and weak lensing mass maps,
in the sense that galaxies in the highest density environments
are redder, and those in low-density regions bluer, than 
expected from the fitted mean relation.  This is closely related to
the morphology--density relation (e.g., Dressler 1980, Postman \etal\ 2005),
but the effect is also found at some level for galaxies within a
given morphological class, except for S0 galaxies in \msname,
which are unusual in this sense.
The scatter in galaxy colors (and, presumably, ages) increases
markedly below a surface mass density of $\sim\,$0.12, in units of the critical
density.  The effect is more dramatic in \rxname, where virtually all
galaxies above this threshold appear to be quiescent.  We also find
some correlation in the color residuals with radius, but with a lower
significance than the underlying density correlation.

There is an offset of $980\pm340$ \kms\ in the mean velocities of the
early- and late-type galaxies in \rxname, in the sense of the
late-type galaxies having preferentially positive peculiar velocities.
This offset causes an apparent trend between galaxy velocities and 
colors, but the trend is not present for either the early- or late-type
galaxy samples individually.  Since these blue, high-velocity, late-type
galaxies are also preferentially located in regions of lower density,
we are likely witnessing the early stages of a diffuse group or association 
of late-type galaxies being accreted from the near-side of the cluster
(thus their infall velocity is away from us).
This apparent infall of late-type galaxies is distinct from
the ongoing major merger of the
two massive northern and southern ``clumps,'' or subclusters.
In particular, although the late-type galaxies have a mean velocity 
closer to that of the northern clump, the region around this clump
has very few blue or late-type galaxies; on the other hand, a
significant fraction of the galaxies within the southern clump
are bluer than the red sequence, including two blue early-type galaxies,
consistent with the southern clump being a lower mass system.
However, the main early-type populations in the two clumps are 
consistent with following the same color-magnitude relation.

We then examined the color--color diagrams for these two clusters 
and compared to predictions from exponentially declining
star-formation models having a range of star-formation time scales
and ages. The loci of red cluster ellipticals are most consistent with 
super-solar metallicity models with short formation time scales
$\tau<0.5$ Gyr and ages in the range $1.5\lta t\lta 3$ Gyr.  
However, solar metallicity models with larger ages may be possible,
given the uncertainties in the bandpass definitions and models.
The bluer galaxies generally follow a model locus
of roughly constant age ($t\approx1$--3 Gyr, depending on metallicity)
but a range in $\tau$, except for a few galaxies in each cluster which
appear to have undergone bursts at $t<1$ Gyr ago.
These galaxies are mainly late-type, but
include one elliptical in \rxname\ and several S0s in \msname,
and appear fairly red in the bluer [\ricolor\ or \vicolor]
color but blue in the redder \izcolor\ color.
There is also one galaxy in each cluster whose colors can only
be explained with by a large amount of internal extinction:
a spiral in \rxname\ with $A_V\sim0.8$ mag, and an S0 in \msname\
with $A_V\sim1.3$ mag.  Both appear to be nearly edge-on systems
with dusty disk components that cut across the central bulges.

We derive transformations from observed to rest-frame colors in order
to better compare the clusters to each other and to previous works.
Based on simple models for the evolution in the color scatter, we
estimate mean ages for the elliptical galaxies of 3.2--3.7 Gyr,
depending on the star-formation history and whether we limit the
sample to the higher density regions.  This implies that the major
epoch of star-formation in these clusters occurred at $z>2$,
in accord with the findings from other recent work at $z\gta0.8$,
as well as the direct evidence from $z>2$ protoclusters.
The age constraints are very similar for both clusters.
We find some evidence that the slope of the color-magnitude
relation in \msname\ may be steeper than in \rxname, which
may indicate an earlier formation epoch for this more massive cluster,
but the significance of the result is less than~2\,$\sigma$.

Much remains to be done.  It would be useful to have
still more redshifts: a complete spectroscopic survey of these fields to
$\iacs=24$, or $\sim\,$0.15 $L^*_B$, would make it possible to study
the ``down-sizing'' effect in more detail and confirm if there is a
lack of fainter red galaxies at these redshifts (Tran \etal\ 2003;
Kodama \etal\ 2004; Tanaka \etal\ 2005).
High-resolution imaging over a larger area will allow us to study
galaxy morphologies in the lower density cluster surroundings and
the very large-scale mass distributions from  weak lensing;
two-band ACS imaging for an
additional eight fields in the outskirts of each of these clusters 
now exists and will be used in this regard.
Accurate colors in additional optical and near-IR bands over the
cluster fields will make it possible to derive reliable masses from
SED fitting, which should be checked and calibrated against good
dynamical masses for a subset of the galaxies.  This will enable large
studies of the star-formation trends as a function of galaxy mass, as
we are currently undertaking.  Of course, additional clusters studied
with the same level of detail are needed to understand the range 
of cluster properties at each redshift and constrain their evolution.

\acknowledgments 
ACS was developed under NASA contract NAS 5-32865, and this research
has been supported by NASA grant NAG5-7697.  We are grateful to all of
our fellow ACS IDT members, and to K.~Anderson, D.~Magee, J.~McCann,
S.~Busching, A.~Framarini, and T.~Allen for their invaluable
contributions to the ACS project, and to Jon McCann for the use of his
{\sc fitscut} software.  JPB thanks Txitxo {Ben\'{\i}tez}, Dan Kelson,
Taddy Kodama, and Guy Worthey for many helpful discussions and John
Tonry for expert plotting advice.

\begin{appendix}
\section{Flat-fields and Galaxy Photometry}
\label{app:pgrad}

In comparing the photometry for galaxies that appear 
in multiple pointings, we 
found some dependence of the measured colors on the position of the
galaxy in the image.  We investigated this in detail and 
traced the most significant positional dependences of the 
colors ($\sim\,$0.025 mag variation peak-to-peak across the image) 
to variations in the F625W and F775W photometry.  We believe
this is due to the flatfields.  The ACS/WFC low-order flat 
fields (``L-flats'') were constructed, in most bandpasses, from 
modeling the variation in the aperture photometry of stars from 
multiple dithered observations of 47~Tucanae (Mack \etal\ 2002).
Among our bandpasses, this was the approach used for F606W
and F850LP.  In F625W and, originally, F775W, the L-flats were
interpolated from the F606W and F814W L-flats based on the
passband `pivot' wavelengths.  The F775W L-flats have since been
reconstructed using the stellar photometry (J.~Mack, private
communication), although based on roughly half as much data.

We attempted to derive flat field corrections using low-order 2-D
spline fits to the sky variations (after liberally masking all
objects) in our \rxname\ observations in all three filters.  We then
applied the normalized `sky flats' (which typically had rms
variations of $\sim\,$1\%, consistent with the rms errors quoted by
Mack \etal\ 2002) to the data and remeasured the photometry for all
objects in all the \rxname\ pointings following the same procedure
(CLEAN, etc) as before.  No significant difference was found for the
F850LP photometry, which was already consistent with no positional
dependence.  In F625W, the systematic variation in photometry with
position (mainly in the $x$-direction) was cut roughly in half to the
$\sim\,$0.01 mag peak-to-peak level.  However, we found no
improvement, and more scatter (perhaps because of scattered
light causing error in the `sky flat'), for F775W, which shows
$\sim\,$0.024 mag top-to-bottom variation along the chip
$y$-direction.  Thus, we applied our correction flat-field in F625W
(this is available by request to the first author),
but not F775W.  Since \msname\ was observed in F606W (which
shows no significant positional dependence in photometry) instead of
F625W, we did not apply any flat field corrections.

\begin{figure}\epsscale{0.9}
\vspace{0.5cm}
\plotone{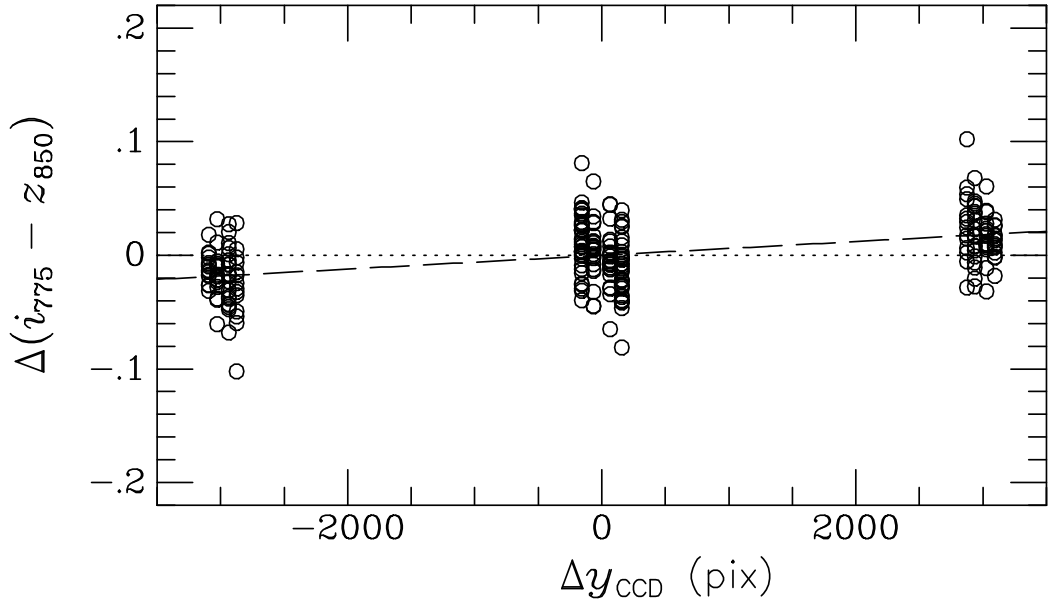}
\vspace{0.2cm}
\caption{Differences in the measured \izcolor\ colors for all
pairs of program galaxies with $\iacs<24$ imaged in multiple
\rxname\ pointings are plotted against the differences in the $y$ positions 
of the paired objects (the image scale is 0\farcs05 pix$^{-1}$).
There are 125 unique pairs, but each is plotted twice, once for
each observation, so that the overall mean of the differences is zero.
The dashed line shows a simple linear least-squares fit, which
gives a slope of 0.0060~mag per 1000~pix, or almost 0.025 mag from
top to bottom in the image.
This is the most significant (more than 4$\,\sigma$)
systematic trend among the different colors in our study.
However, because of the sparse sampling in positional offset
(essentially 2 positions), it is unclear if the trend is truly
linear or if it actually reflects photometric biases near the
edges of the image.
\label{fig:izflatgrad}}
\end{figure}

Figure~\ref{fig:izflatgrad} illustrates the most significant 
photometric gradient remaining in the data, a dependence of
 \izcolor\ color on $y$ position in the image.
While we could remove a simple gradient in the F775W photometry
by fitting a plane to the differences in the multiple measurements,
this would be dangerous because the multiple measurements are
confined to the edges of field of view where the pointings overlapped,
and thus we do not know if it is truly a linear trend.
Instead, we simply accept this additional error (which is $\lta0.01$
mag rms and is effectively included in our error estimates
 based on the multiple measurements) 
for colors that use F775W, and remain cognizant that it represents a $\sim0.02$ mag
systematic effect in the data.  Fortunately, the broader baseline $(\vacs-\zacs)$
and $(\racs-\zacs)$ are more useful for studying stellar population
effects and have been used for assessing intrinsic color variations.

\end{appendix}


\clearpage


\LongTables

\setlength{\tabcolsep}{4pt}
\begin{deluxetable}{rccccccrrrrrrr} 
\tablecaption{Data for \rxname\ Cluster Members}
\tablehead{\colhead{ID} &
\colhead{R.A.} &
\colhead{Dec.} &
\colhead{$i_{775,{\rm tot}}\tablenotemark{a}$} & 
\colhead{$i_{775,{\rm gfit}}\tablenotemark{b}$} & 
\colhead{$R_e$\tablenotemark{c}} &
\colhead{$n_{\rm Ser}$\tablenotemark{d}} &
\colhead{$T$\tablenotemark{e}} & 
\colhead{$b/a\,$\tablenotemark{f}} &
\colhead{$B\,$\tablenotemark{g}} &
\colhead{$\Sigma_L$\tablenotemark{h}} & 
\colhead{\ricolor\ } &
\colhead{\izcolor\ } &
\colhead{\rzcolor\ }
}
\startdata
  543 &  28.17325 & $-$14.00679 & 22.29 & 22.13 & 0.96 & 2.40 &  3 & 0.41 &  0.447 &  0.017 &  1.280 $\pm$ 0.026 &  0.641 $\pm$ 0.016 &  1.921 $\pm$ 0.021  \\ 
 1694 &  28.18989 & $-$14.00385 & 22.08 & 22.36 & 0.50 & 0.79 &  3 & 0.62 &  0.256 & $-$0.027 &  1.178 $\pm$ 0.028 &  0.486 $\pm$ 0.019 &  1.663 $\pm$ 0.026 \\ 
 1142 &  28.17470 & $-$13.99799 & 21.31 & 21.40 & 0.56 & 4.00 & $-$5 & 0.82 &  0.100 &  0.105 &  1.275 $\pm$ 0.018 &  0.695 $\pm$ 0.010 &  1.970 $\pm$ 0.010 \\ 
 1358 &  28.17957 & $-$14.00009 & 22.61 & 22.42 & 0.49 & 2.51 & $-$1 & 0.75 &  0.217 & $-$0.012 &  1.209 $\pm$ 0.029 &  0.590 $\pm$ 0.020 &  1.799 $\pm$ 0.027 \\ 
  950 &  28.17166 & $-$13.99879 & 22.70 & 22.73 & 0.27 & 3.08 & $-$1 & 0.43 &  0.123 &  0.059 &  1.260 $\pm$ 0.032 &  0.654 $\pm$ 0.025 &  1.914 $\pm$ 0.034 \\ 
 1264 &  28.17456 & $-$13.99730 & 22.95 & 23.06 & 0.14 & 4.00 & $-$5 & 0.47 &  0.100 &  0.111 &  1.297 $\pm$ 0.036 &  0.658 $\pm$ 0.029 &  1.955 $\pm$ 0.041 \\ 
 3941 &  28.21666 & $-$13.97119 & 20.71 & 20.79 & 1.41 & 4.00 & $-$5 & 0.80 &  0.263 &  0.104 &  1.284 $\pm$ 0.017 &  0.711 $\pm$ 0.010 &  1.994 $\pm$ 0.010 \\ 
 1895 &  28.18170 & $-$13.99381 & 22.64 & 22.72 & 0.40 & 4.00 &  0 & 0.70 &  0.350 & $-$0.012 &  1.058 $\pm$ 0.032 &  0.494 $\pm$ 0.024 &  1.552 $\pm$ 0.033 \\ 
 3055 &  28.19740 & $-$13.99036 & 20.92 & 20.55 & 2.72 & 4.00 &  1 & 0.50 &  0.395 &  0.055 &  1.323 $\pm$ 0.017 &  0.716 $\pm$ 0.010 &  2.039 $\pm$ 0.010 \\ 
 1290 &  28.16637 & $-$13.98800 & 22.40 & 22.68 & 0.43 & 2.42 &  3 & 0.67 &  0.415 &  0.116 &  0.717 $\pm$ 0.032 &  0.310 $\pm$ 0.024 &  1.027 $\pm$ 0.033 \\ 
 1965 &  28.17909 & $-$13.99060 & 23.43 & 23.44 & 0.27 & 2.29 & $-$2 & 0.50 &  0.127 & $-$0.027 &  1.201 $\pm$ 0.040 &  0.624 $\pm$ 0.035 &  1.825 $\pm$ 0.049 \\ 
  946 &  28.16118 & $-$13.98890 & 22.62 & 22.64 & 0.29 & 2.66 & $-$2 & 0.65 &  0.087 &  0.102 &  1.270 $\pm$ 0.031 &  0.733 $\pm$ 0.023 &  2.003 $\pm$ 0.032 \\ 
 3616 &  28.20164 & $-$13.98910 & 22.30 & 22.28 & 0.31 & 3.87 & $-$4 & 0.61 &  0.085 &  0.027 &  1.266 $\pm$ 0.027 &  0.763 $\pm$ 0.018 &  2.029 $\pm$ 0.024 \\ 
 5292 &  28.20619 & $-$13.97524 & 21.69 & 21.55 & 0.57 & 4.00 & $-$1 & 0.51 &  0.240 &  0.079 &  1.331 $\pm$ 0.014 &  0.794 $\pm$ 0.013 &  2.124 $\pm$ 0.008 \\ 
 1309 &  28.16648 & $-$13.98843 & 22.64 & 23.01 & 0.35 & 1.67 &  3 & 0.97 &  0.129 &  0.118 &  1.184 $\pm$ 0.035 &  0.701 $\pm$ 0.029 &  1.885 $\pm$ 0.040 \\ 
 1341 &  28.16569 & $-$13.98716 & 22.16 & 22.97 & 0.76 & 1.79 &  8 & 0.31 &  0.279 &  0.118 &  1.181 $\pm$ 0.035 &  0.747 $\pm$ 0.028 &  1.929 $\pm$ 0.039 \\ 
  375 &  28.14912 & $-$13.98617 & 22.75 & 22.55 & 0.35 & 4.00 & $-$5 & 0.92 &  0.093 & $-$0.053 &  1.166 $\pm$ 0.070 &  0.663 $\pm$ 0.015 &  1.829 $\pm$ 0.068 \\ 
 1355 &  28.16418 & $-$13.98438 & 21.54 & 22.26 & 0.47 & 0.83 &  3 & 0.69 &  0.191 &  0.120 &  0.876 $\pm$ 0.027 &  0.308 $\pm$ 0.018 &  1.184 $\pm$ 0.024 \\ 
 2570 &  28.18247 & $-$13.98364 & 20.36 & 20.71 & 0.00 & 3.99 & 11 & 0.47 &  1.011 & $-$0.019 & $-$0.244 $\pm$ 0.017 &  0.057 $\pm$ 0.010 & $-$0.187 $\pm$ 0.010 \\ 
 5087 &  28.20723 & $-$13.97437 & 21.85 & 21.53 & 1.46 & 4.00 &  4 & 0.43 &  0.509 &  0.087 &  1.100 $\pm$ 0.028 &  0.676 $\pm$ 0.020 &  1.776 $\pm$ 0.008 \\ 
 1693 &  28.16765 & $-$13.98284 & 22.22 & 22.52 & 0.34 & 1.23 &  8 & 0.55 &  0.414 &  0.034 &  0.180 $\pm$ 0.030 &  0.093 $\pm$ 0.021 &  0.274 $\pm$ 0.029 \\ 
 1571 &  28.16537 & $-$13.98219 & 21.95 & 21.95 & 0.36 & 4.00 & $-$2 & 0.75 &  0.091 &  0.051 &  1.227 $\pm$ 0.040 &  0.646 $\pm$ 0.016 &  1.873 $\pm$ 0.024 \\ 
 1217 &  28.15847 & $-$13.98181 & 22.33 & 22.38 & 0.22 & 4.00 & $-$5 & 0.78 &  0.048 &  0.076 &  1.273 $\pm$ 0.055 &  0.673 $\pm$ 0.033 &  1.947 $\pm$ 0.022 \\ 
 1411 &  28.16105 & $-$13.98176 & 22.56 & 22.56 & 0.13 & 4.00 & $-$5 & 0.94 &  0.064 &  0.092 &  0.888 $\pm$ 0.026 &  0.445 $\pm$ 0.016 &  1.332 $\pm$ 0.021 \\ 
 2263 &  28.17186 & $-$13.97858 & 22.20 & 23.01 & 0.43 & 3.18 &  0 & 0.54 &  0.167 &  0.021 &  1.013 $\pm$ 0.035 &  0.532 $\pm$ 0.029 &  1.545 $\pm$ 0.040 \\ 
 2376 &  28.17340 & $-$13.97881 & 22.15 & 22.32 & 0.22 & 1.39 & $-$2 & 0.39 &  0.082 &  0.028 &  0.162 $\pm$ 0.028 &  0.163 $\pm$ 0.018 &  0.325 $\pm$ 0.025 \\ 
 1808 &  28.16528 & $-$13.97390 & 20.81 & 20.44 & 2.19 & 3.08 & $-$5 & 0.78 &  0.063 &  0.154 &  1.289 $\pm$ 0.017 &  0.727 $\pm$ 0.017 &  2.015 $\pm$ 0.034 \\ 
 1439 &  28.15580 & $-$13.97585 & 22.60 & 22.81 & 0.62 & 0.52 &  3 & 0.52 &  0.292 & $-$0.018 &  1.020 $\pm$ 0.024 &  0.484 $\pm$ 0.018 &  1.504 $\pm$ 0.025 \\ 
  748 &  28.14670 & $-$13.97776 & 23.40 & 23.38 & 0.26 & 3.32 &  0 & 0.40 &  0.113 &  0.063 &  1.182 $\pm$ 0.040 &  0.700 $\pm$ 0.034 &  1.883 $\pm$ 0.047 \\ 
 2185 &  28.16873 & $-$13.97714 & 22.18 & 22.06 & 0.35 & 3.71 & $-$2 & 0.63 &  0.124 &  0.079 &  1.270 $\pm$ 0.018 &  0.715 $\pm$ 0.011 &  1.985 $\pm$ 0.018 \\ 
  899 &  28.14690 & $-$13.97567 & 22.27 & 22.96 & 0.75 & 4.00 &  4 & 0.51 &  0.347 &  0.040 &  1.188 $\pm$ 0.035 &  0.376 $\pm$ 0.028 &  1.564 $\pm$ 0.038 \\ 
 2375 &  28.17046 & $-$13.97571 & 22.55 & 22.47 & 0.44 & 4.00 & $-$4 & 0.62 &  0.072 &  0.081 &  1.251 $\pm$ 0.021 &  0.632 $\pm$ 0.015 &  1.883 $\pm$ 0.020 \\ 
 1826 &  28.16217 & $-$13.97537 & 22.19 & 22.30 & 0.27 & 2.56 & $-$4 & 0.69 &  0.092 &  0.055 &  1.345 $\pm$ 0.058 &  0.773 $\pm$ 0.020 &  2.117 $\pm$ 0.039 \\ 
 3896 &  28.22021 & $-$13.97315 & 22.00 & 22.01 & 0.65 & 4.00 & $-$4 & 0.79 &  0.316 &  0.052 &  1.308 $\pm$ 0.024 &  0.634 $\pm$ 0.014 &  1.941 $\pm$ 0.019 \\ 
 1859 &  28.16589 & $-$13.97334 & 21.32 & 22.68 & 0.23 & 4.00 & $-$4 & 0.65 &  0.053 &  0.159 &  1.313 $\pm$ 0.022 &  0.760 $\pm$ 0.025 &  2.073 $\pm$ 0.023 \\ 
 3168 &  28.17977 & $-$13.97275 & 21.76 & 21.55 & 0.58 & 3.70 & $-$1 & 0.73 &  0.228 &  0.129 &  1.223 $\pm$ 0.020 &  0.697 $\pm$ 0.010 &  1.920 $\pm$ 0.012 \\ 
 2420 &  28.16807 & $-$13.97296 & 22.81 & 22.97 & 0.16 & 2.13 & $-$1 & 0.36 &  0.064 &  0.154 &  1.298 $\pm$ 0.035 &  0.733 $\pm$ 0.020 &  2.031 $\pm$ 0.044 \\ 
 2157 &  28.16315 & $-$13.97209 & 22.92 & 22.93 & 0.20 & 3.48 & $-$5 & 0.86 &  0.033 &  0.088 &  1.293 $\pm$ 0.024 &  0.753 $\pm$ 0.019 &  2.046 $\pm$ 0.027 \\ 
 2569 &  28.16931 & $-$13.97072 & 21.66 & 21.54 & 0.62 & 4.00 & $-$5 & 0.92 &  0.064 &  0.138 &  1.275 $\pm$ 0.031 &  0.719 $\pm$ 0.007 &  1.993 $\pm$ 0.029 \\ 
 3014 &  28.17425 & $-$13.97032 & 22.26 & 22.23 & 0.28 & 3.81 &  0 & 0.34 &  0.096 &  0.176 &  1.273 $\pm$ 0.019 &  0.678 $\pm$ 0.012 &  1.951 $\pm$ 0.017 \\ 
 2412 &  28.16466 & $-$13.96894 & 22.40 & 22.60 & 0.35 & 4.00 & $-$5 & 0.85 &  0.095 &  0.108 &  1.137 $\pm$ 0.022 &  0.668 $\pm$ 0.022 &  1.805 $\pm$ 0.022 \\ 
 2481 &  28.16441 & $-$13.96842 & 22.55 & 22.61 & 0.32 & 4.00 & $-$5 & 0.90 &  0.112 &  0.109 &  1.283 $\pm$ 0.044 &  0.743 $\pm$ 0.028 &  2.027 $\pm$ 0.072 \\ 
 1495 &  28.14917 & $-$13.96888 & 22.67 & 22.74 & 0.14 & 3.07 & $-$4 & 0.67 &  0.074 & $-$0.019 &  1.195 $\pm$ 0.032 &  0.560 $\pm$ 0.025 &  1.754 $\pm$ 0.034 \\ 
 2747 &  28.16794 & $-$13.96799 & 22.09 & 22.11 & 0.22 & 4.00 & $-$5 & 0.92 &  0.034 &  0.138 &  1.264 $\pm$ 0.018 &  0.729 $\pm$ 0.011 &  1.993 $\pm$ 0.015 \\ 
10653 &  28.21253 & $-$13.96470 & 20.98 & 20.78 & 1.21 & 4.00 &  0 & 0.66 &  0.549 &  0.014 &  1.172 $\pm$ 0.017 &  0.613 $\pm$ 0.010 &  1.785 $\pm$ 0.010 \\ 
 3231 &  28.17434 & $-$13.96599 & 22.23 & 22.42 & 0.27 & 2.08 & $-$2 & 0.67 &  0.132 &  0.175 &  1.314 $\pm$ 0.020 &  0.772 $\pm$ 0.026 &  2.086 $\pm$ 0.030 \\ 
 3727 &  28.17868 & $-$13.96516 & 21.99 & 21.95 & 0.41 & 3.03 & $-$4 & 0.69 &  0.096 &  0.146 &  1.270 $\pm$ 0.037 &  0.713 $\pm$ 0.019 &  1.983 $\pm$ 0.018 \\ 
 1898 &  28.15092 & $-$13.96337 & 21.53 & 21.63 & 0.55 & 4.00 &  1 & 0.54 &  0.347 & $-$0.055 &  1.286 $\pm$ 0.021 &  0.637 $\pm$ 0.010 &  1.923 $\pm$ 0.013 \\ 
 2958 &  28.16598 & $-$13.96133 & 20.68 & 20.81 & 0.32 & 4.00 & 11 & 0.75 &  0.234 &  0.177 &  0.556 $\pm$ 0.025 &  0.657 $\pm$ 0.007 &  1.213 $\pm$ 0.030 \\ 
 3605 &  28.17657 & $-$13.96385 & 21.94 & 22.24 & 0.29 & 3.28 & $-$4 & 0.71 &  0.073 &  0.169 &  1.310 $\pm$ 0.019 &  0.755 $\pm$ 0.028 &  2.064 $\pm$ 0.017 \\ 
 6632 &  28.18124 & $-$13.96508 & 24.47 & 24.73 & 0.21 & 0.92 &  8 & 0.42 &  0.398 &  0.113 & $-$0.158 $\pm$ 0.087 & $-$0.174 $\pm$ 0.106 & $-$0.332 $\pm$ 0.038 \\ 
 6165 &  28.18388 & $-$13.96286 & 22.34 & 22.31 & 0.33 & 3.27 & $-$5 & 0.84 &  0.033 &  0.129 &  1.226 $\pm$ 0.014 &  0.672 $\pm$ 0.009 &  1.898 $\pm$ 0.013 \\ 
 2949 &  28.16660 & $-$13.96165 & 20.92 & 21.34 & 0.57 & 2.90 & $-$5 & 0.75 &  0.045 &  0.190 &  1.313 $\pm$ 0.013 &  0.743 $\pm$ 0.007 &  2.056 $\pm$ 0.007 \\ 
10377 &  28.18075 & $-$13.92880 & 22.13 & 22.07 & 0.34 & 2.80 & $-$2 & 0.71 &  0.130 & $-$0.005 &  1.141 $\pm$ 0.025 &  0.574 $\pm$ 0.015 &  1.716 $\pm$ 0.020 \\ 
 4003 &  28.18744 & $-$13.94317 & 22.87 & 22.78 & 0.54 & 1.77 &  3 & 0.80 &  0.346 &  0.112 &  1.086 $\pm$ 0.026 &  0.500 $\pm$ 0.025 &  1.586 $\pm$ 0.025 \\ 
 3729 &  28.15569 & $-$13.94308 & 22.10 & 23.88 & 0.24 & 4.00 &  8 & 0.20 &  0.200 &  0.004 &  0.933 $\pm$ 0.045 &  0.467 $\pm$ 0.042 &  1.401 $\pm$ 0.058 \\ 
10673 &  28.19018 & $-$13.94423 & 21.58 & 21.51 & 0.83 & 3.53 & $-$2 & 0.53 &  0.174 &  0.183 &  1.306 $\pm$ 0.016 &  0.718 $\pm$ 0.009 &  2.024 $\pm$ 0.008 \\ 
 3270 &  28.15093 & $-$13.94368 & 21.49 & 21.45 & 0.56 & 4.00 & $-$2 & 0.52 &  0.174 &  0.011 &  1.299 $\pm$ 0.019 &  0.690 $\pm$ 0.010 &  1.988 $\pm$ 0.011 \\ 
 4126 &  28.18693 & $-$13.94392 & 22.72 & 22.61 & 0.19 & 4.00 & $-$5 & 0.89 &  0.080 &  0.118 &  1.238 $\pm$ 0.022 &  0.655 $\pm$ 0.016 &  1.893 $\pm$ 0.022 \\ 
 4219 &  28.19112 & $-$13.94957 & 20.45 & 20.38 & 1.26 & 3.86 & $-$5 & 0.78 &  0.080 &  0.147 &  1.266 $\pm$ 0.012 &  0.721 $\pm$ 0.007 &  1.986 $\pm$ 0.011 \\ 
 5143 &  28.18168 & $-$13.94880 & 21.60 & 21.61 & 0.66 & 4.00 & $-$5 & 0.91 &  0.065 &  0.150 &  1.257 $\pm$ 0.010 &  0.724 $\pm$ 0.005 &  1.981 $\pm$ 0.007 \\ 
 4680 &  28.18755 & $-$13.95095 & 20.67 & 20.70 & 0.78 & 4.00 & $-$4 & 0.63 &  0.043 &  0.196 &  1.285 $\pm$ 0.008 &  0.713 $\pm$ 0.005 &  1.998 $\pm$ 0.005 \\ 
 1809 &  28.14666 & $-$13.96035 & 22.57 & 22.72 & 0.45 & 1.12 &  3 & 0.50 &  0.203 & $-$0.014 &  1.174 $\pm$ 0.032 &  0.442 $\pm$ 0.024 &  1.616 $\pm$ 0.033 \\ 
 6659 &  28.17711 & $-$13.96147 & 23.97 & 23.89 & 0.37 & 2.50 &  3 & 0.82 &  0.173 &  0.186 &  1.233 $\pm$ 0.064 &  0.529 $\pm$ 0.030 &  1.761 $\pm$ 0.041 \\ 
 5037 &  28.15887 & $-$13.92697 & 21.42 & 21.40 & 0.61 & 2.24 &  4 & 0.95 &  0.225 & $-$0.025 &  0.970 $\pm$ 0.013 &  0.456 $\pm$ 0.008 &  1.426 $\pm$ 0.017 \\ 
 8134 &  28.19529 & $-$13.90882 & 23.25 & 23.46 & 0.39 & 0.43 &  3 & 0.71 &  0.218 &  0.052 &  0.561 $\pm$ 0.041 &  0.283 $\pm$ 0.036 &  0.845 $\pm$ 0.049 \\ 
 8158 &  28.19477 & $-$13.90920 & 24.56 & 24.66 & 0.05 & 3.17 &  9 & 0.58 &  0.132 &  0.058 &  1.217 $\pm$ 0.054 &  0.674 $\pm$ 0.054 &  1.891 $\pm$ 0.075 \\ 
 9517 &  28.17383 & $-$13.90896 & 21.63 & 21.63 & 0.36 & 4.00 & $-$2 & 0.50 &  0.132 &  0.007 &  1.248 $\pm$ 0.021 &  0.626 $\pm$ 0.010 &  1.875 $\pm$ 0.013 \\ 
 9757 &  28.17062 & $-$13.90970 & 21.57 & 22.07 & 0.20 & 4.00 &  0 & 0.73 &  0.178 &  0.037 &  1.404 $\pm$ 0.025 &  0.657 $\pm$ 0.015 &  2.061 $\pm$ 0.020 \\ 
 9503 &  28.17351 & $-$13.90937 & 22.59 & 22.87 & 0.30 & 1.38 & $-$1 & 0.60 &  0.173 &  0.015 &  0.799 $\pm$ 0.034 &  0.233 $\pm$ 0.027 &  1.032 $\pm$ 0.036 \\ 
 9607 &  28.21051 & $-$13.95332 & 22.57 & 22.81 & 0.17 & 4.00 & $-$5 & 0.83 &  0.069 & $-$0.023 &  1.244 $\pm$ 0.033 &  0.638 $\pm$ 0.026 &  1.882 $\pm$ 0.035 \\ 
10190 &  28.17424 & $-$13.91844 & 23.83 & 24.05 & 0.20 & 0.52 &  0 & 0.58 &  0.109 &  0.022 &  0.583 $\pm$ 0.047 &  0.214 $\pm$ 0.045 &  0.797 $\pm$ 0.062 \\ 
 8981 &  28.19994 & $-$13.92527 & 22.47 & 22.83 & 0.38 & 1.01 &  8 & 0.72 &  0.261 & $-$0.056 &  0.914 $\pm$ 0.029 &  0.388 $\pm$ 0.018 &  1.302 $\pm$ 0.025 \\ 
 5366 &  28.15667 & $-$13.92791 & 23.67 & 23.78 & 0.12 & 2.34 & $-$1 & 0.44 &  0.114 &  0.002 &  0.380 $\pm$ 0.031 &  0.327 $\pm$ 0.029 &  0.707 $\pm$ 0.050 \\ 
 4762 &  28.16540 & $-$13.93000 & 22.83 & 22.98 & 0.20 & 1.09 & $-$1 & 0.57 &  0.084 & $-$0.030 &  0.789 $\pm$ 0.109 &  0.271 $\pm$ 0.020 &  1.060 $\pm$ 0.095 \\ 
10456 &  28.18171 & $-$13.93134 & 22.39 & 22.43 & 0.27 & 3.73 & $-$1 & 0.69 &  0.101 &  0.040 &  1.307 $\pm$ 0.029 &  0.682 $\pm$ 0.020 &  1.989 $\pm$ 0.027 \\ 
 5606 &  28.15617 & $-$13.93042 & 21.15 & 21.27 & 0.36 & 3.05 & $-$5 & 0.75 &  0.048 &  0.008 &  1.241 $\pm$ 0.012 &  0.647 $\pm$ 0.016 &  1.887 $\pm$ 0.007 \\ 
10243 &  28.18555 & $-$13.93155 & 21.72 & 21.80 & 0.37 & 4.00 & $-$5 & 0.84 &  0.075 &  0.014 &  1.236 $\pm$ 0.022 &  0.704 $\pm$ 0.011 &  1.940 $\pm$ 0.015 \\ 
 6786 &  28.18197 & $-$13.93418 & 22.44 & 22.44 & 0.24 & 3.56 & $-$1 & 0.38 &  0.090 &  0.089 &  1.212 $\pm$ 0.029 &  0.639 $\pm$ 0.020 &  1.851 $\pm$ 0.028 \\ 
 6612 &  28.15028 & $-$13.93483 & 21.89 & 21.89 & 0.28 & 4.00 & $-$2 & 0.65 &  0.069 &  0.110 &  1.251 $\pm$ 0.023 &  0.661 $\pm$ 0.013 &  1.911 $\pm$ 0.017 \\ 
 5969 &  28.15759 & $-$13.93561 & 22.67 & 23.49 & 0.54 & 0.70 &  8 & 0.81 &  0.230 & $-$0.007 &  0.540 $\pm$ 0.040 &  0.200 $\pm$ 0.187 &  0.741 $\pm$ 0.146 \\ 
 3832 &  28.15066 & $-$13.93553 & 20.85 & 20.88 & 0.51 & 4.00 & $-$4 & 0.76 &  0.099 &  0.110 &  1.260 $\pm$ 0.017 &  0.652 $\pm$ 0.010 &  1.912 $\pm$ 0.010 \\ 
 6434 &  28.15865 & $-$13.94097 & 21.06 & 20.83 & 1.29 & 4.00 &  1 & 0.50 &  0.347 &  0.056 &  1.188 $\pm$ 0.031 &  0.581 $\pm$ 0.030 &  1.769 $\pm$ 0.007 \\ 
 4472 &  28.17677 & $-$13.93832 & 21.17 & 21.21 & 0.44 & 4.00 & $-$2 & 0.80 &  0.108 &  0.091 &  1.277 $\pm$ 0.020 &  0.717 $\pm$ 0.007 &  1.994 $\pm$ 0.014 \\ 
57952 &  28.22004 & $-$13.93984 & 21.26 & 21.52 & 0.74 & 0.55 &  6 & 0.72 &  0.494 &  0.043 &  0.753 $\pm$ 0.019 &  0.324 $\pm$ 0.010 &  1.077 $\pm$ 0.011 \\ 
 6417 &  28.16051 & $-$13.94247 & 21.74 & 21.86 & 0.51 & 4.00 & $-$4 & 0.78 &  0.068 &  0.058 &  1.267 $\pm$ 0.042 &  0.704 $\pm$ 0.031 &  1.971 $\pm$ 0.011 \\ 
10256 &  28.19591 & $-$13.94279 & 23.09 & 23.07 & 0.25 & 4.00 & $-$2 & 0.32 &  0.032 &  0.066 &  1.170 $\pm$ 0.026 &  0.615 $\pm$ 0.031 &  1.785 $\pm$ 0.057 \\ 
 3297 &  28.15014 & $-$13.94219 & 21.92 & 21.69 & 0.82 & 4.00 & $-$2 & 0.58 &  0.197 & $-$0.007 &  1.216 $\pm$ 0.021 &  0.654 $\pm$ 0.010 &  1.869 $\pm$ 0.013 \\ 
 3956 &  28.19088 & $-$13.94592 & 21.94 & 21.92 & 0.46 & 3.64 & $-$5 & 0.91 &  0.043 &  0.209 &  1.252 $\pm$ 0.017 &  0.707 $\pm$ 0.023 &  1.959 $\pm$ 0.012 \\ 
 6259 &  28.17912 & $-$13.95950 & 21.26 & 21.25 & 0.65 & 4.00 & $-$5 & 0.71 &  0.086 &  0.239 &  1.303 $\pm$ 0.009 &  0.738 $\pm$ 0.005 &  2.041 $\pm$ 0.010 \\ 
 5529 &  28.18377 & $-$13.95570 & 22.52 & 22.74 & 0.25 & 2.03 &  1 & 0.48 &  0.055 &  0.282 &  1.294 $\pm$ 0.018 &  0.728 $\pm$ 0.012 &  2.021 $\pm$ 0.017 \\ 
 5897 &  28.17909 & $-$13.95723 & 22.87 & 22.81 & 0.43 & 4.00 & $-$3 & 0.81 &  0.190 &  0.276 &  1.111 $\pm$ 0.023 &  0.658 $\pm$ 0.013 &  1.769 $\pm$ 0.025 \\ 
 5731 &  28.18170 & $-$13.95650 & 22.96 & 22.90 & 0.28 & 4.00 & $-$5 & 0.82 &  0.043 &  0.317 &  1.274 $\pm$ 0.019 &  0.737 $\pm$ 0.013 &  2.011 $\pm$ 0.019 \\ 
 5819 &  28.18067 & $-$13.95721 & 21.94 & 21.81 & 0.47 & 3.99 & $-$4 & 0.78 &  0.072 &  0.305 &  1.287 $\pm$ 0.011 &  0.748 $\pm$ 0.006 &  2.036 $\pm$ 0.008 \\ 
 5495 &  28.18244 & $-$13.95518 & 20.66 & 20.86 & 1.23 & 4.00 & $-$5 & 0.88 &  0.053 &  0.319 &  1.214 $\pm$ 0.008 &  0.679 $\pm$ 0.013 &  1.894 $\pm$ 0.008 \\ 
 5417 &  28.18288 & $-$13.95543 & 20.83 & 20.99 & 0.65 & 4.00 & $-$5 & 0.93 &  0.073 &  0.309 &  1.231 $\pm$ 0.008 &  0.694 $\pm$ 0.008 &  1.925 $\pm$ 0.006 \\ 
 7067 &  28.20945 & $-$13.95258 & 22.38 & 22.39 & 0.44 & 2.03 &  0 & 0.78 &  0.314 & $-$0.030 &  0.945 $\pm$ 0.029 &  0.449 $\pm$ 0.019 &  1.394 $\pm$ 0.027 \\ 
  499 &  28.14386 & $-$13.97821 & 22.22 & 22.10 & 0.60 & 4.00 & $-$1 & 0.45 &  0.281 &  0.019 &  1.249 $\pm$ 0.025 &  0.668 $\pm$ 0.015 &  1.918 $\pm$ 0.021 \\ 
10259 &  28.21236 & $-$13.95833 & 22.49 & 22.55 & 0.28 & 2.86 &  1 & 0.43 &  0.090 &  0.046 &  1.295 $\pm$ 0.030 &  0.739 $\pm$ 0.022 &  2.034 $\pm$ 0.030 \\ 
10164 &  28.21311 & $-$13.95762 & 20.98 & 21.07 & 0.55 & 4.00 &  1 & 0.83 &  0.126 &  0.039 &  1.167 $\pm$ 0.017 &  0.684 $\pm$ 0.010 &  1.851 $\pm$ 0.010 \\ 
 3940 &  28.21666 & $-$13.97074 & 21.09 & 21.58 & 0.49 & 1.53 & $-$1 & 0.50 &  0.138 &  0.102 &  1.278 $\pm$ 0.020 &  0.691 $\pm$ 0.010 &  1.970 $\pm$ 0.012 \\ 
 9877 &  28.18930 & $-$13.92960 & 21.86 & 21.68 & 0.69 & 4.00 & $-$4 & 0.69 &  0.248 &  0.014 &  1.302 $\pm$ 0.021 &  0.654 $\pm$ 0.010 &  1.956 $\pm$ 0.013 \\ 
  900 &  28.18366 & $-$14.01098 & 22.40 & 22.54 & 0.56 & 0.98 &  3 & 0.83 &  0.214 & $-$0.025 &  0.801 $\pm$ 0.030 &  0.353 $\pm$ 0.022 &  1.154 $\pm$ 0.030 \\ 
  607 &  28.12863 & $-$13.96233 & 23.39 & 23.30 & 0.51 & 2.75 & $-$4 & 0.63 &  0.333 & $-$0.022 &  1.161 $\pm$ 0.039 &  0.350 $\pm$ 0.033 &  1.512 $\pm$ 0.046 \\ 
50876 &  28.12637 & $-$13.95348 & 21.14 & 20.95 & 1.14 & 1.70 &  6 & 0.83 &  0.685 &  0.058 &  0.727 $\pm$ 0.017 &  0.283 $\pm$ 0.010 &  1.010 $\pm$ 0.010 \\ 
 8227 &  28.23196 & $-$13.94731 & 21.88 & 22.04 & 0.32 & 4.00 & $-$4 & 0.69 &  0.071 &  0.073 &  1.255 $\pm$ 0.025 &  0.749 $\pm$ 0.015 &  2.003 $\pm$ 0.020 \\ 
 2623 &  28.16858 & $-$13.97001 & 22.54 & 22.56 & 0.38 & 2.48 & $-$1 & 0.32 &  0.187 &  0.136 &  1.286 $\pm$ 0.048 &  0.703 $\pm$ 0.021 &  1.989 $\pm$ 0.027 \\ 
\enddata
\tablenotetext{a}{Adopted total  F775W magnitude, based on SExtractor \magauto\ corrected by 0.2~mag (see text).}
\tablenotetext{b}{Fitted F775W magnitude from GALFIT.}
\tablenotetext{c}{Fitted major axis effective radius}
\tablenotetext{d}{Fitted Sersic index.}
\tablenotetext{e}{Visually classified morphological type from Postman \etal\ (2005).  We assign the two AGNe $T=11$.}
\tablenotetext{f}{Fitted axial ratio.}
\tablenotetext{g}{Galaxy ``bumpiness'' parameter defined in text.}
\tablenotetext{h}{Surface mass density at galaxy position from Jee \etal\ (2005a) weak lensing map (see text).}
\label{tab:dat0152}
\end{deluxetable}

\setlength{\tabcolsep}{5pt}
\begin{deluxetable}{rccccrrrrrrr}
\tabletypesize{\footnotesize}
\tablewidth{0pt}   
\tablecaption{Data for \msname\ Cluster Members}
\tablehead{
\colhead{ID} &
\colhead{$i_{775,{\rm tot}}\tablenotemark{a}$} & 
\colhead{$i_{775,{\rm gfit}}\tablenotemark{b}$} & 
\colhead{$R_e$\tablenotemark{c}} &
\colhead{$n_{\rm Ser}$\tablenotemark{d}} &
\colhead{$T$\tablenotemark{e}} & 
\colhead{$b/a\,$\tablenotemark{f}} &
\colhead{$B\,$\tablenotemark{g}} &
\colhead{$\Sigma_L$\tablenotemark{h}} & 
\colhead{\vicolor\ } &
\colhead{\izcolor\ } &
\colhead{\vzcolor\ }
}
\startdata
   24 & 22.80 & 22.74 & 0.67 & 1.61 & $-$1 & 0.64 &  0.408 &  0.054 &  0.811 $\pm$ 0.038 &  0.279 $\pm$ 0.025 &  1.090 $\pm$ 0.030 \\ 
  156 & 23.54 & 23.44 & 0.65 & 2.41 &  0 & 0.38 &  0.863 &  0.076 &  0.843 $\pm$ 0.044 &  0.284 $\pm$ 0.035 &  1.127 $\pm$ 0.044 \\ 
  215 & 22.16 & 21.87 & 0.92 & 3.91 &  4 & 0.79 &  0.513 &  0.044 &  0.806 $\pm$ 0.027 &  0.214 $\pm$ 0.012 &  1.021 $\pm$ 0.015 \\ 
  308 & 23.77 & 23.88 & 0.18 & 2.37 & $-$5 & 0.76 &  0.027 & $-$0.010 &  1.425 $\pm$ 0.044 &  0.581 $\pm$ 0.060 &  2.007 $\pm$ 0.104 \\ 
  311 & 22.93 & 23.66 & 0.07 & 4.00 &  3 & 0.67 &  0.389 &  0.045 &  0.686 $\pm$ 0.046 &  0.201 $\pm$ 0.039 &  0.887 $\pm$ 0.049 \\ 
  632 & 23.10 & 24.29 & 0.29 & 0.44 &  4 & 0.48 &  0.150 &  0.077 &  0.662 $\pm$ 0.051 &  0.214 $\pm$ 0.048 &  0.876 $\pm$ 0.063 \\ 
  654 & 21.46 & 21.70 & 0.68 & 1.49 &  1 & 0.57 &  0.287 &  0.103 &  1.614 $\pm$ 0.017 &  0.596 $\pm$ 0.007 &  2.210 $\pm$ 0.013 \\ 
  732 & 23.24 & 23.49 & 0.43 & 0.49 &  3 & 0.44 &  0.318 &  0.022 &  0.880 $\pm$ 0.045 &  0.339 $\pm$ 0.036 &  1.219 $\pm$ 0.045 \\ 
  754 & 22.79 & 23.51 & 0.34 & 0.66 &  3 & 0.45 &  0.123 &  0.083 &  1.324 $\pm$ 0.032 &  0.455 $\pm$ 0.026 &  1.778 $\pm$ 0.037 \\ 
  830 & 20.82 & 20.71 & 0.88 & 4.00 & $-$1 & 0.66 &  0.193 &  0.106 &  1.641 $\pm$ 0.014 &  0.691 $\pm$ 0.010 &  2.332 $\pm$ 0.013 \\ 
  846 & 21.32 & 21.32 & 0.44 & 4.00 & $-$5 & 0.80 &  0.053 &  0.163 &  1.628 $\pm$ 0.012 &  0.706 $\pm$ 0.007 &  2.334 $\pm$ 0.009 \\ 
  845 & 22.50 & 22.70 & 0.29 & 2.11 & $-$2 & 0.41 &  0.161 &  0.144 &  1.743 $\pm$ 0.061 &  0.781 $\pm$ 0.017 &  2.524 $\pm$ 0.050 \\ 
  984 & 21.86 & 22.07 & 0.31 & 4.00 & $-$5 & 0.66 &  0.110 &  0.077 &  1.564 $\pm$ 0.030 &  0.603 $\pm$ 0.015 &  2.167 $\pm$ 0.018 \\ 
 1111 & 22.23 & 22.16 & 0.42 & 2.76 & $-$2 & 0.89 &  0.154 &  0.152 &  1.166 $\pm$ 0.022 &  0.501 $\pm$ 0.011 &  1.667 $\pm$ 0.027 \\ 
 1192 & 23.55 & 23.58 & 0.51 & 1.97 &  3 & 0.65 &  0.695 &  0.085 &  0.708 $\pm$ 0.046 &  0.158 $\pm$ 0.037 &  0.866 $\pm$ 0.047 \\ 
 1461 & 21.56 & 22.06 & 0.27 & 4.00 & $-$5 & 0.60 &  0.069 &  0.018 &  1.541 $\pm$ 0.021 &  0.687 $\pm$ 0.011 &  2.228 $\pm$ 0.023 \\ 
 1496 & 21.41 & 21.60 & 0.48 & 4.00 & $-$5 & 0.81 &  0.084 &  0.017 &  1.619 $\pm$ 0.016 &  0.647 $\pm$ 0.020 &  2.267 $\pm$ 0.009 \\ 
 1509 & 22.37 & 22.42 & 0.21 & 4.00 & $-$5 & 0.86 &  0.089 &  0.071 &  1.591 $\pm$ 0.034 &  0.679 $\pm$ 0.020 &  2.270 $\pm$ 0.024 \\ 
 1644 & 22.98 & 22.98 & 0.17 & 4.00 & $-$2 & 0.63 &  0.110 &  0.068 &  0.913 $\pm$ 0.040 &  0.258 $\pm$ 0.028 &  1.171 $\pm$ 0.035 \\ 
 1679 & 22.44 & 22.38 & 0.53 & 4.00 &  1 & 0.69 &  0.332 &  0.057 &  1.125 $\pm$ 0.050 &  0.462 $\pm$ 0.016 &  1.587 $\pm$ 0.034 \\ 
 1978 & 21.95 & 21.67 & 1.22 & 0.90 &  4 & 0.22 &  0.322 &  0.181 &  0.924 $\pm$ 0.024 &  0.303 $\pm$ 0.010 &  1.227 $\pm$ 0.013 \\ 
 1772 & 22.77 & 22.94 & 0.27 & 2.10 & $-$4 & 0.71 &  0.138 &  0.065 &  1.430 $\pm$ 0.040 &  0.413 $\pm$ 0.028 &  1.843 $\pm$ 0.034 \\ 
 1814 & 22.65 & 22.48 & 0.38 & 4.00 & $-$5 & 0.71 &  0.119 &  0.107 &  1.423 $\pm$ 0.063 &  0.558 $\pm$ 0.037 &  1.981 $\pm$ 0.026 \\ 
 1859 & 22.24 & 22.40 & 0.30 & 1.61 &  1 & 0.36 &  0.093 &  0.128 &  1.530 $\pm$ 0.029 &  0.524 $\pm$ 0.021 &  2.054 $\pm$ 0.017 \\ 
 2041 & 23.25 & 23.28 & 0.20 & 3.12 & $-$5 & 0.79 &  0.050 &  0.168 &  1.463 $\pm$ 0.043 &  0.677 $\pm$ 0.033 &  2.140 $\pm$ 0.041 \\ 
 2143 & 20.58 & 20.63 & 1.10 & 4.00 & $-$5 & 0.81 &  0.142 &  0.193 &  1.612 $\pm$ 0.014 &  0.679 $\pm$ 0.010 &  2.292 $\pm$ 0.013 \\ 
 2175 & 22.23 & 22.34 & 0.32 & 4.00 & $-$1 & 0.71 &  0.105 &  0.043 &  1.611 $\pm$ 0.033 &  0.692 $\pm$ 0.019 &  2.303 $\pm$ 0.022 \\ 
 2178 & 22.38 & 22.58 & 0.18 & 3.17 & $-$2 & 0.58 &  0.052 &  0.199 &  1.603 $\pm$ 0.036 &  0.719 $\pm$ 0.022 &  2.322 $\pm$ 0.027 \\ 
 2226 & 25.52 & 25.40 & 0.00 & 4.00 &  9 & 0.82 &  1.080 &  0.124 &  1.144 $\pm$ 0.059 &  0.347 $\pm$ 0.065 &  1.491 $\pm$ 0.087 \\ 
 2218 & 22.76 & 22.97 & 0.39 & 0.98 &  3 & 0.67 &  0.261 & $-$0.012 &  0.922 $\pm$ 0.040 &  0.121 $\pm$ 0.028 &  1.043 $\pm$ 0.035 \\ 
 2225 & 23.44 & 23.48 & 0.40 & 2.92 & $-$1 & 0.41 &  0.415 &  0.053 &  0.626 $\pm$ 0.045 &  0.059 $\pm$ 0.036 &  0.684 $\pm$ 0.045 \\ 
 2240 & 23.67 & 24.17 & 0.48 & 0.50 &  0 & 0.20 &  0.568 &  0.009 &  0.789 $\pm$ 0.050 &  0.270 $\pm$ 0.046 &  1.059 $\pm$ 0.060 \\ 
 2257 & 21.65 & 21.96 & 0.32 & 4.00 & $-$5 & 0.85 &  0.055 &  0.176 &  1.616 $\pm$ 0.028 &  0.606 $\pm$ 0.013 &  2.222 $\pm$ 0.016 \\ 
 2301 & 22.20 & 22.38 & 0.24 & 4.00 & $-$5 & 0.71 &  0.083 &  0.150 &  1.558 $\pm$ 0.034 &  0.670 $\pm$ 0.019 &  2.228 $\pm$ 0.023 \\ 
 2315 & 21.68 & 21.66 & 0.50 & 4.00 & $-$2 & 0.48 &  0.131 &  0.072 &  1.702 $\pm$ 0.024 &  0.790 $\pm$ 0.010 &  2.492 $\pm$ 0.013 \\ 
 2328 & 22.01 & 22.21 & 0.35 & 2.39 & $-$5 & 0.86 &  0.059 &  0.163 &  1.643 $\pm$ 0.032 &  0.720 $\pm$ 0.017 &  2.363 $\pm$ 0.020 \\ 
 2333 & 22.40 & 22.54 & 0.27 & 3.17 & $-$1 & 0.31 &  0.103 &  0.236 &  1.386 $\pm$ 0.035 &  0.335 $\pm$ 0.022 &  1.721 $\pm$ 0.026 \\ 
 2385 & 22.75 & 22.90 & 0.30 & 1.81 & $-$1 & 0.24 &  0.124 &  0.037 &  1.702 $\pm$ 0.039 &  0.685 $\pm$ 0.027 &  2.388 $\pm$ 0.033 \\ 
 2356 & 23.66 & 24.08 & 0.46 & 0.58 &  8 & 0.35 &  0.575 &  0.153 &  0.499 $\pm$ 0.035 & $-$0.116 $\pm$ 0.169 &  0.383 $\pm$ 0.201 \\ 
 2353 & 21.17 & 21.27 & 0.42 & 3.49 & $-$5 & 0.75 &  0.053 &  0.185 &  1.641 $\pm$ 0.016 &  0.732 $\pm$ 0.010 &  2.373 $\pm$ 0.013 \\ 
 2390 & 21.23 & 21.13 & 0.79 & 4.00 & $-$5 & 0.81 &  0.155 &  0.121 &  1.633 $\pm$ 0.014 &  0.734 $\pm$ 0.010 &  2.367 $\pm$ 0.013 \\ 
 2403 & 21.31 & 21.20 & 1.35 & 4.00 &  1 & 0.90 &  0.522 &  0.132 &  1.555 $\pm$ 0.020 &  0.610 $\pm$ 0.007 &  2.165 $\pm$ 0.013 \\ 
 2391 & 21.82 & 21.83 & 0.56 & 4.00 & $-$5 & 0.71 &  0.165 &  0.130 &  1.593 $\pm$ 0.026 &  0.702 $\pm$ 0.012 &  2.295 $\pm$ 0.015 \\ 
 2468 & 21.05 & 21.15 & 0.21 & 4.00 & $-$1 & 0.23 &  0.098 &  0.228 &  1.487 $\pm$ 0.014 &  0.426 $\pm$ 0.010 &  1.914 $\pm$ 0.013 \\ 
 2484 & 23.74 & 24.57 & 0.30 & 0.44 &  3 & 0.16 &  0.187 &  0.072 &  0.688 $\pm$ 0.053 &  0.058 $\pm$ 0.053 &  0.746 $\pm$ 0.069 \\ 
 2476 & 22.64 & 22.53 & 0.44 & 4.00 & $-$1 & 0.39 &  0.142 &  0.163 &  1.643 $\pm$ 0.035 &  0.635 $\pm$ 0.022 &  2.278 $\pm$ 0.026 \\ 
 2547 & 22.60 & 23.40 & 0.26 & 0.57 &  3 & 0.49 &  0.137 &  0.223 &  1.026 $\pm$ 0.044 &  0.415 $\pm$ 0.035 &  1.441 $\pm$ 0.044 \\ 
 2577 & 22.45 & 22.45 & 0.32 & 4.00 & $-$4 & 0.63 &  0.100 &  0.151 &  1.551 $\pm$ 0.034 &  0.645 $\pm$ 0.020 &  2.196 $\pm$ 0.024 \\ 
 2589 & 21.81 & 21.73 & 0.47 & 4.00 & $-$5 & 0.71 &  0.111 &  0.132 &  1.544 $\pm$ 0.025 &  0.698 $\pm$ 0.010 &  2.241 $\pm$ 0.014 \\ 
 2639 & 22.35 & 22.61 & 0.26 & 4.00 & $-$5 & 0.77 &  0.101 &  0.156 &  1.549 $\pm$ 0.026 &  0.650 $\pm$ 0.016 &  2.199 $\pm$ 0.025 \\ 
 2651 & 22.39 & 22.60 & 0.19 & 2.05 & $-$1 & 0.32 &  0.130 &  0.151 &  1.252 $\pm$ 0.028 &  0.248 $\pm$ 0.016 &  1.500 $\pm$ 0.040 \\ 
 2503 & 21.41 & 21.60 & 0.70 & 0.99 &  4 & 0.84 &  0.269 &  0.089 &  1.005 $\pm$ 0.023 &  0.492 $\pm$ 0.010 &  1.497 $\pm$ 0.013 \\ 
 2649 & 20.63 & 20.49 & 1.21 & 2.32 &  1 & 0.49 &  0.335 &  0.159 &  1.460 $\pm$ 0.010 &  0.461 $\pm$ 0.029 &  1.921 $\pm$ 0.037 \\ 
 2709 & 21.88 & 21.86 & 0.40 & 4.00 & $-$5 & 0.64 &  0.078 &  0.194 &  1.597 $\pm$ 0.027 &  0.667 $\pm$ 0.012 &  2.264 $\pm$ 0.015 \\ 
 2741 & 21.39 & 21.58 & 0.43 & 2.39 & $-$2 & 0.61 &  0.178 &  0.143 &  1.645 $\pm$ 0.016 &  0.717 $\pm$ 0.007 &  2.362 $\pm$ 0.010 \\ 
 2787 & 22.77 & 22.83 & 0.31 & 2.55 & $-$4 & 0.89 &  0.087 &  0.078 &  1.634 $\pm$ 0.038 &  0.606 $\pm$ 0.026 &  2.240 $\pm$ 0.032 \\ 
 2830 & 19.87 & 19.80 & 2.15 & 4.00 & $-$5 & 0.65 &  0.078 &  0.274 &  1.705 $\pm$ 0.035 &  0.751 $\pm$ 0.015 &  2.456 $\pm$ 0.020 \\ 
 2844 & 22.48 & 22.56 & 0.40 & 2.26 & $-$1 & 0.43 &  0.193 &  0.157 &  0.905 $\pm$ 0.036 &  0.114 $\pm$ 0.022 &  1.019 $\pm$ 0.026 \\ 
 2896 & 22.97 & 23.22 & 0.14 & 2.66 & $-$4 & 0.40 &  0.092 &  0.068 &  1.446 $\pm$ 0.042 &  0.524 $\pm$ 0.032 &  1.970 $\pm$ 0.040 \\ 
 2873 & 23.12 & 23.25 & 0.35 & 1.77 &  3 & 0.94 &  0.236 &  0.058 &  1.445 $\pm$ 0.043 &  0.595 $\pm$ 0.032 &  2.041 $\pm$ 0.041 \\ 
 2894 & 21.56 & 21.85 & 0.28 & 2.24 & $-$5 & 0.70 &  0.046 &  0.166 &  1.643 $\pm$ 0.027 &  0.708 $\pm$ 0.012 &  2.351 $\pm$ 0.015 \\ 
 2898 & 23.05 & 23.20 & 0.27 & 0.92 &  1 & 0.94 &  0.206 &  0.017 &  0.740 $\pm$ 0.042 &  0.216 $\pm$ 0.032 &  0.955 $\pm$ 0.039 \\ 
 2874 & 22.15 & 22.18 & 0.23 & 4.00 & $-$2 & 0.44 &  0.100 &  0.097 &  1.644 $\pm$ 0.031 &  0.670 $\pm$ 0.017 &  2.315 $\pm$ 0.020 \\ 
 2926 & 24.04 & 23.87 & 0.62 & 3.63 &  8 & 0.52 &  1.081 &  0.030 &  0.254 $\pm$ 0.048 &  0.116 $\pm$ 0.042 &  0.370 $\pm$ 0.054 \\ 
 2901 & 22.28 & 22.38 & 0.26 & 3.64 & $-$5 & 0.77 &  0.064 &  0.161 &  1.551 $\pm$ 0.024 &  0.677 $\pm$ 0.023 &  2.229 $\pm$ 0.016 \\ 
 3035 & 21.77 & 21.80 & 0.35 & 4.00 & $-$3 & 0.45 &  0.053 &  0.242 &  1.719 $\pm$ 0.018 &  0.755 $\pm$ 0.013 &  2.474 $\pm$ 0.024 \\ 
 2937 & 21.33 & 21.49 & 0.28 & 2.15 & $-$4 & 0.91 &  0.088 &  0.159 &  1.519 $\pm$ 0.021 &  0.521 $\pm$ 0.010 &  2.040 $\pm$ 0.013 \\ 
 2991 & 22.45 & 22.70 & 0.29 & 1.19 &  1 & 0.65 &  0.281 &  0.050 &  0.719 $\pm$ 0.037 &  0.258 $\pm$ 0.024 &  0.977 $\pm$ 0.029 \\ 
 3003 & 22.50 & 22.84 & 0.57 & 0.58 &  4 & 0.53 &  0.308 &  0.005 &  1.366 $\pm$ 0.039 &  0.507 $\pm$ 0.026 &  1.873 $\pm$ 0.032 \\ 
 3148 & 22.43 & 22.67 & 0.79 & 1.91 & 10 & 0.49 &  0.226 &  0.267 &  1.494 $\pm$ 0.026 &  0.501 $\pm$ 0.021 &  1.994 $\pm$ 0.034 \\ 
 3031 & 21.95 & 22.15 & 0.34 & 2.10 & $-$2 & 0.52 &  0.060 &  0.156 &  1.593 $\pm$ 0.031 &  0.681 $\pm$ 0.016 &  2.274 $\pm$ 0.019 \\ 
 3052 & 21.54 & 21.70 & 0.37 & 3.04 & $-$4 & 0.48 &  0.100 &  0.073 &  1.597 $\pm$ 0.024 &  0.718 $\pm$ 0.010 &  2.315 $\pm$ 0.013 \\ 
 3070 & 21.63 & 21.73 & 0.50 & 2.45 & $-$2 & 0.35 &  0.203 &  0.153 &  1.649 $\pm$ 0.025 &  0.731 $\pm$ 0.010 &  2.381 $\pm$ 0.014 \\ 
 3131 & 22.58 & 22.59 & 0.31 & 2.91 & $-$4 & 0.70 &  0.078 &  0.157 &  1.630 $\pm$ 0.036 &  0.643 $\pm$ 0.022 &  2.273 $\pm$ 0.027 \\ 
 3113 & 22.19 & 22.23 & 0.34 & 4.00 & $-$5 & 0.76 &  0.064 &  0.097 &  1.649 $\pm$ 0.032 &  0.651 $\pm$ 0.017 &  2.300 $\pm$ 0.020 \\ 
 3180 & 21.25 & 21.28 & 0.47 & 4.00 &  3 & 0.95 &  0.208 &  0.119 &  1.513 $\pm$ 0.016 &  0.422 $\pm$ 0.010 &  1.935 $\pm$ 0.013 \\ 
 3185 & 22.14 & 22.27 & 0.29 & 4.00 & $-$5 & 0.72 &  0.138 &  0.185 &  1.621 $\pm$ 0.032 &  0.662 $\pm$ 0.018 &  2.283 $\pm$ 0.021 \\ 
 3224 & 21.85 & 21.97 & 0.45 & 4.00 & $-$4 & 0.74 &  0.227 &  0.042 &  1.588 $\pm$ 0.028 &  0.679 $\pm$ 0.013 &  2.267 $\pm$ 0.016 \\ 
 5116 & 22.32 & 22.45 & 0.22 & 4.00 & $-$4 & 0.86 &  0.054 &  0.205 &  1.644 $\pm$ 0.079 &  0.743 $\pm$ 0.036 &  2.387 $\pm$ 0.043 \\ 
 3241 & 22.57 & 22.71 & 0.25 & 2.24 & $-$1 & 0.44 &  0.109 &  0.158 &  1.610 $\pm$ 0.037 &  0.673 $\pm$ 0.024 &  2.282 $\pm$ 0.029 \\ 
 5082 & 22.70 & 22.83 & 0.75 & 0.88 &  8 & 0.56 &  0.404 &  0.045 &  0.442 $\pm$ 0.038 &  0.022 $\pm$ 0.026 &  0.464 $\pm$ 0.032 \\ 
 5134 & 23.22 & 23.18 & 0.16 & 4.00 & $-$4 & 0.62 &  0.148 &  0.195 &  1.211 $\pm$ 0.042 &  0.365 $\pm$ 0.031 &  1.576 $\pm$ 0.039 \\ 
 3386 & 22.67 & 22.77 & 0.55 & 1.80 &  4 & 0.86 &  0.174 &  0.136 &  0.830 $\pm$ 0.038 &  0.232 $\pm$ 0.025 &  1.062 $\pm$ 0.031 \\ 
 3652 & 22.01 & 21.94 & 0.34 & 4.00 & $-$1 & 0.56 &  0.168 &  0.085 &  1.578 $\pm$ 0.028 &  0.684 $\pm$ 0.013 &  2.263 $\pm$ 0.016 \\ 
 3686 & 22.67 & 22.93 & 0.15 & 3.22 & $-$4 & 0.47 &  0.037 &  0.149 &  1.587 $\pm$ 0.040 &  0.706 $\pm$ 0.028 &  2.293 $\pm$ 0.034 \\ 
 3692 & 22.21 & 22.40 & 0.79 & 1.13 &  6 & 0.61 &  0.288 &  0.056 &  1.167 $\pm$ 0.034 &  0.392 $\pm$ 0.020 &  1.560 $\pm$ 0.024 \\ 
 3718 & 21.76 & 21.78 & 0.43 & 3.38 & $-$4 & 0.67 &  0.032 &  0.144 &  1.736 $\pm$ 0.026 &  0.757 $\pm$ 0.011 &  2.493 $\pm$ 0.014 \\ 
 3739 & 23.16 & 23.20 & 0.11 & 4.00 & $-$5 & 0.81 &  0.085 &  0.155 &  1.493 $\pm$ 0.042 &  0.584 $\pm$ 0.032 &  2.078 $\pm$ 0.039 \\ 
 3795 & 23.83 & 24.04 & 0.24 & 1.69 & $-$1 & 0.36 &  0.047 &  0.146 &  1.426 $\pm$ 0.048 &  0.211 $\pm$ 0.031 &  1.637 $\pm$ 0.051 \\ 
 3812 & 21.64 & 21.95 & 0.45 & 4.00 &  8 & 0.81 &  0.380 &  0.026 &  0.274 $\pm$ 0.020 &  0.125 $\pm$ 0.019 &  0.399 $\pm$ 0.029 \\ 
 4015 & 22.33 & 22.28 & 0.28 & 4.00 & $-$1 & 0.49 &  0.154 &  0.091 &  1.473 $\pm$ 0.027 &  0.469 $\pm$ 0.015 &  1.942 $\pm$ 0.015 \\ 
 4049 & 22.80 & 22.82 & 0.21 & 4.00 & $-$2 & 0.49 &  0.028 &  0.059 &  1.574 $\pm$ 0.036 &  0.663 $\pm$ 0.018 &  2.237 $\pm$ 0.026 \\ 
 4063 & 23.09 & 23.21 & 0.35 & 2.77 & $-$2 & 0.34 &  0.022 &  0.179 &  1.468 $\pm$ 0.030 &  0.602 $\pm$ 0.027 &  2.070 $\pm$ 0.028 \\ 
 4137 & 22.95 & 22.92 & 0.31 & 3.08 & $-$4 & 0.66 &  0.109 &  0.176 &  1.538 $\pm$ 0.028 &  0.607 $\pm$ 0.019 &  2.146 $\pm$ 0.024 \\ 
 4169 & 22.93 & 23.47 & 0.15 & 1.97 & $-$5 & 0.65 &  0.030 &  0.141 &  1.544 $\pm$ 0.035 &  0.633 $\pm$ 0.038 &  2.178 $\pm$ 0.026 \\ 
 4254 & 23.44 & 23.32 & 0.27 & 3.89 & $-$5 & 0.72 &  0.032 &  0.097 &  1.591 $\pm$ 0.034 &  0.678 $\pm$ 0.042 &  2.269 $\pm$ 0.076 \\ 
 4305 & 22.55 & 22.56 & 0.26 & 4.00 & $-$5 & 0.78 &  0.082 &  0.132 &  1.573 $\pm$ 0.018 &  0.704 $\pm$ 0.011 &  2.277 $\pm$ 0.013 \\ 
 4308 & 23.48 & 23.60 & 0.33 & 3.91 & $-$4 & 0.42 &  0.050 &  0.200 &  1.403 $\pm$ 0.128 &  0.587 $\pm$ 0.027 &  1.990 $\pm$ 0.128 \\ 
 4324 & 21.84 & 21.78 & 0.50 & 2.99 & $-$5 & 0.75 &  0.096 &  0.165 &  1.539 $\pm$ 0.018 &  0.649 $\pm$ 0.009 &  2.188 $\pm$ 0.010 \\ 
 4353 & 21.69 & 21.85 & 0.50 & 2.52 & $-$2 & 0.48 &  0.092 &  0.087 &  1.653 $\pm$ 0.019 &  0.725 $\pm$ 0.008 &  2.378 $\pm$ 0.011 \\ 
 4358 & 21.68 & 21.67 & 0.57 & 3.58 & $-$5 & 0.61 &  0.046 &  0.321 &  1.603 $\pm$ 0.017 &  0.698 $\pm$ 0.022 &  2.302 $\pm$ 0.020 \\ 
 4213 & 22.07 & 22.03 & 0.66 & 4.00 & $-$2 & 0.46 &  0.304 &  0.325 &  1.613 $\pm$ 0.021 &  0.688 $\pm$ 0.038 &  2.301 $\pm$ 0.038 \\ 
 4363 & 23.58 & 23.78 & 0.08 & 3.13 & $-$5 & 0.59 &  0.091 &  0.328 &  1.680 $\pm$ 0.111 &  0.703 $\pm$ 0.035 &  2.383 $\pm$ 0.076 \\ 
 4366 & 21.78 & 22.47 & 0.40 & 0.76 &  8 & 0.93 &  0.181 &  0.147 &  1.131 $\pm$ 0.017 &  0.330 $\pm$ 0.018 &  1.460 $\pm$ 0.024 \\ 
 4374 & 22.01 & 22.03 & 0.21 & 4.00 & $-$4 & 0.82 &  0.072 &  0.029 &  1.538 $\pm$ 0.059 &  0.591 $\pm$ 0.029 &  2.129 $\pm$ 0.030 \\ 
 4214 & 22.08 & 22.18 & 0.42 & 4.00 & $-$1 & 0.47 &  0.213 &  0.320 &  1.522 $\pm$ 0.022 &  0.630 $\pm$ 0.012 &  2.153 $\pm$ 0.014 \\ 
 4167 & 20.31 & 20.63 & 0.99 & 2.60 & $-$5 & 0.87 &  0.115 &  0.324 &  1.593 $\pm$ 0.010 &  0.711 $\pm$ 0.016 &  2.303 $\pm$ 0.011 \\ 
 4364 & 21.38 & 21.46 & 0.38 & 4.00 & $-$5 & 0.90 &  0.044 &  0.281 &  1.612 $\pm$ 0.037 &  0.695 $\pm$ 0.007 &  2.307 $\pm$ 0.036 \\ 
 4488 & 21.58 & 21.36 & 0.59 & 4.00 &  0 & 0.82 &  0.304 &  0.126 &  1.454 $\pm$ 0.013 &  0.546 $\pm$ 0.008 &  2.000 $\pm$ 0.021 \\ 
 4212 & 23.18 & 23.18 & 0.26 & 4.00 & $-$5 & 0.81 &  0.068 &  0.331 &  1.607 $\pm$ 0.030 &  0.676 $\pm$ 0.033 &  2.283 $\pm$ 0.028 \\ 
 4620 & 21.42 & 23.36 & 0.18 & 4.00 &  6 & 0.78 &  0.057 &  0.221 &  0.970 $\pm$ 0.025 &  0.368 $\pm$ 0.020 &  1.338 $\pm$ 0.032 \\ 
 4717 & 23.40 & 24.47 & 0.36 & 0.36 &  8 & 0.28 &  0.113 &  0.068 &  1.358 $\pm$ 0.245 &  0.275 $\pm$ 0.135 &  1.633 $\pm$ 0.111 \\ 
 4735 & 21.41 & 21.53 & 0.36 & 4.00 & $-$2 & 0.41 &  0.203 &  0.216 &  1.975 $\pm$ 0.015 &  0.514 $\pm$ 0.007 &  2.489 $\pm$ 0.009 \\ 
 4766 & 21.22 & 21.41 & 0.42 & 3.76 & $-$5 & 0.98 &  0.060 &  0.070 &  1.628 $\pm$ 0.013 &  0.729 $\pm$ 0.007 &  2.357 $\pm$ 0.009 \\ 
 4778 & 22.77 & 22.87 & 0.24 & 3.05 & $-$2 & 0.41 &  0.093 &  0.104 &  1.518 $\pm$ 0.039 &  0.691 $\pm$ 0.027 &  2.209 $\pm$ 0.033 \\ 
 4942 & 21.82 & 21.72 & 0.41 & 4.00 & $-$4 & 0.57 &  0.120 &  0.155 &  1.520 $\pm$ 0.025 &  0.619 $\pm$ 0.010 &  2.139 $\pm$ 0.013 \\ 
 4954 & 21.22 & 21.47 & 0.67 & 1.52 &  1 & 0.28 &  0.234 &  0.237 &  1.289 $\pm$ 0.014 &  0.419 $\pm$ 0.014 &  1.709 $\pm$ 0.009 \\ 
 4987 & 21.02 & 21.39 & 0.32 & 2.08 & $-$1 & 0.42 &  0.181 &  0.119 &  1.544 $\pm$ 0.019 &  0.594 $\pm$ 0.010 &  2.138 $\pm$ 0.013 \\ 
 4988 & 21.96 & 22.05 & 0.40 & 4.00 & $-$5 & 0.82 &  0.121 &  0.121 &  1.604 $\pm$ 0.029 &  0.682 $\pm$ 0.015 &  2.286 $\pm$ 0.018 \\ 
 5006 & 21.01 & 21.09 & 0.56 & 3.40 & $-$5 & 0.68 &  0.089 &  0.200 &  1.674 $\pm$ 0.014 &  0.747 $\pm$ 0.010 &  2.421 $\pm$ 0.013 \\ 
 5048 & 23.61 & 23.77 & 0.19 & 1.54 & $-$4 & 0.81 &  0.081 &  0.214 &  1.153 $\pm$ 0.047 &  0.420 $\pm$ 0.040 &  1.573 $\pm$ 0.051 \\ 
 5062 & 22.18 & 22.30 & 0.32 & 3.20 & $-$2 & 0.27 &  0.085 &  0.262 &  1.579 $\pm$ 0.023 &  0.629 $\pm$ 0.020 &  2.208 $\pm$ 0.015 \\ 
 5072 & 22.17 & 22.32 & 0.19 & 4.00 & $-$2 & 0.51 &  0.063 &  0.118 &  1.619 $\pm$ 0.033 &  0.689 $\pm$ 0.018 &  2.308 $\pm$ 0.022 \\ 
 5099 & 22.49 & 22.48 & 0.21 & 3.84 & $-$2 & 0.65 &  0.060 &  0.271 &  1.639 $\pm$ 0.026 &  0.724 $\pm$ 0.031 &  2.363 $\pm$ 0.018 \\ 
 5472 & 23.37 & 23.39 & 0.30 & 2.21 & $-$4 & 0.55 &  0.140 &  0.110 &  1.620 $\pm$ 0.044 &  0.582 $\pm$ 0.035 &  2.202 $\pm$ 0.043 \\ 
 5532 & 22.88 & 22.90 & 0.26 & 4.00 & $-$1 & 0.58 &  0.133 &  0.054 &  1.662 $\pm$ 0.039 &  0.651 $\pm$ 0.027 &  2.313 $\pm$ 0.033 \\ 
 8107 & 23.37 & 23.23 & 0.41 & 2.73 & $-$1 & 0.55 &  0.377 &  0.037 &  0.665 $\pm$ 0.042 &  0.227 $\pm$ 0.032 &  0.892 $\pm$ 0.040 \\ 
 6011 & 22.57 & 23.06 & 0.53 & 1.75 &  6 & 0.53 &  0.270 &  0.001 &  0.857 $\pm$ 0.041 &  0.260 $\pm$ 0.029 &  1.117 $\pm$ 0.036 \\ 
 6304 & 21.74 & 21.84 & 0.26 & 4.00 & $-$2 & 0.59 &  0.134 &  0.063 &  1.712 $\pm$ 0.026 &  0.769 $\pm$ 0.012 &  2.481 $\pm$ 0.015 \\ 
 6736 & 22.22 & 22.42 & 0.63 & 0.88 &  3 & 0.50 &  0.375 &  0.026 &  1.041 $\pm$ 0.034 &  0.370 $\pm$ 0.020 &  1.411 $\pm$ 0.024 \\ 
 6995 & 21.82 & 21.72 & 0.95 & 1.87 &  6 & 0.81 &  0.315 &  0.024 &  0.875 $\pm$ 0.025 &  0.243 $\pm$ 0.010 &  1.117 $\pm$ 0.013 \\ 
 7431 & 22.78 & 22.76 & 0.24 & 4.00 & $-$1 & 0.33 &  0.176 &  0.134 &  1.279 $\pm$ 0.038 &  0.579 $\pm$ 0.025 &  1.858 $\pm$ 0.030 \\ 
 7835 & 22.13 & 22.15 & 0.31 & 4.00 & $-$2 & 0.55 &  0.154 &  0.116 &  1.600 $\pm$ 0.031 &  0.555 $\pm$ 0.016 &  2.154 $\pm$ 0.019 \\ 
 8222 & 21.89 & 21.86 & 0.54 & 4.00 & $-$4 & 0.85 &  0.145 &  0.100 &  1.643 $\pm$ 0.027 &  0.711 $\pm$ 0.012 &  2.354 $\pm$ 0.015 \\ 
 8262 & 21.46 & 21.60 & 0.25 & 4.00 & $-$5 & 0.91 &  0.085 &  0.066 &  1.586 $\pm$ 0.023 &  0.597 $\pm$ 0.010 &  2.183 $\pm$ 0.013 \\ 
 8312 & 22.72 & 22.76 & 0.32 & 2.37 & $-$1 & 0.64 &  0.121 &  0.128 &  1.635 $\pm$ 0.038 &  0.664 $\pm$ 0.025 &  2.299 $\pm$ 0.030 \\ 
 8362 & 22.20 & 22.08 & 0.37 & 4.00 & $-$1 & 0.63 &  0.097 &  0.064 &  1.511 $\pm$ 0.046 &  0.568 $\pm$ 0.011 &  2.079 $\pm$ 0.048 \\ 
 8456 & 21.11 & 21.21 & 0.51 & 4.00 & $-$4 & 0.75 &  0.084 &  0.072 &  1.650 $\pm$ 0.014 &  0.733 $\pm$ 0.010 &  2.383 $\pm$ 0.013 \\ 
 8422 & 21.86 & 22.10 & 0.42 & 1.76 &  1 & 0.86 &  0.293 &  0.084 &  1.359 $\pm$ 0.030 &  0.538 $\pm$ 0.015 &  1.897 $\pm$ 0.018 \\ 
 8438 & 21.46 & 21.34 & 0.76 & 4.00 & $-$2 & 0.54 &  0.282 &  0.067 &  1.604 $\pm$ 0.018 &  0.652 $\pm$ 0.010 &  2.256 $\pm$ 0.013 \\ 
 5447 & 22.46 & 22.15 & 1.02 & 1.44 &  3 & 0.66 &  0.327 &  0.080 &  0.847 $\pm$ 0.031 &  0.192 $\pm$ 0.016 &  1.038 $\pm$ 0.019 \\ 
 5324 & 20.77 & 21.36 & 0.32 & 3.64 &  0 & 0.75 &  0.080 &  0.176 &  1.675 $\pm$ 0.018 &  0.732 $\pm$ 0.010 &  2.406 $\pm$ 0.013 \\ 
\enddata
\vspace{-6pt}
\tablenotetext{a}{Adopted total  F775W magnitude, based on SExtractor \magauto\ corrected by 0.2~mag (see text).}
\tablenotetext{b}{Fitted F775W magnitude from GALFIT.}
\tablenotetext{c}{Fitted major axis effective radius}
\tablenotetext{d}{Fitted Sersic index.}
\tablenotetext{e}{Visually classified morphological type from Postman \etal\ (2005).}
\tablenotetext{f}{Fitted axial ratio.}
\tablenotetext{g}{Galaxy ``bumpiness'' parameter defined in text.}
\tablenotetext{h}{Surface mass density at galaxy position from Jee \etal\ (2005b) weak lensing map (see text).}
\tablecomments{Coordinates will be published with the spectroscopic
membership catalogue (K.-V.~Tran \etal, in preparation).}
\label{tab:dat1054}
\end{deluxetable}

\clearpage

\begin{deluxetable}{cccccrrccccc}
\tabletypesize{\small}
\tablewidth{0pt}   
\tablecaption{Offsets and Scatters for \rxname\ Color--Magnitude Relations}
\tablehead{
\colhead{Color} &
\colhead{Type} &
\colhead{$i_{775,{\rm lim}}$} &
\colhead{$R_{\rm max}$\tablenotemark{a}} & 
\colhead{$\Sigma_{\rm min}\tablenotemark{b}$} & 
\colhead{$N_g$\tablenotemark{c}} &
\colhead{Offset\tablenotemark{d}} &
\colhead{$\pm$} & 
\colhead{$\sigma_{\rm obs}$\tablenotemark{e}} &
\colhead{$\pm$} &
\colhead{$\sigma_{\rm int}$\tablenotemark{f}} &
\colhead{$\pm$} 
}
\startdata
\riclr &      E &  23.0 & \dots & \dots &  39 &  0.002 & 0.006 &  0.036 & 0.006 &    0.027 & 0.009 \\ 
\riclr &   E+S0 &  23.0 & \dots & \dots &  64 &$-$0.001 & 0.006 &  0.043 & 0.006 &    0.034 & 0.008 \\ 
\rzclr &      E &  23.0 & \dots & \dots &  39 &  0.003 & 0.012 &  0.075 & 0.010 &    0.072 & 0.010 \\ 
\rzclr &   E+S0 &  23.0 & \dots & \dots &  64 &$-$0.005 & 0.011 &  0.086 & 0.009 &    0.083 & 0.009 \\[5pt]
\riclr &      E &  23.0 &   1.1 & \dots &  25 &  0.006 & 0.007 &  0.035 & 0.008 &    0.028 & 0.011 \\ 
\riclr &   E+S0 &  23.0 &   1.1 & \dots &  34 &  0.006 & 0.008 &  0.042 & 0.009 &    0.034 & 0.011 \\ 
\rzclr &      E &  23.0 &   1.1 & \dots &  25 &  0.022 & 0.017 &  0.074 & 0.012 &    0.072 & 0.012 \\ 
\rzclr &   E+S0 &  23.0 &   1.1 & \dots &  34 &  0.020 & 0.015 &  0.081 & 0.011 &    0.078 & 0.012 \\[5pt]
\riclr &      E &  23.0 & \dots &   0.1 &  24 &  0.005 & 0.006 &  0.030 & 0.006 &    0.023 & 0.009 \\ 
\riclr &   E+S0 &  23.0 & \dots &   0.1 &  33 &  0.007 & 0.006 &  0.032 & 0.007 &    0.025 & 0.009 \\ 
\rzclr &      E &  23.0 & \dots &   0.1 &  24 &  0.012 & 0.013 &  0.063 & 0.010 &    0.060 & 0.011 \\ 
\rzclr &   E+S0 &  23.0 & \dots &   0.1 &  33 &  0.013 & 0.013 &  0.069 & 0.010 &    0.067 & 0.010 \\[5pt]
\riclr &      E &  22.5 & \dots & \dots &  29 &  0.004 & 0.006 &  0.033 & 0.006 &    0.025 & 0.009 \\ 
\riclr &   E+S0 &  22.5 & \dots & \dots &  46 &  0.001 & 0.006 &  0.039 & 0.006 &    0.033 & 0.007 \\ 
\rzclr &      E &  22.5 & \dots & \dots &  29 &  0.008 & 0.011 &  0.065 & 0.011 &    0.063 & 0.012 \\ 
\rzclr &   E+S0 &  22.5 & \dots & \dots &  46 &  0.001 & 0.011 &  0.076 & 0.010 &    0.074 & 0.011 \\[5pt]
\izclr &      E &  23.0 & \dots & \dots &  39 &  0.002 & 0.008 &  0.047 & 0.007 &    0.045 & 0.007 \\ 
\izclr &   E+S0 &  23.0 & \dots & \dots &  64 &$-$0.005 & 0.007 &  0.052 & 0.006 &    0.050 & 0.006 \\ 
\rzclr &     S0 &  23.0 & \dots & \dots &  25 &$-$0.023 & 0.023 &  0.105 & 0.020 &    0.102 & 0.020 \\ 
\rzclr &   Late &  23.0 & \dots & \dots &  31 &$-$0.362 & 0.086 &  0.399 & 0.062 &    0.398 & 0.062 \\ 
\enddata
\vspace{-6pt}
\tablenotetext{a}{Maximum radius (arcminutes) from the center of the cluster (1\farcm1 = 0.5 Mpc).} 
\tablenotetext{b}{Minimum surface mass density, based on weak lensing.}
\tablenotetext{c}{Number of galaxies in subsample.}
\tablenotetext{d}{Color offset in magnitude, based on the biweight location estimator, of galaxies in the subsample
from the elliptical galaxy relations given in Eqs.~\ref{eq:cmrs0152}}
\tablenotetext{e}{Observed color scatter for the subsample based on the biweight scale estimator.}
\tablenotetext{f}{Estimated internal color scatter, corrected for measurement error.}
\label{tab:stats0152}
\end{deluxetable}

\begin{deluxetable}{cccccrrccccc}
\tabletypesize{\small}
\tablewidth{0pt}   
\tablecaption{Offsets and Scatters for \msname\ Color--Magnitude Relations}
\tablehead{
\colhead{Color} &
\colhead{Type} &
\colhead{$i_{775,{\rm lim}}$} &
\colhead{$R_{\rm max}$\tablenotemark{a}} & 
\colhead{$\Sigma_{\rm min}\tablenotemark{b}$} & 
\colhead{$N_g$\tablenotemark{c}} &
\colhead{Offset\tablenotemark{d}} &
\colhead{$\pm$} & 
\colhead{$\sigma_{\rm obs}$\tablenotemark{e}} &
\colhead{$\pm$} &
\colhead{$\sigma_{\rm int}$\tablenotemark{f}} &
\colhead{$\pm$} 
}
\startdata
\viclr &      E &  23.0 & \dots & \dots &  46 &  0.004 & 0.007 &  0.049 & 0.008 &    0.040 & 0.010 \\ 
\viclr &   E+S0 &  23.0 & \dots & \dots &  84 &  0.012 & 0.007 &  0.068 & 0.010 &    0.061 & 0.011 \\ 
\vzclr &      E &  23.0 & \dots & \dots &  46 &$-$0.000 & 0.014 &  0.082 & 0.015 &    0.080 & 0.015 \\ 
\vzclr &   E+S0 &  23.0 & \dots & \dots &  84 &  0.014 & 0.012 &  0.114 & 0.018 &    0.112 & 0.019 \\[5pt]
\viclr &      E &  23.0 &   1.1 & \dots &  25 &  0.000 & 0.010 &  0.050 & 0.009 &    0.040 & 0.012 \\ 
\viclr &   E+S0 &  23.0 &   1.1 & \dots &  43 &  0.010 & 0.010 &  0.063 & 0.013 &    0.057 & 0.014 \\ 
\vzclr &      E &  23.0 &   1.1 & \dots &  25 &$-$0.008 & 0.016 &  0.079 & 0.019 &    0.076 & 0.020 \\ 
\vzclr &   E+S0 &  23.0 &   1.1 & \dots &  43 &  0.016 & 0.015 &  0.090 & 0.021 &    0.088 & 0.022 \\[5pt]
\viclr &      E &  23.0 & \dots &   0.1 &  32 &  0.004 & 0.008 &  0.048 & 0.009 &    0.039 & 0.012 \\ 
\viclr &   E+S0 &  23.0 & \dots &   0.1 &  57 &  0.009 & 0.008 &  0.066 & 0.014 &    0.060 & 0.015 \\ 
\vzclr &      E &  23.0 & \dots &   0.1 &  32 &  0.003 & 0.015 &  0.082 & 0.017 &    0.080 & 0.017 \\ 
\vzclr &   E+S0 &  23.0 & \dots &   0.1 &  57 &  0.010 & 0.015 &  0.113 & 0.023 &    0.111 & 0.024 \\[5pt]
\viclr &      E &  22.5 & \dots & \dots &  38 &  0.002 & 0.007 &  0.044 & 0.006 &    0.036 & 0.008 \\ 
\viclr &   E+S0 &  22.5 & \dots & \dots &  66 &  0.007 & 0.007 &  0.058 & 0.008 &    0.052 & 0.009 \\ 
\vzclr &      E &  22.5 & \dots & \dots &  38 &$-$0.002 & 0.014 &  0.080 & 0.012 &    0.078 & 0.013 \\ 
\vzclr &   E+S0 &  22.5 & \dots & \dots &  66 &  0.005 & 0.014 &  0.110 & 0.017 &    0.108 & 0.017 \\[5pt]
\izclr &      E &  23.0 & \dots & \dots &  46 &  0.002 & 0.007 &  0.046 & 0.007 &    0.043 & 0.008 \\ 
\izclr &   E+S0 &  23.0 & \dots & \dots &  84 &  0.006 & 0.007 &  0.061 & 0.010 &    0.058 & 0.011 \\ 
\vzclr &     S0 &  23.0 & \dots & \dots &  38 &  0.037 & 0.032 &  0.172 & 0.062 &    0.171 & 0.063 \\ 
\vzclr &   Late &  23.0 & \dots & \dots &  29 &$-$0.793 & 0.108 &  0.493 & 0.064 &    0.493 & 0.064 \\ 
\enddata
\vspace{-6pt}
\tablenotetext{a}{Maximum radius (arcminutes) from the center of the cluster (1\farcm1 = 0.5 Mpc).} 
\tablenotetext{b}{Minimum surface mass density, based on weak lensing.}
\tablenotetext{c}{Number of galaxies in subsample.}
\tablenotetext{d}{Color offset in magnitude, based on the biweight location estimator, of galaxies in the subsample
from the elliptical galaxy relations given in Eqs.~\ref{eq:cmrs1054}}
\tablenotetext{e}{Observed color scatter for the subsample based on the biweight scale estimator.}
\tablenotetext{f}{Estimated internal color scatter, corrected for measurement error.}
\label{tab:stats1054}
\end{deluxetable}


\begin{deluxetable}{lccc}
\centering
\tabletypesize{\small}
\tablecaption{Residual Correlations with Radius and Density\label{tab:rescorrels}}
\tablewidth{0pt} 
\tablehead{ 
\colhead{Sample} & 
\colhead{Cluster radius} &
\colhead{Number density} &
\colhead{Mass density~$\Sigma$}
}\startdata
\multicolumn{4}{c}{\rxname\ $\Delta$\rzcolor\ Correlations} \\[2pt] \tableline\\
  All & 0.999, 0.899 & 0.999, 0.983 & 1.000, 1.000 \\
 Late & 0.304, 0.199 & 0.512, 0.503 & 0.891, 0.828 \\
 E+S0 & 0.992, 0.800 & 0.998, 0.911 & 0.999, 0.983 \\
    E & 0.945, 0.432 & 0.904, 0.750 & 0.957, 0.801 \\
   S0 & 0.916, 0.478 & 0.953, 0.734 & 0.993, 0.933 \\[2pt]
\tableline\\
\multicolumn{4}{c}{\msname\ $\Delta$\vzcolor\ Correlations} \\[2pt] \tableline\\
  All & 0.957, 0.999 & 0.999, 1.000 & 0.995, 0.999 \\
 Late & 0.938, 0.839 & 0.998, 0.999 & 0.984, 0.967 \\
 E+S0 & 0.442, 0.720 & 0.798, 0.555 & 0.557, 0.566 \\
    E & 0.256, 0.365 & 0.812, 0.748 & 0.858, 0.883 \\
   S0 & 0.170, 0.405 & 0.270, 0.186 & 0.171, 0.258 \\
\enddata
\tablecomments{\small Fractional significances of the correlations of the 
residuals from the color--magnitude relations with radius,
local number density, and local mass density.  For each correlation,
we give the significances based on (1)~Spearman rank ordering,~ 
(2)~Pearson's~$r$.
}
\end{deluxetable}

\end{document}